\documentclass[aps,showpacs,preprint,epsf,epsfig, nofootinbib]{revtex4-2}
\usepackage{mathrsfs}
\usepackage{dcolumn}
\usepackage{bm}
\usepackage{hyperref}
\usepackage{epsfig,graphicx,color,appendix}
\usepackage{placeins}
\usepackage{multirow}
\usepackage{tabularx}
\usepackage{capt-of}
\usepackage{subcaption}
\usepackage[countmax]{subfloat}

\usepackage{threeparttable}
\usepackage{array}
\usepackage{amsmath}
\usepackage[format=hang]{caption}

\graphicspath{ ./Figures/ }

\usepackage[english]{babel}[2020/02/03]
\makeatletter\AtBeginDocument{\let\@elt\relax}\makeatother
\begin{document}
    \count\footins=1000
	\def\ben{\begin{eqnarray}}
		\def\ben{\end{eqnarray}}
	\def\non{\nonumber}
	\def\obtp{\frac{1^{\prime}}{2}}
	\def\la{\langle}
	\def\ra{\rangle}
	\def\t{\times}
	\def\ve{\varepsilon}
	\def\p{{\prime}}
	\def\pp{{\prime\prime}}
	\def\nc{N_c^{\rm eff}}
	\def\hep{\hat{\varepsilon}}
	\def\J{{J/\psi}}
	\def\ov{\overline}
	
	\def\q2{q^2}
	\def\kb{{\bf k}_\bot}
	\def\pb{{\bf p}_\bot}
	\def\r{\gamma}
	\def\w{\omega} 
	\def\u{\mu}
	\def\v{\nu}
	\def\e{\epsilon}
	\def \d {{\rm d}}
	\def\plus{\texttt{+}}
	\def\minus{\texttt{-}}
	
	\long\def\symbolfootnote[#1]#2{\begingroup%
		\def\thefootnote{\fnsymbol{footnote}}\footnote[#1]{#2}\endgroup}
	\def\lsim{ {\ \lower-1.2pt\vbox{\hbox{\rlap{$<$}\lower5pt\vbox{\hbox{$\sim$}
			}}}\ } }
	\def\gsim{ {\ \lower-1.2pt\vbox{\hbox{\rlap{$>$}\lower5pt\vbox{\hbox{$\sim$}
			}}}\ } }
	
	\font\el=cmbx10 scaled \magstep2{\obeylines\hfill \today}
	\vskip 1.5 cm
	
	\centerline{\large\bf Weak decays of $\bm{B_c}$ involving vector mesons in self-consistent }
	\centerline{\large\bf covariant light-front approach}
	
	\small
	\vskip 1.0 cm
	
	\centerline{\bf Thejus Mary S.$^{1}$\symbolfootnote[1]{\href{mailto:thejusmarys@gmail.com}{thejusmarys@gmail.com}}, Avijit Hazra$^{1}$\symbolfootnote[2]{\href{mailto:hazra_avijit@outlook.com}{hazra\_avijit@outlook.com}}, Neelesh Sharma$^{2}$\symbolfootnote[3]{\href{mailto:nishu.vats@gmail.com}{nishu.vats@gmail.com}}, and Rohit Dhir$^{1}$\symbolfootnote[4]{Corresponding author: \href{mailto:dhir.rohit@gmail.com}{dhir.rohit@gmail.com}}}
	
	\centerline{\it $^{1}$Department of Physics and Nanotechnology,}
	\centerline{\it SRM Institute of Science and Technology, Kattankulathur 603203, India.}
	\centerline{\it $^{2}$Paradigm of Science Cultivation and Ingenious, Kangra 176032, India.}
	
    \bigskip
    \begin{center}{\large \bf Abstract}\end{center}
    We present a comprehensive analysis of weak transition form factors, semileptonic decays, and nonleptonic decays of $B_c$ meson involving pseudoscalar ($P$) and vector ($V$) meson for bottom-conserving and bottom-changing decay modes. We employ self-consistent covariant light-front quark model (CLFQM), termed as Type-II correspondence, to calculate the $B_c$ to $P(V)$ transition form factors. The Type-II correspondence in the CLF approach gives self-consistent results associated with the $B^{(i)}_j$ functions, which vanish numerically after the replacement $M^{\p(\p\p)} \to M_0^{\p(\p\p)}$ in traditional Type-I correspondence, and the covariance of the matrix elements is also restored. We investigate these effects on bottom-conserving $B_c$ to $P(V)$ form factors that have not yet been studied in CLFQM Type-II correspondence. In addition, we quantify the implications of self-consistency propagating to weak decays involving both bottom-conserving and bottom-changing $B_c$ transition form factors. We use two different parameterizations, the usual three-parameter function of $\q2$ and the model-independent $z$-series expansion, to establish a clear understanding of $\q2$ dependence. Using the numerical values of the form factors, we predict the branching ratios and other physical observables, such as forward-backward asymmetries, polarization fractions, etc., of the semileptonic $B_c$ decays. Subsequently, we predict the branching ratios of two-body nonleptonic weak decays using the factorization hypothesis in self-consistent CLFQM. We also compare our results with those of other theoretical studies.
	
    \newpage 	
    \section {Introduction} \label{S1}
    The $B_c$ meson is a quark-antiquark bound-state composed of two heavy quarks ($b \text{~and~} c$) with distinct flavors that decay solely via weak interactions~\cite{Gershtein:1994jw}. The study of $B_c$ meson decays provides valuable insights into the fundamental aspects of the Standard Model (SM) and offers a unique platform to explore the underlying heavy flavor dynamics, which is of immense experimental and theoretical significance. A peculiarity of $B_c$ decays, compared to $B$ and $B_s$ decays, is that both constituent quarks are involved in weak decays, \textit{i.e.}, $b$ quark decays with ${c}$ quark as spectator, and $c$ quark transitions with spectator $b$ quark, in addition to weak annihilation of constituent quarks. The weak annihilation processes decay to leptons or lighter mesons that are relatively suppressed and are, therefore, ignored in the current analysis. The phase space available for $c$ quark decays is significantly smaller compared to $b$ quark decays, but the Cabibbo-Kobayashi-Maskawa (CKM) matrix elements strongly favor $c$ quark decays~\cite{Gershtein:1994jw, Gouz:2002kk}. The study of heavy flavor weak decays is a powerful tool to test SM and search for new physics (NP) beyond SM. The semileptonic decays are governed by tree-level processes in the SM, which provides a relatively simple theoretical description to capture the effects of the weak interaction in terms of Lorentz invariant form factors. In addition, these decays are of immense importance for extracting the CKM matrix elements (and their phases), and studying lepton flavor universality (LFU). On the other hand, the study of two-body weak decays of $B_c$ mesons offers an excellent opportunity to explore quantum chromodynamics (QCD) in both perturbative and nonperturbative regimes to understand the interplay of strong and electroweak interactions. Additionally, these decays allow for testing QCD-motivated effective theories and models within and beyond the SM.
 
    Currently, modern experimental collaborations such as LHCb, CMS, ATLAS, and CDF have been exploring the $B_c$ meson to provide valuable insights into heavy flavor physics in the SM and NP. The Large Hadron Collider (LHC) and Relativistic Heavy Ion Collider (RHIC) are expected to produce a sizable number of $B_c$ meson events (about $10^{6}$) via the proton-nucleus and nucleus-nucleus collision modes~\cite{Chen:2018obq}. Therefore, in the near future, it would be possible to study the $B_c$ meson properties by using more collision modes other than the usually considered proton-proton collision mode. In the recent past, the LHCb has reported precise measurement of $B_c$ meson mass and lifetime as $M_{B_c}=(6274.47 \pm 0.27 \pm 0.17)$ MeV and $\tau_{B_c} = (0.5134 \pm 0.011 \pm 0.0057)$ ps, respectively~\cite{LHCb:2020ayi, LHCb:2014glo}. Although the spectroscopy and decays of $B_c$ meson are being probed extensively, their experimental observations and measurements are scarce~\cite{LHCb:2019bem, CMS:2019uhm, ATLAS:2014lga}. So far, the LHCb has reported the observation of two-body nonleptonic $B_c^+ \to B_s^0 \pi^+$ decay~\cite{LHCb:2013xlg} and their experimental efforts have resulted in the observation of $B_c$ decays involving two charm mesons, such as, $B_c^+ \to D_{(s)}^{(*)+}\ov{D}^{(*)0}$ and $B_c^+ \to D_{(s)}^{(*)+}{D}^{(*)0}$~\cite{LHCb:2021azb, LHCb:2017ogk, Workman:2022ynf}. Recently, LHCb and ATLAS reported the ratios of branching fractions of two-body nonleptonic $B_c$ decays involving a $J/\psi$ meson in the final state, \textit{i.e.}, $\frac{\mathcal{B}(B_c^+ \to J/\psi D_s^{(*)+})}{\mathcal{B}(B_c^+ \to J/\psi \pi^+)}$, $\frac{\mathcal{B}(B_c^+ \to J/\psi D_s^{*+})}{\mathcal{B}(B_c^+ \to J/\psi D_s^+)}$, and $\frac{\mathcal{B}(B_c^+ \to J/\psi K^+)}{\mathcal{B}(B_c^+ \to J/\psi \pi^+)}$~\cite{HFLAV:2022esi, ATLAS:2022aiy, LHCb:2016vni, ATLAS:2015jep}. Even though observations exist of a few semileptonic and nonleptonic decays of the $B_c$ meson, more efforts are required for precise experimental measurements. Interestingly, the LHCb collaboration reported the LFU ratio for $J/\psi$ in the final state as $\mathcal{R}_{J/\psi} = 0.71 \pm 0.18 \pm 0.17$~\cite{LHCb:2017vlu}. However, this ratio significantly exceeds the theoretical estimates, including the lattice QCD (LQCD) results~\cite{Harrison:2020nrv}. Such discrepancies between theory and experiment garner significant attention to physics beyond the SM.
	
    The aforementioned, theoretical studies of the semileptonic and nonleptonic decays of heavy flavor $b$-mesons provide valuable insights into the weak interaction and allow us to measure fundamental parameters within the SM. Additionally, they offer information about quark mixing, CP violation, and heavy quark physics. Furthermore, investigations of semileptonic decays are essential not only for precise theoretical predictions but also for probing physics beyond the SM. Therefore, considering the imminent advancements in precision measurements of the $B_c$ meson at hadron colliders and $B$-factories, several theoretical models, such as LQCD~\cite{Harrison:2020gvo, Cooper:2020wnj, Cooper:2021bkt}, QCD sum rules (QCDSR)~\cite{Kiselev:2000pp, Kiselev:2001zb, Kiselev:2002vz}, Bethe-Salpeter (BS) model~\cite{AbdEl-Hady:1998uiq, AbdEl-Hady:1999jux}, covariant light-front quark model (CLFQM)~\cite{Zhang:2023ypl, Wang:2008xt, Wang:2007sxa, Li:2023wgq, Sun:2023iis}, relativistic quark model (RQM)~\cite{Faustov:2022ybm, Ebert:2003wc, Ebert:2003cn}, relativistic constituent quark model (RCQM)~\cite{Ivanov:2006ni, Ivanov:2006ib, Ivanov:2002un}, relativistic independent quark model (RIQM)~\cite{Naimuddin:2012dy, Nayak:2022qaq}, perturbative QCD (pQCD) approach~\cite{Rui:2012qq, Rui:2014tpa}, QCD factorization (QCDF) approach~\cite{Sun:2015exa}, etc., studied the semileptonic and nonleptonic $B_c$ meson decays involving pseudoscalar ($P$) and vector ($V$) mesons. Current theoretical research has predominantly concentrated on the semileptonic weak decays of the $B_c$ meson to ground-state and orbitally excited charmonium states. It is noteworthy that studies examining bottom-conserving and bottom-changing semileptonic decays of $B_c$ that result in $B^*$, $B_s^*$, $D^*$, or $D_s^*$ mesons in the final state (excluding decays to charmonia) remain relatively limited in the literature. Moreover, among these studies, analyses based on the CLFQM are particularly scarce and require re-investigation in light of recent issues pertaining to self-consistency and covariance in some of the involved form factors. Thus, in the present work, we focus on comprehensive investigations into the effects of self-consistency and covariance on bottom-conserving and bottom-changing semileptonic and nonleptonic decays within the CLFQM framework. Our main objectives are twofold: first, to examine the impact of self-consistency on weak semileptonic and nonleptonic decays using modified form factors within a CLFQM approach; second, to establish self-consistency in bottom-conserving transition form factors, which have not yet been explored, and to quantify these effects on bottom-conserving weak decays. Additionally, we address the ambiguities related to the $\q2$ parameterization in our analysis to provide a more robust understanding of these decay processes. 
 
    The CLFQM, apart from providing a relativistic treatment of physical quantities, has several advantages over the traditional light-front quark model (LFQM)~\cite{Jaus:1989au, Jaus:1991cy, Ji:1992yf, Cheng:1996if, Cheng:1997au}. In the traditional LFQM, the Lorentz covariance of the matrix element is violated due to the spurious contributions, and it does not provide any systematic approach to determine the zero-mode contributions~\cite{Cheng:1996if, Cheng:1997au}. Jaus~\cite{Jaus:1999zv} proposed the CLFQM to provide resolution of these ambiguities by using the manifestly covariant BS approach~\cite{Salpeter:1951sz, Salpeter:1952ib}. The CLFQM ensures covariance of the matrix elements by the inclusion of zero-mode contributions, which make the spurious contributions proportional to the light-like four-vector $\omega^\mu = (0, 2, 0_\perp$) irrelevant~\cite{Jaus:1999zv, Choi:1998nf, Cheng:1997au}. Following this, CLFQM has been extensively used to investigate the semileptonic and nonleptonic decays of bottom mesons~\cite{Hazra:2023zno, Zhang:2023ypl, Choi:2021mni, Chang:2020wvs, Chang:2019mmh, Chang:2018zjq, Choi:2013mda, Verma:2011yw, Wang:2008xt, Wang:2007sxa, Choi:2005fj, Bakker:2003up, Cheng:2003sm, Jaus:2002sv, Bakker:2002mt, Bakker:2000pk}.
	
    In this study, we employ the recent advancements in CLFQM, termed as self-consistent CLFQM, to calculate the $B_c$ to $P$ and $V$ meson transition form factors. The $B_c$ meson decays involve $c$ quark transitions, $c \to s(d)$ and $b$ quark transitions, $b \to c(u)$. These quark-level transitions are categorized as bottom-conserving $(\Delta b = 0)$ and bottom-changing $(\Delta b = -1)$ CKM-favored and -suppressed modes (their selection rules are defined in Sec.~\ref{S2} and~\ref{S3}), respectively. It should be noted that the self-consistent CLFQM is termed as Type-II correspondence in CLFQM on account of the challenges associated with Type-I correspondence~\cite{Jaus:1999zv, Choi:2013mda, Cheng:2003sm}. In the traditional Type-I scheme, the CLF predictions for the $P$ to $V$ transition form factors suffer from the self-consistency problem, for example, the results obtained via the longitudinal ($\lambda=0$) and transverse ($\lambda= \pm$) polarization states are different from each other, due to the additional contributions characterized by the coefficients $B_1^{(2)}$ and $B_3^{(3)}$. These additional contributions affect $f(\q2)$ and $a_{-}(\q2)$ form factors only\footnote{The form factors $f(\q2)$ and $a_{-}(\q2)$ can be related to the Bauer-Stech-Wirbel (BSW) form factors $A_1(\q2)$ and $A_0(\q2)$, respectively, and their transformation relations are given in Eq.~\eqref{e17}.}. Moreover, the manifest covariance of the matrix element in CLFQM is also violated within the Type-I scheme due to the residual $\omega$-dependencies associated with $B^{(i)}_{j}$ functions that are independent of zero-mode contributions. Therefore, both these issues originate from the same source, which can be remarkably resolved by incorporating Type-II correspondence~\cite{Chang:2019mmh}. The CLFQM with Type-II correspondence can, however, give self-consistent results because integration over the terms associated with the coefficient $B^{(i)}_{j}$ vanish numerically after the replacement $M^{\p(\p\p)} \to M_{0}^{\p(\p\p)}$ and the covariance of the matrix elements is also restored. It should be noted that Type-II correspondence scheme has been employed to calculate the bottom-changing $B_c \to D_{(s)}^* (J/\psi)$ transition form factors~\cite{Chang:2019mmh}; however, the bottom-conserving $B_c \to B_{(s)}^*$ form factors have not yet been studied. Furthermore, the implications of self-consistency have not been investigated on the decays involving both bottom-conserving as well as bottom-changing $B_c$ transition form factors. It should be emphasized that the study of semileptonic and nonleptonic weak decays is necessary to quantitatively assess the effect of self-consistency on these decays. The self-consistency issues originating from form factors $A_{0}(\q2)$ and $A_1(\q2)$, affects the semileptonic decays of the $B_c$ meson. On the other hand, $B_c \to PV$ decays explicitly involve $A_0(\q2)$ (other than $F_1(\q2)$) form factor and provide an excellent scenario for quantitative analysis of self-consistency issues that are expected to be more serious in these decays. We further investigate the implication of $\q2$ dependence on the $B_c$ to $P(V)$ transition form factors over the available momentum range. In order to establish a clear understanding of $\q2$ dependence, we utilize two different parameterizations, \textit{i.e.}, the usual three-parameter function of $\q2$ influenced by vector meson dominance (VMD) and model-independent $z$-series expansion. Furthermore, we plot these $B_c$ to $P(V)$ transition form factors to analyze their behavior with respect to the available $\q2$ range. Using the numerical values of the form factors, we predict the physical observables, such as branching ratios, forward-backward (FB) asymmetries, polarization fractions, etc., of the semileptonic $B_c$ decays. In addition, we analyze the $\q2$ dependence of these physical observables by plotting them. Later, we extend our analysis to predict the branching ratios of two-body nonleptonic weak decays using the factorization hypothesis in self-consistent CLFQM. In addition, we also compare our results with existing results from other models. 
	
    Our paper is organized as follows. In Sec.~\ref{S2}, we present the methodology for the calculation of form factors and its $\q2$ dependence in self-consistent CLFQM. In addition, we provide the decay rate expressions for semileptonic $B_c$ to $P(V)$ and nonleptonic $B_c$ to $PV$ decays. In Sec.~\ref{S3}, we give the numerical results and detailed discussions of the form factors, as well as decay rates of semileptonic and nonleptonic $B_c$ to $PV$ decays. We summarize and conclude in Sec.~\ref{S4}.
	
    \section{Methodology} \label{S2}
    \subsection {Self-consistent covariant light-front approach} \label{S2_A}
	\begin{figure}[h]
		\centering
		\includegraphics[width=.6\textwidth]{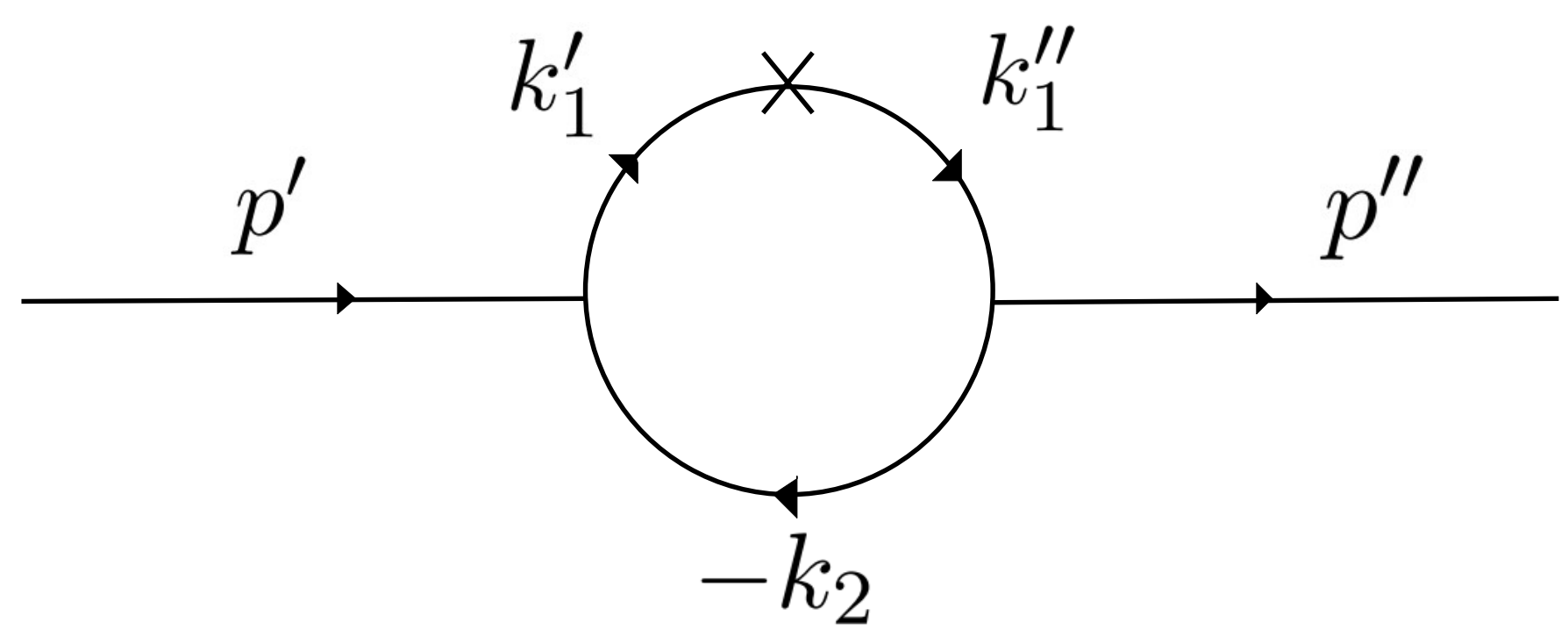}
		\caption{Feynman diagram for meson transition amplitudes, where $\bm{\t}$ denotes the vector or axial vector current vertex}
		\label{f1}
	\end{figure}
    In this work, we focus on the self-consistent CLF approach~\cite{Jaus:1999zv, Cheng:2003sm, Choi:2013mda, Chang:2018zjq, Chang:2019mmh, Chang:2020wvs} and summarize the theoretical framework to calculate the $B_c$ to $P(V)$ form factors. In CLFQM, a meson transition, as shown in Figure~\ref{f1}, is represented in terms of the four momenta of the parent and daughter mesons, \textit{i.e.}, $p^{\p} = k_{1}^{\p} + k_2$ and $p^{\pp} = k_1^{\pp} + k_2$, respectively. Here, $k_1^{\p(\pp)}$ and $k_2$ represent the momenta of the quark and the antiquark of the incoming (outgoing) meson with masses, $m^{\p(\pp)}_1$ and $m_2$, respectively. These momenta can further be expressed using internal variables, namely, momentum fraction ($x_{1(2)}$) and transverse momentum (${\bm{k}}_{\perp}^{\p}$) of the quark, as follows:
	\begin{equation} \label{e1}
		k_{1(2)}^{\p +} = x_{1(2)} p^{\p +},~~ {\bm{k}}_{1(2)\perp}^{\p} = x_{1(2)} {\bm{p}}_{\perp}^{\p} \pm {\bm {k}}_{\perp}^{\p},~ 
	\end{equation}
	where they must satisfy the relation $x_1 + x_2 = 1$. The meson momentum is defined as $p^\p = (p^{\p -}, p^{\p +}, \bm{p}_{\perp}^{\p})$ with $p^{\p \pm} = p^{\p 0} \pm p^{\p 3}$, such that $(p^{\p})^2 = p^{\p +}p^{\p -} - {p}_{\perp}^{\p 2}= M^{\p2}$, where $M^{\p}$ is the mass of the parent meson. The transverse momenta of the quark and meson are given by ${\bm {k}}_{\perp}^{\p} = (k^{\p x},k^{\p y})$ and ${\bm {p}}_{\perp}^{\p} = (p^{\p x},p^{\p y})$, respectively. The definition of the internal quantities for the outgoing meson can be obtained by replacing the prime notation with a double-prime. 
	
    Conventionally, a meson bound-state ($q_1^{\p}, \bar{q}_2$) can be represented as 
	\begin{eqnarray}\label{e2}
		|M(p, ^{2S+1}L_J, J_z)\ra
		=\int\{\d^3\tilde{k}_1\}\{\d^3\tilde{k}_2\} ~2(2\pi)^3 \delta^3 ({\tilde{p}-\tilde{k}_1-\tilde{k}_2})~\nonumber \\
		\t\sum_{h_1,h_2}\Psi^{JJ_z}_{LS}(\tilde k_1,\tilde k_2,h_1,h_2)
		|q_1^{\p}(k_1^{\p},h_1) \bar q_2(k_2,h_2)\ra,
	\end{eqnarray}
    where $L$ and $J$ are orbital angular and total spin quantum numbers, respectively~\cite{Cheng:2003sm}. Further, $\tilde{p}=(p^{\p +},\bm{p}_{\perp}^{\p})$, and $\tilde{k}_{1,2}=(k_{1,2}^{\p +},\bm{k}^{\p}_{1,2\perp})$ represent the on-mass-shell LF momenta, and $\{\d^3\tilde{k}\} \equiv {1\over 2(2\pi)^3}\d k^{\p +}\d^2\bm{k}_{\perp}^{\p}$. The wave function $\Psi^{JJ_z}_{LS}(\tilde k_1,\tilde k_2,h_1,h_2)$, which describes the distribution of momentum in space for $^{2S+1}L_J$ meson, satisfies the normalization condition
	\begin{eqnarray}\label{e3}
		\sum_{h_1,h_2} \int \frac{\d x_1 \d^2\bm{k}_{\perp}^{\p}}{2(2\pi)^3}|\Psi^{JJ_z}_{LS}(x_1,\bm{k}_\perp^{\p},h_1,h_2)|^2 =1,
	\end{eqnarray}
    and can be written as
	\begin{equation} \label{e4}
		\Psi^{JJ_z}_{LS}(x_1,\bm{k}_\perp^{\p},h_1,h_2)= R^{SS_z}_{h_1h_2}(x_1,\bm{k}_{\perp}^{\p})~ \psi_{LL_z}(x_1, \bm{k}_{\perp}^{\p}).
	\end{equation}
    The radial wave function $\psi_{LL_z}(x_1,\bm{k}_\perp^{\p})$ characterizes how the constituent quarks' momenta are distributed in a bound-state that possesses orbital angular momentum $L$~\cite{Cheng:2003sm}. The spin-orbital LF wave function ($R^{SS_Z}_{h_1h_2}$) represents the definite spin state ($S, S_Z$) corresponding to the LF helicity ($h_1,h_2$) eigenstates. Additional details for the treatment of spin, polarization, and complete normalization procedure are discussed in Refs.~\cite{Jaus:1989au, Cheng:1996if, Cheng:2003sm}. A suitable choice for the radial wave function is the phenomenological Gaussian-type wave function, \textit{i.e.},
	\begin{equation} \label{e5}
		\psi(x_1,{\bm{k}}_{\perp}^{\p}) = 4 \frac{\pi^{\frac{3}{4}}}{\beta^{\frac{3}{2}}} \sqrt{\frac{\partial k_z^{\p}}{\partial x_1}} {\rm exp} \Big[- \frac{k_{z}^{\p2} + {k}_{\perp}^{\p2}}{2 \beta^2}\Big],
	\end{equation} 
    for $s$-wave mesons~\cite{Jaus:1989au}. The shape parameter (also called Gaussian parameter), $\beta$, in Eq.~\eqref{e5}, describes the momentum distribution and is expected to be of the order $\Lambda_{QCD}$~\cite{Verma:2011yw}. The relative momentum $k_z^{\p}$ (in the $z$-direction) is given by
	\begin{equation} \label{e6}
		k_z^{\p} = \Big(x_1 - \frac{1}{2}\Big)M_0^{\p} + \frac{m_{2}^{2} - m_{1}^{\p2}}{2M_0^{\p}},
	\end{equation} 
    which yields~\cite{Choi:2013mda} 
	\begin{equation} \label{e7}
		\frac{\partial k_z^{\p}}{\partial x_1} = \frac{M_0^{\p}}{4x_1 (1-x_1)}\Big\{1-\Big[\frac{m_1^{\p2} - m_2^2}{M_0^{\p2}}\Big]^2\Big\},
	\end{equation}
    where,
	\begin{equation} \label{e8}
		M_{0}^{\p} = \sqrt{\frac{m_{1}^{\p2} + {k}_{\perp}^{\p2}}{x_1} + \frac{m_{2}^{2} + {k}_{\perp}^{\p2}}{x_2}},
	\end{equation} 
    is the kinetic invariant mass of the incoming meson. In addition, the kinetic invariant mass of the outgoing meson is denoted as
	\begin{equation} \label{e9}
		M_{0}^{\p\p} = \sqrt{\frac{m_{1}^{\p\p2} + {k}_{\perp}^{\p\p2}}{x_1} + \frac{m_{2}^{2} + {k}_{\perp}^{\p\p2}}{x_2}},
	\end{equation}
    with ${\bm{k}}_{\perp}^{\pp} = {\bm{k}}_{\perp}^{\p} -x_2{\bm{q}}_{\perp}$. The detailed formalism for the CLFQM is described in Refs.~\cite{Jaus:1999zv, Jaus:2002sv, Cheng:2003sm, Choi:2013mda, Chang:2018zjq, Chang:2019mmh, Chang:2020wvs}.
	
    In general, the transition form factors $B_c \to M^{\p\p}$ (where $M^{\p\p} = P,~V$), corresponding to the Feynman diagram of Figure~\ref{f1}, are obtained from explicit expressions for matrix elements of currents between meson states~\cite{Jaus:1989au}, 
	\begin{align} \label{e10}
		{\cal B} ~\equiv~ < M^{\p\p}(p^{\p\p}) | V_{\mu}-A_{\mu}|B_c(p^{\p})>,
	\end{align} 
    where $V_{\mu}$ and $A_{\mu}$ are the vector and axial vector ($A$) currents, respectively. The	form factors for $B_c$ meson to $P$ and $V$ transitions are defined by the following matrix elements~\cite{Cheng:2003sm},
	\begin{align} \label{e11}
		<P(p^{\pp})|V_{\mu}|B_c(p^{\p})> & = p_{\mu}f_{+}(q^{2}) + q_{\mu}f_{-}(q^{2}),\\
		\label{e12}
		<V(p^{\pp}, \varepsilon^{\pp})|V_{\mu}|B_c(p^{\p})> & = \epsilon_{\mu \nu \alpha \beta}\varepsilon^{\pp*\nu}p^{\alpha}q^{\beta}g(q^{2}),\\
		\label{e13}
		<V(p^{\pp}, \varepsilon^{\pp})|A_{\mu}|B_c(p^{\p})> & = -i\{\varepsilon^{\pp*}_{\mu}f(q^{2}) + \varepsilon^{\pp*}\cdot \bm{p}[p_{\mu}a_{+}(q^{2}) + q_{\mu}a_{-}(q^{2})]\},
	\end{align}
    where, $p_\mu = p^{\p} + p^{\pp}$ and $q_{\mu} = p^{\p} - p^{\pp}$. The polarization of the outgoing vector meson is denoted by $\varepsilon_{\mu}$ and the convention $\epsilon_{0123} = 1$ is adopted. The matrix element expressions, Eqs.~\eqref{e11}-\eqref{e13}, are conventionally represented in terms of the BSW~\cite{Wirbel:1985ji} form factors as,
	\begin{align} \label{e14}
		<P(p^{\pp})|V_{\mu}|B_c(p^{\p})> & = (p_{\mu} - \frac{M_{B_c}^{2}-M_{P}^{2}}{q^{2}}q_{\mu})F^{B_cP}_{1}(q^{2}) + \frac{M_{B_c}^{2}-M_{P}^{2}}{q^{2}}q_{\mu}F^{B_cP}_{0}(q^{2}),\\
		\label{e15}
		<V(p^{\pp}, \varepsilon^{\pp})|V_{\mu}|B_c(p^{\p})> & = -\frac{1}{M_{B_c}+M_{V}}\epsilon_{\mu \nu \alpha \beta}\varepsilon^{\pp*\nu}p^{\alpha}q^{\beta}V^{B_cV}(q^{2}), \\
		\label{e16}
		<V(p^{\pp}, \varepsilon^{\pp})|A_{\mu}|B_c(p^{\p})> & = i\{(M_{B_c} + M_{V})\varepsilon_{\mu}^{\pp*}A^{B_cV}_{1}(q^{2}) - \frac{\varepsilon^{\pp*} \cdot \bm{p}}{M_{B_c} + M_{V}}p_{\mu}A^{B_cV}_{2}(q^{2}) \nonumber\\
		& - 2M_{V}\frac{\varepsilon^{\pp*} \cdot \bm{p}}{q^{2}}q_{\mu}[A^{B_cV}_{3}(q^{2})- A^{B_cV}_{0}(q^{2})]\},
	\end{align}
    where the meson masses are denoted by $M_{B_c}$ and $M_{P(V)}$. The BSW-type form factors can be related to the CLFQM form factors as~\cite{Cheng:2003sm},
	\begin{equation} \label{e17}
		\begin{gathered}
			F^{B_cP}_{1}(q^{2}) = f_{+}(q^{2}),\hspace{0.5cm}
			F^{B_cP}_{0}(q^{2}) = f_{+}(q^{2}) + \frac{q^{2}}{q\cdot p}f_{-}(q^{2}),\\
			V^{B_cV}(q^{2}) = -(M_{B_c} +M_{V})g(q^{2}),\hspace{0.5cm}
			A^{B_cV}_{1}(q^{2}) = -\frac{f(q^{2})}{M_{B_c} +M_{V}},\\
			A^{B_cV}_{2}(q^{2}) = (M_{B_c} +M_{V})a_{+}(q^{2}),\hspace{0.5cm}
			A^{B_cV}_{3}(q^{2}) - A^{B_cV}_{0}(q^{2}) = \frac{q^{2}}{2M_{V}}a_{-}(q^{2}),
		\end{gathered}
	\end{equation}
    with
	\begin{equation} \label{e18}
		\begin{gathered}
			F^{B_cP}_{1}(0) = F^{B_cP}_{0}(0), \\
			A^{B_cV}_{3}(0) = A^{B_cV}_{0}(0), ~~\text{and}\\
			A^{B_cV}_{3}(q^{2}) = \frac{M_{B_c} +M_{V}}{2 M_{V}}A^{B_cV}_{1}(q^{2}) - \frac{M_{B_c} -M_{V}}{2 M_{V}}A^{B_cV}_{2}(q^{2}).
		\end{gathered}
	\end{equation}
	
    In contrast to LFQM, the quark and antiquark within a meson system are off-shell in CLFQM. As mentioned before, the CLFQM provides a systematic way to handle zero-mode contributions. The light-front matrix element obtained in CLFQM receives additional spurious contributions proportional to the light-like vector $\omega^\mu = (0, 2, 0_\perp$) which violates the covariance~\cite{Jaus:1999zv}. However, these spurious contributions are canceled out by the addition of zero-mode contributions, restoring the covariance of current matrix elements in CLFQM. Thus, allowing the calculation of physical quantities in terms of manifestly covariant Feynman momentum loop-integrals. Customarily, for the $B_c (p^{\p})\to M^{\p\p}(p^{\p\p})$ transition, it is convenient to use the Drell–Yan–West frame, $q^+ = 0$, which implies that the form factors are known only for space-like momentum transfer, $\q2=-{q}_{\perp}^2 \leq 0$, and for the time-like region ($\q2=-{q}_{\perp}^2 \geq 0$), an additional $\q2$ extrapolation is needed. Furthermore, we consider a Lorentz frame in which $\bm{p}_{\perp}^{\p}=0$ and $\bm{p}_{\perp}^{\p\p}=-\bm{q}_{\perp}$ leads to $\bm{k}_{\perp}^{\p\p}=\bm{k}_{\perp}^{\p}-x_2\bm{q}_{\perp}$~\cite{Hazra:2023zno}. Note that $\q2 = \q2_{max} = (M_{B_c} -M_{P(V)})^2$ corresponds to zero-recoil of the final meson in the initial meson rest frame and the $\q2 = 0$ indicates the maximum recoil of the final meson~\cite{Choi:2021mni}. Following the CLF approach~\cite{Jaus:1999zv, Chang:2019mmh, Chang:2020wvs}, the form factors in Eqs.~\eqref{e11},~\eqref{e12}, and~\eqref{e13} can be extracted from one-loop approximation as a momentum integral given by
	\begin{eqnarray} \label{e19}
		{\cal B}=N_c \int \frac{\d^4 k_1^{\p}}{(2\pi)^4} \frac{H_{M^{\p}}H_{M^{\p\p}}}{N_1^{\p}\,N_1^{\p\p}\,N_2}iS_{\cal B}\,,
	\end{eqnarray} 
    where $N_c$ denotes the number of colors, $\d^4 k_1^{\p}=\frac{1}{2} \d k_1^{\p-} \d k_1^{\p+} \d^2 \bm{k}_{\perp}^{\p}$, and $H_{M^{\p(\p\p)}}$ is the bound-state vertex functions. The terms $N_1^{\p(\p\p)}=k_{1}^{\p(\p\p)2} - m_{1}^{\p(\p\p)2}+i\varepsilon$ and $N_2=k_{2}^{2} - m_{2}^{2}+i\varepsilon$, arise from the quark propagators, and the trace $S_{\cal B}$ can be directly obtained by using the Lorentz contraction,
	\begin{eqnarray}
		\label{e20}
		S_{\cal B} = \text{Tr}[\Gamma (\not{\!k_1^{\p}} + m_1^{\p})(i\Gamma_{M^\p})(-\not{\!k_2} + m_2)(i\gamma^0 \Gamma^\dagger_{M^{\p\p}}\gamma^0)(\not{\!k_1^{\p\p}} + m_1^{\p\p})]\,,
	\end{eqnarray}
    where the vertex operator $\Gamma_{M^{\p(\p\p)}}$ corresponds to the relevant meson, and have the forms
	\begin{equation} \label{e21}
		i\Gamma_{P} = -i\gamma_5 ~\text{and}~~~ i\Gamma_{V} = i\Big[\gamma^{\mu}- \frac{(k_1 -k_2)^{\mu}}{D_{V,con}}\Big],
	\end{equation}
    for $P$ and $V$ mesons, respectively~\cite{Chang:2019mmh}.
	
    The method proposed by Jaus~\cite{Jaus:1999zv} would be most effective if vertex functions could be determined by solving the QCD bound state equation. However, in practice, phenomenological vertex functions similar to those in the conventional light-front model are often employed. The covariant approach represents hadronic matrix elements of one-body currents as one-loop diagrams, evaluable using standard space-time formalism. This yields a covariant matrix element expressed as a Feynman momentum loop integral. Alternatively, light-front matrix elements can be obtained through light-front decomposition of the loop momentum and integration over the minus component ($k_1^{\p-}$) using contour methods~\cite{Jaus:1999zv}. This integration technique requires vertex functions free of singularities, with only quark propagator singularities contributing within the contour. A class of covariant meson vertex functions exhibits this property, characterized by asymmetry in the constituent quark-antiquark pair variables. The integration over the negative component of loop momentum defines the corresponding light-front vertex functions. This approach eliminates the spurious contributions that are proportional to the vector $\omega^\mu=(0, 2, 0_{\bot})$. Consequently, transforming the covariant BS approach to the standard LFQM necessitates a light-front decomposition of the loop momentum and integration over its minus component. This transformation entails the following replacements:
	\begin{equation} \label{e22}
		N_1^{\p(\p\p)} \to \hat{N}_1^{\p(\p\p)} = x_1(M^{\p(\p\p)2}-M_0^{\p(\p\p)2}),
	\end{equation}
    and
	\begin{equation} \label{e23}
		\chi_{M^{\p(\p\p)}} = \frac{H_M^{\p(\p\p)}}{N_1^{\p(\p\p)}} \to \frac{h_M^{\p(\p\p)}}{\hat{N}_1^{\p(\p\p)}},~~~~~~~~
		D_{V,con}^{\p(\p\p)} \to D_{V,LF}^{\p(\p\p)}~,~~~~~~~~~\text{(Type-I)}
	\end{equation}
    where the $D$ factor $D_{V,con}^{\p(\p\p)} = M^{\p(\p\p)} + m_1^{\p(\p\p)} + m_2$ present in the vertex operator are substituted with $D_{V,LF}^{\p(\p\p)} = M_0^{\p(\p\p)} + m_1^{\p(\p\p)} + m_2$~\cite{Choi:2013mda, Chang:2018zjq}. The LF forms of vertex functions, $h_{M^{\p}}$ for $P$ and $V$ mesons are given by
	\begin{equation} \label{e24}
		\frac{h_{P}}{\hat{N}_1^{\p(\p\p)}} = \frac{h_{V}}{\hat{N}_1^{\p(\p\p)}} =\frac{1}{\sqrt{2 N_c}} \sqrt{\frac{x_2}{x_1}}\frac{\psi(x_1, {\bm{k}}_{\perp}^{\p(\p\p)})}{\hat{M}_0^{\p(\p\p)}},
	\end{equation}
    where $\hat{M}_0^{\p(\p\p)} \equiv \sqrt{M_0^{\p(\p\p)}-(m_1^{\p(\p\p)}-m_2)^2}$. It should be noted that there are some debates regarding the self-consistency of the CLFQM~\cite{Cheng:2003sm, Choi:2013mda, Chang:2018zjq}. The explicit validity of replacing $D_{V, con}^{\p(\p\p)}$ with $D_{V, LF}^{\p(\p\p)}$ leads to inconsistency issues in Type-I correspondence. Qin Chang \textit{et al.}~\cite{Chang:2019mmh} found that the resulting $P \to V$ form factors extracted with the longitudinal ($\lambda=0$) and transverse ($\lambda= \pm$) polarization states are not consistent with each other. This is because the $P \to V$ form factors obtained from the longitudinal polarization state receive an additional contribution characterized by the coefficients $B_1^{(2)}$ and $B_3^{(3)}$, which is noticeable in the $B_c$ to $V$ form factor expressions of $f(\q2)$ and $a_{-}(\q2)$. Furthermore, the manifest covariance of the matrix element in CLFQM is also violated in the Type-I correspondence scheme because of the residual $\omega$-dependencies associated with $B^{(i)}_{j}$ functions, which are independent of zero-mode contributions. Therefore, a proposed solution to address these inconsistencies observed in the Type-I CLF form factors is to modify the relationship between the manifestly covariant BS approach and the standard LFQM~\cite{Jaus:1999zv, Cheng:2003sm, Choi:2013mda}. In regard to this, Choi and Ji~\cite{Choi:2013mda} suggested the replacement of $M^{\p(\p\p)}$ with kinetic invariant mass $M_{0}^{\p(\p\p)}$ in every term that contains $M^{\p(\p\p)}$ within the integrand, in addition to the $D$ factor. As a result, the correspondence given by Eq.~\eqref{e23} can be generalized to
	\begin{equation} \label{e25}
		\chi_{M^{\p(\p\p)}} = \frac{H_M^{\p(\p\p)}}{N_1^{\p(\p\p)}} \to \frac{h_M^{\p(\p\p)}}{\hat{N}_1^{\p(\p\p)}},~~~~~~~~
		M^{\p(\p\p)} \to M_0^{\p(\p\p)}.~~~~~~~~~\text{(Type-II)}
	\end{equation}
    Thus, by employing Type-II correspondence from Eq.~\eqref{e25}, the matrix element ${\cal B}$ in Eq.~\eqref{e19} shall reduce to the LF form,
	\begin{equation}
		\label{e26}
		\hat{\cal B}=N_c \int \frac{\d x_1 \d^2{\bm{k}_{\perp}^{\p}}}{2(2 \pi)^3} \frac{h_{M^{\p}}h_{M^{\p\p}}}{x_2\hat{N}_1^{\p} \hat{N}_1^{\p\p}}\hat{S}_{\cal B}.
	\end{equation} 
    Essentially, by embracing the Type-II correspondence described by Eq.~\eqref{e25}, the manifest covariance of the CLFQM can be restored, which in turn should yield numerically equal form factors for $\lambda=0$ and $\lambda=\pm$ polarization states. Therefore, it can be inferred that Type-II correspondence offers a potentially self-consistent framework that resolves the issues connected to the covariance of the matrix elements and the inconsistencies.
	
    The determination of transition form factors for the $B_c$ to ground-state $s$-wave meson for $\q2 = -{q}_{\perp}^2 \leq 0$ is a straightforward process since the calculation of the zero-mode contribution is obtained in a frame where the momentum transfer $q^+$ becomes zero. As a result, the form factors are only known for space-like momentum transfer $\q2 = (p^{\p} - p^{\p\p})^2 =-{q}_{\perp}^2 \leq 0$~\cite{Jaus:1999zv}. Nevertheless, the transition form factors in the time-like region can be obtained through extrapolation, which will be discussed in the following subsection.
	
    Furthermore, the $B_c$ to $P(V)$ transition form factors are explicitly expressed as~\cite{Chang:2019mmh}, 
	\begin{equation} \label{e27}
		F(\q2) = N_c \int \frac{\d x_1 \d^2{\bm{k}_{\perp}^{\p}}}{(2 \pi)^3}\frac{\chi_{B_c}^{\p}\chi_{P(V)}^{\pp}} {2 x_{2}} \widetilde{F}(x_1, {\bm{k}}_{\perp}^{\p}, \q2),
	\end{equation}
	where \begin{align} \label{e28}
		\chi_{B_c}^{\p} = \frac{1}{\sqrt{2 N_c}} \sqrt{\frac{x_2}{x_1}}\frac{\psi(x_1, {\bm{k}}_{\perp}^{\p})}{\hat{M}_0^{\p}},~\text{and} ~~~\chi_{P(V)}^{\pp} = \frac{1}{\sqrt{2 N_c}} \sqrt{\frac{x_2}{x_1}}\frac{\psi(x_1, {\bm{k}}_{\perp}^{\pp})}{\hat{M}_0^{\pp}}.
	\end{align}
    It should be noted that the integration is carried out within the limits of $[0,1]$ and $[0, \infty]$ for $x_1$ and $\bm{k}_{\perp}^{\p}$, respectively, in Eq.~\eqref{e27}. The form factor function $\widetilde{F}(x_1, {\bm{k}}_{\perp}^{\p}, \q2) \equiv \{ \widetilde{f}_{\pm}(x_1, {\bm{k}}_{\perp}^{\p}, \q2),~\widetilde{g}(x_1, {\bm{k}}_{\perp}^{\p}, \q2),\\~\widetilde{f}(x_1, {\bm{k}}_{\perp}^{\p}, \q2), ~\widetilde{a}_{\pm}(x_1, {\bm{k}}_{\perp}^{\p}, \q2)\}$ corresponding to $B_c$ to $P(V)$ transitions are defined as follows: 
	\begin{itemize}
		\item[(i)] $B_c$ to $P$ form factors~\cite{Cheng:2003sm, Chang:2018zjq},
		\begin{align}
			\label{e29}
			\widetilde{f}_{+}(x_1, {\bm{k}}_{\perp}^{\p}, \q2) = x_1 M_0^{\p 2} + x_1 M_0^{\pp 2} +x_2\q2 - x_1 (m_1^{\p} -m_2)^2 - x_1 (m_1^{\pp} -m_2)^2 - x_2 (m_1^{\p} -m_1^{\pp})^2, 
		\end{align}
		\begin{align}
			\label{e30}
			\widetilde{f}_{-}(x_1, {\bm{k}}_{\perp}^{\p}, \q2) = & -2 x_1 x_2 M^{\p 2} - 2 k_{\perp}^{\p 2} - 2 m_1^{\p} m_2 +2 (m_1^{\pp} - m_2)(x_2 m_1^{\p} +x m_2) \nonumber\\& - 2 \frac{{\bm{k}}_{\perp}^{\p} \cdot {\bm{q}}_{\perp}}{\q2}\Big[(x_1 - x_2)M^{\p 2} +M^{\pp 2} + x_2 (\q2 +q\cdot p) +2 x_1 M_0^{\p 2} \nonumber\\&- 2(m_1^{\p} + m_1^{\pp})(m_1^{\p} -m_2)\Big] + a \frac{p\cdot q}{\q2}\Big[k_{\perp}^{\p 2} + \frac{2 ({\bm{k}}_{\perp}^{\p} \cdot {\bm{q}}_{\perp})^2}{\q2}\Big] + 4 \frac{({\bm{k}}_{\perp}^{\p} \cdot {\bm{q}}_{\perp})^2}{\q2}.
		\end{align}
		
		\item[(ii)] $B_c$ to $V$ form factors~\cite{Cheng:2003sm, Chang:2019mmh},
		\begin{align} 
			\label{e31}
			\widetilde{g}(x_1, {\bm{k}}_{\perp}^{\p}, \q2) = -2\Big\{x_2m_1^{\p} + x_1m_2 + (m_1^{\p} - m_1^{\pp})\frac{{\bm{k}}_{\perp}^{\p}\cdot {\bm{q}}_\perp}{\q2} + \frac{2}{D_{V, con}^{\pp}}\Big[{k}_{\perp}^{\p 2} +\frac{({\bm{k}}_{\perp}^{\p} \cdot {\bm{q}}_{\perp})^2}{\q2} \Big] \Big\},	\end{align}
		\begin{align}
			\label{e32}
			\widetilde{f}(x_1, {\bm{k}}_{\perp}^{\p}, \q2) = & -2\Big\{
			-(m_1^{\p}+m_1^{\pp})^2 (m_1^{\p}-m_2) +(x_1m_2-x_2m_1^{\p}) M^{\p 2} +(x_1m_2+x_2m_1^{\p}) M^{\pp 2} \nonumber\\& -x_1(m_2-m_1^{\p} )(M_0^{\p 2}+M_0^{\pp 2}) + 2x_1m_1^{\pp}M_0^{\p 2} -4 (m_1^{\p}-m_2)\Big({k}_{\perp}^{\p 2} +\frac{({\bm{k}}_{\perp}^{\p} \cdot {\bm{q}}_{\perp})^2}{\q2}\Big) \nonumber\\& - m_2 \q2 -(m_1^{\p}+m_1^{\pp})(\q2+q\cdot p)\frac{{\bm{k}}_{\perp}^{\p}\cdot {\bm{q}}_{\perp}}{\q2} +4 (m_1^{\p}-m_2) B_1^{(2)} +\frac{2}{D_{V, con}^{\pp}} \Big[ \Big({k}_{\perp}^{\p 2} \nonumber\\& +\frac{({\bm{k}}_{\perp}^{\p} \cdot {\bm{q}}_{\perp})^2}{\q2}\Big) \Big((x_1-x_2)M^{\p 2} + M^{\pp 2} -2(m_1^{\p}-m_1^{\pp})(m_1^{\p}-m_2) +2x_1M_0^{\p 2} \nonumber\\& -\q2 -2 (\q2 +q\cdot p)\frac{{\bm{k}}_{\perp}^{\p}\cdot {\bm{q}}_{\perp}}{\q2}\Big) - \Big(M^{\p 2} +M^{\pp 2} -\q2 + 2(m_1^{\p}-m_2)(m_1^{\pp} \nonumber\\& +m_2)\Big) B_1^{(2)} +2 B_3^{(3)}\Big]\Big\}, 	\end{align}
		\begin{align}
			\label{e33}
			\widetilde{a}_{+}(x_1, {\bm{k}}_{\perp}^{\p}, \q2) = & 2\Big\{(m_1^{\pp} -2x_1m_1^{\p} +m_1^{\p} +2x_1m_2)\frac{{\bm{k}}_{\perp}^{\p}\cdot {\bm{q}}_{\perp}}{{q}_{\perp}^2} + (x_1-x_2)(x_2m_1^{\p} +x_1m_2) \nonumber\\& +\frac{2}{D_{V, con}^{\pp}} \frac{{\bm{k}}_{\perp}^{\pp}\cdot {\bm{q}}_{\perp}}{x_2 {q}_{\perp}^2} \Big[{\bm{k}}_{\perp}^{\p}\cdot{\bm{k}}_{\perp}^{\pp}+ (x_1m_2 - x_2m_1^{\pp})(x_1m_2+ x_2m_1^{\p})\Big] \Big\},
		\end{align}
		\begin{align}
			\label{e34}
			\widetilde{a}_{-}(x_1, {\bm{k}}_{\perp}^{\p}, \q2) = & -2\Big\{(3-2x_1)(x_2m_1^{\p}+x_1m_2) -\Big[(6x_1-7)m_1^{\p}+(4 -6x_1)m_2 +m_1^{\pp}\Big]\frac{{\bm{k}}_{\perp}^{\p} \cdot {\bm{q}}_{\perp}}{\q2}\nonumber\\& + 4(m_1^{\p} - m_2) \Big[ 2 \Big( \frac{{\bm{k}}_{\perp}^{\p} \cdot {\bm{q}}_{\perp}}{\q2}\Big)^2 + \frac{{k}_{\perp}^{\p 2}}{\q2}\Big] {- 4 \frac{(m_1^{\p} - m_2)}{\q2} B_1^{(2)}} + \frac{1}{D_{V, con}^{\pp}} \Big[-2\big(M^{\p 2} \nonumber\\& +M^{\pp 2} -\q2 + 2(m_1^{\p}-m_2)(m_1^{\pp}+m_2)\big) \big(A_3^{(2)} + A_4^{(2)} - A_2^{(1)} \big) + \Big(2M^{\p 2} -\q2 \nonumber\\& -x_1(M^{\p2} - M_0^{\p2}) +x_1(M^{\pp2} - M_0^{\pp2})- 2(m_1^{\p} -m_2)^2 +(m_1^{\p} +m_1^{\pp})^2 \Big)\big(A_1^{(1)}\nonumber\\& + A_2^{(1)} -1\big) +2Z_2\big(2A_4^{(2)} -3A_2^{(1)} +1\big) + 2 \frac{q\cdot p}{\q2} \big(4 A_2^{(1)}A_1^{(2)} - 3A_1^{(2)}\big) \nonumber\\& +\frac{2}{\q2}\Big(\big(M^{\p 2} + M^{\pp 2} -\q2 + 2(m_1^{\p}-m_2)(m_1^{\pp}+m_2)\big) B_1^{(2)} -2 B_3^{(3)}\Big)\Big] \Big\}.
		\end{align}
	\end{itemize}
    The coefficients $A^{(i)}_{j}$ and $B^{(i)}_{j}$ are given as~\cite{Cheng:2003sm, Chang:2019mmh},
	\begin{equation} \label{e35}
		\begin{gathered}
			A_1^{(1)} = \frac{x_1}{2}, \hspace{0.5cm}
			A_1^{(2)} = -{k}_{\perp}^{\p 2} - \frac{({\bm{k}}_{\perp}^{\p}\cdot{\bm{q}}_{\perp})^2}{\q2}, \hspace{0.5cm}
			A_2^{(1)} = A_1^{(1)} - \frac{{\bm{k}}_{\perp}^{\p}\cdot{\bm{q}}_{\perp}}{\q2}, \\
			A_3^{(2)} = A_1^{(1)}A_2^{(1)}, \hspace{0.5cm}
			A_4^{(2)} = (A_2^{(1)})^2 - \frac{1}{\q2}A_1^{(2)}, \\
			B_1^{(2)} = A_1^{(1)}Z_2 - A_2^{(1)},\hspace{0.5cm}
			B_3^{(3)} = B_1^{(2)}Z_2 + (p^2 -\frac{(q\cdot p)^2}{\q2})A_1^{(1)}A_1^{(2)}, ~\text{ and}
		\end{gathered}
	\end{equation}
	\begin{equation} \label{e36}
		Z_2 = x_1(M^{\p2} - M_0^{\p2}) + m_1^{\p 2} -m_2^2 +(1-2x_1)M^{\p 2} + (\q2 + q\cdot p)\frac{{\bm{k}}_{\perp}^{\p}\cdot{\bm{q}}_{\perp}}{\q2}. 
	\end{equation}
    It should be noted that the above given expressions for the form factors correspond to the traditional Type-I scheme, for which Type-II correspondence can be obtained by an additional replacement of $M^{\p(\p\p)}$ to $M_0^{\p(\p\p)}$~\cite{Chang:2019mmh}. Moreover, the above form factor expressions are for the case of $\lambda = 0$ (\textit{i.e.}, longitudinal polarization state); and the results for the case of $\lambda = \pm$ (\textit{i.e.}, transverse polarization states) can be obtained from these expressions by omitting the terms associated with $B^{(i)}_{j}$ functions. 
	
    \subsection {$\pmb{\q2}$ dependence of the form factors} 
    The numerical evaluation of $B_c$ to $P(V)$ transition form factors requires an understanding of the momentum dependence of these form factors over the entire $\q2$ region in the CLFQM. Conventionally, the meson transition in the Drell-Yan-West frame with $q^+ = 0$ restricts the evaluation of the form factors for the momentum transfer $\q2 = -q_{\perp}^2 \leq 0$, \textit{i.e.}, space-like region~\cite{Jaus:1989au, Jaus:1996np, Jaus:1999zv, Jaus:2002sv}. However, only the form factors in the time-like region ($\q2 = -q_{\perp}^2 \geq 0$) are relevant for physical decay processes~\cite{Jaus:1999zv, Cheng:2003sm}. Therefore, to evaluate the total decay rate of $B_c$ decays, the momentum dependence of the form factors should be reproduced in the space-like region and extrapolated to the time-like region using simplified parameterizations. 
 
    Jaus~\cite{Jaus:1989au} suggested to estimate the invariant form factors as functions of $\q2$, extending them analytically from space-like ($\q2 \leq 0$) to time-like regions ($\q2 \geq 0$)~\cite{Wang:2008xt, Cheng:2003sm, Chang:2019mmh, Jaus:1996np, Choi:2021mni}. This reformulation relies on the assumption that the form factors are continuously differentiable with respect to $\q2$, emphasizing the importance of understanding their behavior near $\q2 = 0$~\cite{Jaus:1989au}. Therefore, understanding wave function overlaps between the initial and final state mesons near $\q2 = 0$ is significant. Furthermore, it has been argued that the form factors obtained directly in the time-like region ($q^+ > 0$) are equivalent to those from analytic continuation from the space-like region~\cite{Bakker:2003up}. A more refined approach to computing form factors at $\q2 > 0$ involves calculations in a frame where the momentum transfer is purely longitudinal ($\bm{q}_{\perp} = 0$), covering the entire range of momentum transfer, as shown in Refs.~\cite{Cheng:2003sm, Bakker:2003up}, and more recently in Ref.~\cite{Tang:2020org}. However, it introduces additional complexity: beyond the conventional valence-quark contribution, one must also consider nonvalence configurations. These include phenomena such as the Z-graph, which arises from quark-pair creation from the vacuum. Consequently, uncertainties arise in transition form factors calculated for $\q2 \geq 0$ (with $\bm{q}_{\perp} = 0$) due to nonvalence configurations~\cite{Li:2019kpr, Tang:2020org}. However, the estimations of these Z-graph contributions are still lacking within the CLFQM formalism. Recent efforts~\cite{Heger:2021gxt} show that the Z-graph contributions to form factors become more significant in the time-like regime $(\q2 > 0)$. In the annihilation process of the emitted quark-antiquark system into a $W$-boson, intermediate vector-meson states dominate. This allows approximation of the Z-graph contributions using a VMD-like decay mechanism~\cite{Heger:2021gxt}. Parameterizing form factors as meromorphic functions of $\q2$, with analytic continuation from $\q2 < 0$ to $\q2 > 0$, is proposed to reasonably describe form factors at time-like momentum transfers. However, considering a frame with purely transverse momentum transfer ($q^+ = 0$) is suggested to reduce nonvalence contributions~\cite{Cheng:2003sm}. In addition, zero-mode contributions affecting these transition form factors are addressed by the Type-II self-consistent CLF approach. 
    
    In continuation of the previous section, it is well-established that the theoretical expressions formulated within the $q^+=0$ frame are specifically applicable for calculating form factors exclusively in the space-like domain. However, to extend our understanding to the time-like region, we require parameterization as explicit functions of $\q2$ to describe the form factors~\cite{Jaus:1996np}. These descriptions of form factors in both space-like and time-like regions complement each other, providing valuable insights into the complete decay dynamics across the entire $\q2$ range. The literature suggests numerous functions of $\q2$ dependence influenced by the VMD approach, which has been used to parameterize and reproduce the transition form factors in space-like region and then extrapolate to physical form factors for $\q2 \geq 0$~\cite{Richman:1995wm, Becirevic:1999kt, Melikhov:1995xz, Melikhov:2000yu}. The conventional form factor dependence on $\q2$ is often expressed as a BSW-type monopole approximation~\cite{Wirbel:1985ji}, $F(\q2) = F(0)/(1-\frac{\q2}{M_{pole}^2})$, based on VMD. However, this approach is not sufficient to explain the experimental observations. Moreover, higher resonance contributions are likely necessary beyond the monopole form. The nearest pole dominance assumption may not always apply because multiple resonances can be significant. Furthermore, given the complexity of nonperturbative physics governing $\q2$ dependence, no single parameterization is universally accurate. A more general approach involves using a simple pole and summing effective poles, though this requires multiple parameters to be determined experimentally~\cite{Hill:2005ju, Hill:2006ub}.
 
    In our analysis, the $\q2$ dependence of form factors in the space-like region can be effectively parametrized and reproduced using a three-parameter form~\cite{Melikhov:2000yu} as follows:
	\begin{equation}
		\label{e37}
		F(\q2) = \frac{F(0)}{\Big(1-\frac{\q2}{M_{pole}^2}\Big)\Big(1-a\frac{\q2}{M_{pole}^2}+b\frac{q^4}{M_{pole}^4}\Big)}~,
	\end{equation}
    where $M_{pole}$ is the transition pole mass. The parameters $a$, $b$, and $F(0)$ are determined by fitting Eq.~\eqref{e37} in the space-like region and extrapolated to the physical region $\q2 \geq 0$. In Type-II correspondence scheme, the numerical results obtained using the parameterization Eq.~\eqref{e37} are referred to as $``$Type-II$"$ throughout the manuscript. Typically, the parameterization presented in Eq.~\eqref{e37} is characterized as a four-parameter fit, wherein the parameters $F(0)$, $a$, $b$, and $M_{pole}$ are ideally determined from the available experimental data. In order to maintain the validity of our calculations and select appropriate quark-model parameters due to the lack of experimental data, we utilize the mass of the nearest pole (listed in Table~\ref{t2}) as the pole mass ($M_{pole}$) to describe it as a three-parameter fit~\cite{Melikhov:2001zv, Melikhov:2000yu}. The parameterization (Eq.~\eqref{e37}) incorporates slope parameters $a$ and $b$ to account for effective poles. These poles deviate from the single resonance typically observed in the $q^{\p}_1 \to q^{\p\p}_1$ transition. In simpler terms, slope parameters represent additional poles beyond the pole mass ($M_{pole}$), reflecting the influence of higher-order resonances~\cite{HFLAV:2022esi}. The phenomenological accuracy and reliability of $\q2$ dependence, given in Eq.~\eqref{e37}, have been extensively discussed in Refs.~\cite{Hill:2005ju, Hill:2006ub, Melikhov:2000yu, Melikhov:2001zv}.
 
    It is worth mentioning that the available $\q2$ range for the bottom-conserving $B_c \to P(V)$ transitions is $0 \leq \q2 \lesssim 1$ GeV${}^2$. However, for bottom-changing transitions, the $\q2$ range is considerably larger, \textit{i.e.}, $0 \leq \q2 \lesssim 20$ GeV${}^2$. Since the $M_{pole}^2$ is greater than available $\q2$ in heavy to heavy meson transitions, the contributing poles lie farther from the kinematic region. Therefore, it is important to accurately determine the $\q2$ dependence in decay amplitudes across the entire kinematic range~\cite{Melikhov:2001zv}. The implementation of the aforementioned parameterization is particularly relevant in bottom-changing decays due to the extensive $\q2$ range, wherein contributions from bottom, bottom-strange, and bottom-charmed resonances may be substantial. This can be explained through the confining interaction, for example, between $b$ and $\bar{u}$ to produce $B$ meson resonances that fluctuate into $W$-boson. In transitions involving significant momentum transfer ($\q2_{max} \simeq 20$ GeV${}^2$), the incorporation of higher-order contributions becomes imperative for accurate modeling of physical decay processes. Form factors spanning such extensive $\q2$ ranges cannot be adequately described by considering only a limited number of initial physical poles~\cite{Becirevic:2014kaa}. Consequently, the poles associated with these form factors are situated at $\q2 = M_{pole}^2$ (as detailed in Table~\ref{t2}), typically at unphysical values of time-like momentum transfer, distinct from $\q2_{max}$. The parameterization outlined in Eq.~\eqref{e37} offers a viable solution for such scenarios. This parameterization (Eq.~\eqref{e37}) is also applicable to $B_c \to B_{(s)}^{(*)}$ transitions. Note that the production threshold for mesons (e.g., $D_{(s)}$ resonances being lightest) from the $c \to d(s)$ current occurs at $\q2$ values where the poles are significantly far from the physical region of $\q2_{max} \simeq 1$ GeV${}^2$. This integration enables a comprehensive exploration of the entire physical momentum transfer range, potentially leading to a significant enhancement in the accuracy of our predictions.
 
    Furthermore, the $\q2$ dependence of the form factors defined by Eq.~\eqref{e37} involves contributions from the said resonances of particular spin in the available $\q2$ range, for example, the form factors $F_1(\q2)$ and $V(\q2)$ exhibit a pole at $\q2 = M_{1^-}^2$, while $A_0(\q2)$ contains a pole at $\q2 = M_{0^-}^2$. It is important to note that the remaining form factors, namely, $F_0(\q2)$, $A_1(\q2)$, and $A_2(\q2)$, do not receive contributions from the lowest-lying negative parity states~\cite{Bauer:1988fx}. The form factor $F_0(\q2)$ include the pole mass corresponding to $0^+$ state, whereas $A_1(\q2)$ and $A_2(\q2)$ incorporates $1^+$ state; interestingly, both have significantly higher masses~\cite{Hill:2005ju, Hill:2006ub, Melikhov:2000yu}, as shown in Table~\ref{t2}. As a result, these form factors are expected to show less variation in the decay region for the available $\q2$.

    Furthermore, it should be noted that for the calculation of transition form factors, several other theoretical studies have employed the following $\q2$ dependence~\cite{Cheng:2003sm, Chang:2019mmh},
	\begin{equation} \label{e40}
		F(\q2) = \frac{F(0)}{1-a\frac{\q2}{M_{B_c}^2}+b\frac{q^4}{M_{B_c}^4}},
	\end{equation}
    where mass of the parent meson $M_{B_c} = 6274.47$~MeV~\cite{Workman:2022ynf} is taken as the pole mass. We use this parameterization in the Type-I correspondence, denoted as ``Type-I'' in numerical results, for the sake of comparison. It is expected that the parameterization presented in Eq.~\eqref{e40} is also valid for the physical decay region~\cite{Melikhov:2001zv}.
	
    Alternatively, many experimental and lattice observations are made using a model-independent parameterization following the general QCD constraints, which is known as $z$-series (expansion) parameterization. The utilization of forms such as Eqs.~\eqref{e37} and~\eqref{e40} for data fitting, while mathematically feasible, presents interpretative challenges due to the absence of clear physical significance for the resulting fit parameters. This ambiguity raises concerns that different experimental (small $\q2$) or lattice (large $\q2$) determinations may not converge to a single value. Therefore, discrepancies arising from fitting different datasets to models like single pole model or modified pole model become ambiguous~\cite{Wiss:2006ih, Poling:2006da}. This issue becomes especially challenging when comparing lattice and experimental data due to differing emphasized ranges of the parameter $\q2$ (usually represented as $t$). To navigate these challenges, it is advisable to use a general parameterization like $z$-series parameterization, which ensures the inclusion of the true form factor. This approach facilitates more robust comparisons of physical quantities, ensuring that the analysis remains grounded in observable phenomena rather than potentially arbitrary fitting parameters~\cite{Hill:2006ub}. 
    
    With the aim to establish a clear understanding of $\q2$ dependence and comparison among different $\q2$ formulations, we also incorporate $z$-series expansion form. Furthermore, the $z$-series parameterization is given in terms of a complex parameter $z$, which is the analytic continuation of $\q2$ into the complex plane~\cite{HFLAV:2022esi}. This parametrization of the form factor is based on the power series expansion around the value $\q2 = t_0$. Thus, the form factor is expressed as~\cite{Gubernari:2022hrq},
	\begin{equation} \label{e38}
		F(\q2) = \frac{1}{1-\frac{\q2}{M_{pole}^2}}\sum_{k=0}^{K} a'_{k}\big[z(\q2)-z(0)\big]^k,
	\end{equation}
	where $a_k$ are real coefficients and $z(\q2) \equiv z(\q2, t_0)$ is the function
	\begin{equation} \label{e39}
		z(\q2) = \frac{\sqrt{t_+ - \q2}- \sqrt{t_+ - t_0}}{\sqrt{t_+ - \q2}+ \sqrt{t_+ - t_0}}~,
	\end{equation}
    which maps the $\q2$-plane cut for $\q2 \geq t_+$ onto the disk $|z(\q2, t_0)| < 1$ in the $z$-complex plane, such that $|z(t_+, t_0)| = -1$ and $|z(\infty, t_0)| = 1$. The arbitrary parameter $t_0 < t_+$ determines the point $\q2$ mapped onto the origin in the $z$-plane, \textit{i.e.}, $|z(t_0, t_0)| = 0$ corresponding to $\q2 = t_0$, and the physical region extends in either direction up to $\pm |z|_{max}$~\cite{Bourrely:2008za}.	The parameters $t_+$ and $t_0$ are $(M_{B_c} + M_{P(V)})^2$ and $(M_{B_c} + M_{P(V)})(\sqrt{M_{B_c}} - \sqrt{M_{P(V)}})^2$, respectively~\cite{Bourrely:2008za, Gubernari:2022hrq}. In comparison to other phenomenological approaches, the fitted coefficients $a'_k$ have no physical interpretation~\cite{HFLAV:2022esi}. Since the higher order terms in the $z$-series parameterization given in Eq.~\eqref{e38} have trivial contributions, we restrict ourselves to the power $K=2$, which contains the free parameters $a'_0~(\approx F(0))$, $a'_1$, and $a'_2$. Unlike Eq.~\eqref{e37}, the numerical results corresponding to parameterization by Eq.~\eqref{e38} in the Type-II correspondence are designated as $``$Type-II*$"$.

    \subsection{Semileptonic decay widths and other physical observables} \label{S2_C}	
    The differential decay width of $B_c$ to $P(V)$ semileptonic decays is expressed in terms of the helicity components as~\cite{Faustov:2022ybm, Korner:1989qb},
	\begin{equation} \label{e41}
		\frac{d \Gamma(B_c^{+} \to P(V)l^+ \nu_l)}{d\q2} = \frac{G_F^2}{(2\pi)^3}|V_{q_1q_2}|^2 \frac{\q2\sqrt{\lambda}}{24 M_{B_{c}}^3} (1-\frac{m_l^2}{\q2})^2 \mathcal{H}_{\rm total},
	\end{equation}
    where $G_F$ is the Fermi constant and $V_{q_1 q_2}$ is the relevant CKM matrix element for $q_1 \to q_2$ transition. The term $\lambda \equiv \lambda (M_{B_c}^2,M_{P(V)}^2,\q2) = (M_{B_c}^2 + M_{P(V)}^2 + \q2)^2 - 4M_{B_c}^2M_{P(V)}^2$ is the K\"all\'en function, and $m_l$ is the lepton mass ($l = e,~\mu,~\tau$). The total helicity structure, $\mathcal{H}_{\rm total}$, is given by,
	\begin{equation} \label{e42}
		\mathcal{H}_{\rm total} = (\mathcal{H}_U + \mathcal{H}_L)(1+\frac{m_l^2}{2\q2})+\frac{3m_l^2}{2\q2}\mathcal{H}_S,
	\end{equation}
    and the helicity components $\mathcal{H}_U$, $\mathcal{H}_L$, and $\mathcal{H}_S$ can be defined as,
	\begin{equation}\label{e43}
		\begin{gathered}
			\mathcal{H}_U = |{H}_+|^2 + |{H}_-|^2, \hspace{0.5cm}
			\mathcal{H}_L = |{H}_0|^2,~~\text{ and }~~ \mathcal{H}_S = |{H}_t|^2,
		\end{gathered}
	\end{equation}
    where ${H}_{\pm}$, ${H}_{0}$, and ${H}_{t}$ are the helicity amplitudes. These helicity amplitudes are related to the corresponding invariant form factors by the following relations:
	\begin{itemize}
		\item[(i)] For $B_c$ to $P$ meson transitions,
		\begin{equation}\label{e44}
			\begin{gathered}
				{H}_{\pm}(\q2) = 0, \hspace{0.5cm}
				{H}_{0}(\q2) = \frac{\sqrt{\lambda}}{\sqrt{\q2}}F_1(\q2),~~\text{ and }~~ {H}_{t}(\q2) = \frac{1}{\sqrt{\q2}}(M_{B_c}^2-M_P^2)F_0(\q2).
			\end{gathered}
		\end{equation}
		
		\item[(ii)] For $B_c$ to $V$ meson transitions,
		\begin{align}
			\label{e45}
			{H}_{\pm}(\q2) = & (M_{B_c} + M_V) A_1(\q2) \mp \frac{\sqrt{\lambda}}{M_{B_c} + M_V}V(\q2), \\
			\label{e46}
			{H}_{0}(\q2) = &\frac{1}{2 M_V \sqrt{\q2}}(M_{B_c} + M_V)(M_{B_c}^2 - M_V^2 -\q2) A_1(\q2) - \frac{\lambda}{M_{B_c} + M_V}A_2(\q2), \\
			\label{e47}
			{H}_{t}(\q2) = & \frac{\sqrt{\lambda}}{\sqrt{\q2}}A_0(\q2).
		\end{align}
	\end{itemize}
	
    Following Eq.~\eqref{e41} the longitudinal and transverse differential decay widths are given by 
	\begin{align} 
		\label{e48}
		\frac{d \Gamma_{L}(B_c^{+} \to Vl^+ \nu_l)}{d\q2} = & \frac{G_F^2}{(2\pi)^3}|V_{q_1q_2}|^2 \frac{\q2\sqrt{\lambda}}{24 M_{B_{c}}^3} (1-\frac{m_l^2}{\q2})^2 [\mathcal{H}_L(1+\frac{m_l^2}{2\q2})+\frac{3m_l^2}{2\q2}\mathcal{H}_S], \text{ and} \\
		\label{e49}
		\frac{d \Gamma_{T}(B_c^{+} \to Vl^+ \nu_l)}{d\q2} = & \frac{G_F^2}{(2\pi)^3}|V_{q_1q_2}|^2 \frac{\q2\sqrt{\lambda}}{24 M_{B_{c}}^3} (1-\frac{m_l^2}{\q2})^2 [\mathcal{H}_U(1+\frac{m_l^2}{2\q2})],
	\end{align} 
	respectively.
	
    In order to gain a deeper understanding of semileptonic decays beyond just the branching ratios, it is valuable to investigate the influence of the lepton mass. Moreover, by defining additional physical observables that are experimentally measurable, we can obtain a more comprehensive and intricate depiction of the underlying physics in these decays. Some of these physical observables are FB asymmetry ($A_{\rm FB}(\q2)$), leptonic convexity parameter ($C_F^l(\q2)$), longitudinal (transverse) ($P_{L(T)}^l(\q2)$) polarization of the charged lepton, and asymmetry parameter ($\alpha^*(\q2)$). These observables can be expressed by the above helicity structure functions as~\cite{Faustov:2022ybm, Ivanov:2005fd}
	\begin{align} 
		\label{e50}
		A_{FB}(\q2) = & \frac{3}{4}\frac{\mathcal{H}_P - 2\frac{m_l^2}{\q2} \mathcal{H}_{SL}}{\mathcal{H}_{\rm total}}, \\
		\label{e51}
		C_{F}^l(\q2) = & \frac{3}{4}(1- \frac{m_l^2}{\q2})\frac{\mathcal{H}_U - 2\mathcal{H}_{L}}{\mathcal{H}_{\rm total}}, \\
		\label{e52}
		P_{L}^l(\q2) = & \frac{(\mathcal{H}_U + \mathcal{H}_L)(1-\frac{m_l^2}{2\q2})-\frac{3m_l^2}{2\q2}\mathcal{H}_S}{\mathcal{H}_{\rm total}}, \\
		\label{e53}
		P_{T}^l(\q2) = & -\frac{3 \pi m_l}{8 \sqrt{\q2}}\frac{\mathcal{H}_P + 2 \mathcal{H}_{SL}}{\mathcal{H}_{\rm total}}, \text{~and}\\
		\label{e54}
		\alpha^*(\q2) = & \frac{\mathcal{H}_U + \widetilde{\mathcal{H}}_U - 2(\mathcal{H}_L + \widetilde{\mathcal{H}}_L + 3\widetilde{\mathcal{H}}_S)}{\mathcal{H}_U + \widetilde{\mathcal{H}}_U + 2(\mathcal{H}_L + \widetilde{\mathcal{H}}_L + 3\widetilde{\mathcal{H}}_S)},
	\end{align}
    where $\widetilde{\mathcal{H}}_i = \frac{m_l^2}{2\q2}\mathcal{H}_i$ for ($i =~U,~L,~S$). The helicity components $\mathcal{H}_P$ and $\mathcal{H}_{SL}$ are defined by $\mathcal{H}_P = |{H}_+|^2 - |{H}_-|^2$ and $\mathcal{H}_{SL} = \mathcal{R}({H}_0{H}_t^{\dagger})$.
	
    For $B_c^{-} \to Vl^- \ov{\nu}_l$ decays, the physical observables like FB asymmetry, longitudinal and transverse polarization of the charged lepton are altered due to the opposite sign in the leptonic tensor~\cite{Faustov:2022ybm}. However, there is no change in the expression for other observables. In this study, we calculate the mean values of all the above mentioned physical observables by separately integrating the numerator and denominator over $\q2$, with the inclusion of a kinematic factor $\q2\sqrt{\lambda}(1-\frac{m_l^2}{\q2})^2$, where $(1-\frac{m_l^2}{\q2})$ represents the velocity-type parameter.
	
    \subsection{Nonleptonic decay widths} \label{S2_D}
    The QCD modified weak Hamiltonian generating the $B_c^+$ decay involving $b\to c(u)$ transitions is expressed as follows~\cite{Dhir:2013zia}:
	\begin{eqnarray}\label{e55}
		H_{w}^{(\Delta b=-1)} = \frac{G_F}{\sqrt{2}}\displaystyle\sum\limits_{Q(q)=u,c}\displaystyle\sum\limits_{q^{'}=d,s}V^*_{Qb}V_{qq^{'}} \Big(a_1(\mu) O^{qq^{'}}_1(\mu) + a_2(\mu) O^{qq^{'}}_2(\mu)\Big) + h.c.,
	\end{eqnarray}
    where $a_{1}$ and $a_{2}$ are the standard perturbative QCD coefficients, evaluated at renormalization scale $\mu \approx m_b^2 $. Local tree-level operators $O_{1,2}$ involving $b \to q$ transition can be expressed as products of color-singlet currents are given below:
	\begin{eqnarray} \label{e56}
		O^{qd}_1 = (\bar b_\alpha q_\alpha)_{V-A} \cdot (\bar q_\beta d_\beta)_{V-A},~~O^{qd}_2 = (\bar b_\alpha q_\beta)_{V-A} \cdot (\bar q_\beta d_\alpha)_{V-A}, \nonumber \\
		O^{qs}_1 = (\bar b_\alpha q_\alpha)_{V-A} \cdot (\bar q_\beta s_\beta)_{V-A},~~O^{qs}_2 = (\bar b_\alpha q_\beta)_{V-A} \cdot (\bar q_\beta s_\alpha)_{V-A}, 
	\end{eqnarray}
    where $(\bar{q}q')_{V-A} \equiv \bar{q} \gamma_{\mu}(1-\gamma_5)q'$, $\alpha$ and $\beta$ are $SU(3)$ color indices. Selection rules for various decay modes corresponding to the Hamiltonian, Eq.~\eqref{e55}, are:
	\begin{itemize}
		\item[(i)] CKM-enhanced modes $\Delta b = -1, \Delta C =-1, \Delta S=0 ;~\Delta b = -1, \Delta C =0,\Delta S =1$;
		\item[(ii)] CKM-suppressed modes $\Delta b = -1, \Delta C=-1, \Delta S=1; ~\Delta b = -1, \Delta C=0, \Delta S=0$;
		\item[(ii)] CKM-doubly-suppressed modes $\Delta b = -1, \Delta C=1, \Delta S=1; ~\Delta b = -1, \Delta C=1, \Delta S=0$.
	\end{itemize}
    In addition to the bottom-changing decays, $B_c^+$ meson can exhibit bottom-conserving decay modes for the $c$ quark decaying to an $s$ or $d$ quark. The weak Hamiltonian generating the $c$ quark decays, $H_{w}^{(\Delta C=-1)}$, is expressed by replacing $b$ with $c$, $Q(q)=d,s$, and $q'=u$ in Eq.~\eqref{e55}. The selection rules for various bottom-conserving decay channels are given as,
	\begin{itemize}
		\item[(i)] CKM-enhanced mode $\Delta b = 0, \Delta C =-1, \Delta S=-1$;
		\item[(ii)] CKM-suppressed mode $\Delta b = 0, \Delta C =-1, \Delta S=0$;
		\item[(ii)] CKM-doubly-suppressed mode $\Delta b = 0, \Delta C =-1, \Delta S=1$.
	\end{itemize}
	
    The factorization scheme expresses the decay amplitudes as a product of the matrix elements of weak currents, \textit{i.e.},
	\begin{equation} 
		\label{e57}
		\mathcal{A}(B_c \to P V) \simeq <P|J^{\mu}|0><V|J_{\mu}|B_c>+<V|J^{\mu}|0><P|J_{\mu}|B_c>,
	\end{equation} 
    where $J^{\mu}$ stands for $V-A$ current. The matrix element of the $J^{\mu}$ between vacuum and final meson ($P$ or $V$) is parameterized by the decay constants $f_{P(V)}$ as,
	\begin{align}
		\label{e58}
		<0|J_{\mu}|P(p^{\p})> & = ~<0|A_{\mu}|P(p^{\p})>     \hspace{0.4cm} =~ if_{P}p^{\p}_{\mu},\\ \nonumber
		<0|J_{\mu}|V(p^{\p},\varepsilon^{\p})> & =~ <0|V_{\mu}|V(p^{\p},\varepsilon^{\p})> ~          =~ M^{\p}_{V}f_{V}\varepsilon^{\p}_{\mu}.
	\end{align}
    The values of the decay constants used in our calculations are given in Table~\ref{t7}. 
	
    The nonleptonic $B_c$ decays can be categorized based on the color-favored and -suppressed contribution into three classes, as follows~\cite{Wirbel:1988ft, Browder:1995gi, Colangelo:1999zn}:
	\begin{itemize}
		\item[(i)] Class I: Decays primarily governed by color-favored diagrams, which can be generated from the color singlet current and their decay amplitudes are proportional to $a_1$, given by $a_1(\mu) = c_1(\mu) + \frac{1}{N_c}c_2(\mu)$, where $N_c$ represents the number of colors, and $c_1(\mu)$ and $c_2(\mu)$ are the QCD coefficients.
		\item[(ii)] Class II: Decays primarily influenced by color-suppressed diagrams, which can be generated from the neutral current and their decay amplitudes are proportional to $a_2$, defined as $a_2(\mu) = c_2(\mu) + \frac{1}{N_c}c_1(\mu)$.
		\item[(iii)] Class III: Decays resulting from a combination of both color-favored and color-suppressed diagrams, which can be generated from the interference of color singlet and color neutral currents, \textit{i.e.}, the $a_1$ and $a_2$ amplitudes interfere.
	\end{itemize}
	
    In general, the color-favored decay amplitude can be expressed as~\cite{Dhir:2008hh},
	\begin{align} 
		\label{e59}
		\mathcal{A}(B_c \to P V) = & \frac{G_F}{\sqrt{2}} \t \text{CKM factors} \t 2M_V a_1 \nonumber\\ & \t (\text{CG Coeff.} f_V F_1^{B_cP}(M_V^2) + \text{CG Coeff.} f_P A_0^{B_cV}(M_P^2)).
	\end{align}
    For the color-suppressed modes, the QCD factor $a_1$ is replaced by $a_2$. It is important to note that $a_1$ and $a_2$ are undetermined coefficients assigned to the effective charged current and effective neutral current, respectively~\cite{Browder:1994zy}. For the sake of consistency with the large $N_c$ limit (\textit{i.e.}, $N_c = \infty$), we adopt the convention of setting the QCD coefficients $a_1 \approx c_1$ and $a_2 \approx c_2$, as suggested in Refs.~\cite{Wirbel:1988ft, Browder:1995gi}. The numerical values we employ are as follows:
	\begin{align}
		\label{e60}
		\text{For $c$ decays (\textit{i.e.}, $\mu \approx m_c^2$)}: & ~~ c_1(\mu) = 1.26;~ c_2(\mu) = -0.51, \nonumber\\
		\text{For $b$ decays (\textit{i.e.}, $\mu \approx m_b^2$)}: & ~~ c_1(\mu) = 1.12;~ c_2(\mu) = -0.26.
	\end{align}
    The relatively smaller magnitudes of $a_2$ imply that, unlike in the charm sector, one anticipates a more pronounced pattern of color suppression in $B_c$ meson decays~\cite{Browder:1995gi}. Since $B_c$ decays primarily occur through tree diagrams or are tree-dominated, we neglect the anticipated small nonfactorizable and penguin contributions within our formalism. It may be noted that $N_c$ may be treated as a phenomenological parameter in weak meson decays, which account for nonfactorizable contributions~\cite{Jugeau:2005yr, Dhir:2012sv}. Therefore, we also use $N_c = 3$ to obtain the effective coefficients $a_1(\mu) = c_1(\mu)+ \frac{1}{3}c_2(\mu)$ and $a_2(\mu) = c_2(\mu)+ \frac{1}{3}c_1(\mu)$,
	\begin{align}
		\label{e61}
		\text{for $c$ decays (at $N_c = 3$)}: & ~~ a_1(\mu) = 1.09;~ a_2(\mu) = -0.09, \nonumber\\
		\text{for $b$ decays (at $N_c = 3$)}: & ~~ a_1(\mu) = 1.03;~ a_2(\mu) = 0.11.
	\end{align}
    We have calculated nonleptonic branching ratios of $B_c \to PV$ decays both at $N_c = \infty$ and $N_c = 3$. It is worth noting that for bottom-conserving decays, experimental charm decay studies have provided a parameterization for $a_1$ and $a_2$. These results suggest that considering the large $N_c$ limit is appropriate for $c$ quark decays~\cite{Cheng:2010ry}. On the other hand, for bottom-changing decays, phenomenological analyses~\cite{Cheng:2014rfa} indicate variations in the magnitudes of the Wilson coefficients $a_1$ and $a_2$, as well as sub-leading contributions from the $1/N_c$ term. This can be accounted for by allowing a certain range of values for these coefficients, as shown in Eq.~\eqref{e61}. We would like to emphasize that the decay amplitudes can be expressed as factorizable contributions multiplied by their respective $a_i$ values, which are independent of the (renormalization) scale and process.
	
    Using the decay amplitude defined in Eq.~\eqref{e59}, the decay rate for the $B_c$ to $P V$ decay is given by
	\begin{equation} 
		\label{e62}
		\Gamma(B_c \to P V)= \frac{{\textbf{k}}^3}{8 \pi M_{V}^{2}} |\mathcal{A}(B_c \to P V)|^2,
	\end{equation}
    where \textbf{k} is the three-momentum of the final-state particle in the rest frame of $B_c$ meson and is expressed as,
	\begin{equation} 
		\label{e63}
		\textbf{k} = \frac{1}{2 M_{B_c}}\sqrt{[M_{B_c}^{2}-(M_P + M_V)^2][M_{B_c}^{2}-(M_P - M_V)^2]}.
	\end{equation}
    The numerical results for semileptonic and nonleptonic weak decays of $B_c$ meson are discussed in the following section. 
	
    \section {Numerical Results and discussions} \label{S3}
    In the present work, we calculate the transition form factors for $B_c$ to $P$ and $V$ using the Type-II self-consistent CLFQM across the available range of momentum transfer. Furthermore, we provide a comprehensive investigation into their dependence on $\q2$ and compare our results with other formalisms. We compute the transition form factors for $B_c$ to $P$ and $V$ mesons, using the constituent quark masses and $\beta$ values provided in Table~\ref{t1}. The variation in quark masses introduces uncertainties in form factor calculations. Therefore, we incorporate a range of values based on established literature as the default input~\cite{Verma:2011yw, Zhang:2023ypl, Chang:2019mmh, Chang:2018zjq, Li:2023wgq, Sun:2023iis, Wang:2008xt}. It may be noted that the Gaussian parameter $\beta$, which characterizes the momentum distribution, is commonly determined by fitting the meson decay constant. In our work, we use the $\beta$ values from Ref.~\cite{Verma:2011yw} for the majority of $s$-wave mesons (corresponding to the input quark masses), which typically match with the latest decay constants provided in the Particle Data Group (PDG)~\cite{Workman:2022ynf}, and other analysis based on experimental results\footnote{The experimental averages for $b$-meson decay constants are not available in PDG, however, recent LQCD predictions yield, $f_{B}=(190.0 \pm 1.3)~\text{MeV}$~\cite{FLAG:2021npn}, $f_{B_s}=(230.3\pm 1.3)~\text{MeV}$~\cite{FLAG:2021npn}.}~\cite{Bharucha:2015bzk, BESIII:2023zjq, Dowdall:2013rya, Chakraborty:2017hry, Lubicz:2017asp}, as shown in Table~\ref{t7}. Furthermore, the used values for $\beta$ parameters are reasonably close to the latest results obtained in the self-consistent CLFQM approach~\cite{Chang:2020wvs, Chang:2019mmh, Chang:2018zjq}. However, the theoretical uncertainties used in our work correspond to a wider range as compared to the Refs.~\cite{Chang:2020wvs, Chang:2019mmh, Chang:2018zjq}. On the other hand, for the $B_c$ meson, the scenario is relatively different due to the lack of experimental data and a wide range of decay constant estimates available in the literature~\cite{Narison:2020guz, Narison:2019tym, Shi:2016gqt, Chang:2018zjq, Zhang:2023ypl, Colquhoun:2015oha}. Thus, we have used $\beta_{b \bar{c}} =(0.9207\pm 0.0921)~\text{GeV}$, where the central value (as obtained in Ref.~\cite{Chang:2019mmh}) reproduces the LQCD estimates for decay constants\footnote{The LQCD predicts the decay constant for $B_c$ as $f_{B_c} = (434 \pm 15)~\text{MeV}$~\cite{Colquhoun:2015oha}, for which the values of $\beta_{b\bar{c}}$ can be obtained.}. In addition, we have allowed larger uncertainties typically to address wide domain of decay constant predictions that range from $f_{B_c}= (371 - 489)~\text{MeV}$, for various theoretical models~\cite{Narison:2020guz, Narison:2019tym, Chang:2018zjq, Shi:2016gqt, Zhang:2023ypl, Colquhoun:2015oha}. In this work, we have investigated the variation of the form factors and their slope parameters for $\q2$ dependence concerning changes in constituent quark masses and $\beta$ values. We use three different $\q2$ formulations, namely, Type-II, Type-II*, and Type-I following Eqs.~\eqref{e37},~\eqref{e38}, and~\eqref{e40}, respectively. The transition pole masses given in Table~\ref{t2} are used for the calculation of the form factors of $\q2$ for both Type-II and Type-II*, while we fix the mass of the parent meson as the pole for Type-I. The obtained form factors for bottom-conserving and bottom-changing transitions are tabulated in Tables~\ref{t3} and~\ref{t4}, respectively. We plot their $\q2$ dependence for the available range $0 \leq \q2 \leq \q2_{max}= (M_{B_c}-M_{P(V)})^2$, as shown in Figures~\ref{f2}$-$\ref{f5}. We also plot corresponding wave function overlap (Eq.~\eqref{e5}) and overlap integrand (Eq.~\eqref{e27}) at $\q2=0$, as shown in Figures~\ref{f6}$-$\ref{f9}. Using the numerical values of the form factors, we predict the branching ratios for semileptonic decays of the $B_c$ meson\footnote{The branching ratio is calculated from the decay rate expression given in Eq.~\eqref{e41} by multiplying by $\frac{\tau_{B_c}}{\hbar}$.}, as shown in Tables~\ref{t5} and~\ref{t6}. In our calculations, we use the following values for the lepton mass: $m_e = 0.511$~MeV, $m_\mu = 105.66$~MeV and $m_\tau = 1776.86$~MeV; CKM matrix elements: $|V_{ub}| = (3.82 \pm 0.20) \t 10^{-3}$, $|V_{cd}| = 0.221 \pm 0.004$, $|V_{cs}| = 0.975 \pm 0.006$ and $|V_{cb}| = (40.8 \pm 1.4) \t 10^{-3}$, and lifetime of $B_c$ meson: $\tau_{B_c} = 0.51$ ps~\cite{Workman:2022ynf}. It should be noted that the uncertainties in the masses of mesons (leptons) and other parameters have been neglected due to their considerably smaller magnitude in comparison to the uncertainties in both quark masses and $\beta$ parameters. Also, we compare our results of semileptonic branching ratios with the existing literature, as shown in Table~\ref{t6}. Besides determining the branching ratios, we also calculate the numerical values of various physical observables, such as $A_{\rm FB}(\q2)$, $C_F^l(\q2)$, $P_{L(T)}^l(\q2)$, and $\alpha^*(\q2)$, as listed in Table~\ref{t5}. Additionally, we plot the differential decay rates and FB asymmetries for $B_c^+ \to Vl^+\nu_l$ decays in Figures~\ref{f10} and~\ref{f11}, respectively. Finally, we utilize the obtained form factors and the decay constants listed in Table~\ref{t7} to predict the branching ratios of nonleptonic $B_c$ to $PV$ decays\footnote{Noted that for $\eta$ and $\eta^\p$ pseudoscalar states, we use $\eta=\frac{1}{\sqrt{2}}(u\bar{u}+d\bar{d})\text{sin} \phi_{P} -(s\bar{s})\text{cos}\phi_{P}$, $\eta^\p= \frac{1}{\sqrt{2}}(u\bar{u}+d\bar{d})\text{cos} \phi_{P} +(s\bar{s})\text{sin}\phi_{P}$, with $\phi_{P} = \theta_{ideal}-\theta_{physical}$, where $\theta_{physical}= -15.4^{\circ}$; for $\omega$ and $\phi$ vector states, we consider ideal mixing, \textit{i.e.}, $\omega=\frac{1}{\sqrt{2}}(u\bar{u}+d\bar{d})$ and  $\phi= \frac{1}{\sqrt{2}}(s\bar{s})$~\cite{Workman:2022ynf}.}. The obtained results are presented in Tables~\ref{t8}$-$\ref{t11}. We also compare our predictions for nonleptonic branching ratios with other theoretical works, as shown in Tables~\ref{t12}$-$\ref{t14}. We discuss our numerical results as follows.
	
    \subsection{Form factors} \label{S3_A}
    In this subsection, we discuss the results for the self-consistent $B_c$ to $V$ transition form factors along with $B_c$ to $P$ for bottom-conserving CKM-enhanced $(\Delta b = 0, \Delta C =-1, \Delta S = -1)$ and suppressed $(\Delta b = 0, \Delta C =-1, \Delta S = 0)$ modes, as well as bottom-changing CKM-enhanced $(\Delta b = -1, \Delta C =-1, \Delta S = 0;~~\Delta b = -1, \Delta C =0, \Delta S = 1)$ and suppressed $(\Delta b = -1, \Delta C =0, \Delta S = 0)$ modes. We also contrast the form factors in Type-I and Type-II schemes corresponding to different $\q2$ dependence formulations, as presented in Tables~\ref{t3} and~\ref{t4}. The form factors are presented at $\q2 =0$ and at the maximum $\q2$. The first and second uncertainties on the form factors and slope parameters ($a$, $b$, $a_1^{\p}$, and $a_2^{\p}$) are from the constituent quark masses and the $\beta$ values, respectively. Aforementioned, to observe the variation of both Type-II and Type-II* form factors with respect to $\q2$, we plot these transition form factors, as shown in Figures~\ref{f2}$-$\ref{f5}. We list our observations as follows.
	
    \subsubsection{\textbf{Bottom-conserving transition form factors}}
	\begin{itemize}
        \item[(i)] The bottom-conserving $B_c \to B_{(s)}$ transitions are governed by $c$ quark decays, for which the observed $\q2$ range is limited to a narrow interval of $0 \leq \q2\leq (M_{B_c}-M_{B_{(s)}})^2 \simeq 1 ~\text{GeV}^2$. As a result, we expect these form factors to show minimal variations corresponding to the available $\q2$ range, as shown in Figure~\ref{f2}. The Type-II* form factors, corresponding to $z$-series parameterization, show more deviation than Type-II form factors. This is because difference $\q2$ formulations (Eqs.~\eqref{e37} and~\eqref{e38}) used in the analyses. It must be noted that $B_c \to P$ form factors are free from self-consistency issues, by replacement of $M^{\p(\p\p)} \to M_0^{\p(\p\p)}$ in Type-II correspondence, which results in modified numerical values. In addition, the choice of $\q2$ dependence between the two correspondences, \textit{i.e.}, Eqs.~\eqref{e37} and~\eqref{e38} in Type-II correspondence and Eq.~\eqref{e40} in Type-I correspondence, will also lead to changes in the numerical values of form factors and parameters ($a,~b,~a'_1,~\text{and}~a'_2$). It is important to note that for the Type-I correspondence scheme, the numerical values are computed using the parent pole mass in Eq.~\eqref{e40}, as recommended in previous studies~\cite{Chang:2018zjq, Cheng:2003sm}. This approach contrasts with the Type-II correspondence, where we employ transition pole masses utilizing Eqs.~\eqref{e37} and~\eqref{e38}. We observe that $B_c \to B_{(s)}$ form factors in Type-I scheme show marginal change in $F(0)$ values as compared to Type-II scheme. However, the slope parameters in both the correspondences are significantly different\footnote{Note that the sign and magnitude of the slope parameters signify how sharply the form factor varies with respect to allowed $\q2$.}. The form factors within Type-I scheme have substantially larger values for slope parameter $b$, which decreases on account of transition pole masses, as reported in our previous work~\cite{Hazra:2023zno}. However, the slope parameters $a$ and $b$ in Type-II correspondence are less than one. On the other hand, the parameters $a'_1$ and $a'_2$ also take significantly larger values for $z$-series parameterization (in Type-II*), unfortunately, there cannot be any physical interpretation associated with these coefficients~\cite{HFLAV:2022esi}. In addition, it is important to note that the form factors in Type-I correspondence show a decreasing trend with respect to $\q2$ variation, in contrast to Type-II correspondence. This observed trend is the opposite of what has been expected based on LQCD predictions~\cite{Cooper:2020wnj}. 
        
        \item[(ii)] As said before, we analyze the effect of the variation of quark masses and $\beta$ parameters on these form factors, and we observe that the form factors are less sensitive to the variation in constituent quark masses and $\beta$ values, which produce a collective uncertainty up to $\sim 10\%$ (for both Type-II and Type-II*). In contrast, the corresponding slope parameters $a$ and $b$ demonstrate substantially higher uncertainties. Notably, the uncertainties for Type-II and Type-II* show broadly similar patterns in response to quark mass and $\beta$ parameter variations, with a few exceptions. For the sake of comparison, we list numerical results for $P \to P$ form factors in Table~\ref{t3}. We observe that the numerical values of the form factors in Type-II correspondence are larger than those of Type-I scheme. Among the Type-II and Type-II* results, we observe that form factors are marginally different but uniformly larger numerical values for Type-II* form factors, as shown in Table~\ref{t3}. Furthermore, we wish to emphasize that the Type-II$^{(}$*${}^{)}$ numerical results for the form factors $F_{0}^{B_cB_{(s)}}(\q2)$ and $F_{1}^{B_cB_{(s)}}(\q2)$ are in very good agreement with the LQCD observations~\cite{Cooper:2020wnj}, for both at $\q2=0$ and $\q2_{max}$. The form factors for lattice results for both at $\q2 = 0$ and $\q2_{max}$ are as follows~\cite{Cooper:2020wnj}: $F_{0[1]}^{B_cB}(0) = 0.555 \pm 0.016 ~[0.555 \pm 0.016]$, $F_{0[1]}^{B_cB}(\q2_{max}) = 0.756 \pm 0.016 ~[0.910 \pm 0.028]$; $F_{0[1]}^{B_cB_s}(0) = 0.621 \pm 0.010 ~[0.621 \pm 0.010]$, $F_{0[1]}^{B_cB_s}(\q2_{max}) = 0.817 \pm 0.011~ [0.911 \pm 0.018]$. For Type-II and Type-II*, the numerical values of $B_c \to B_{(s)}$ form factors differ by $\sim 8\% (14\%)$ and $\sim 12\% (16\%)$ at $\q2 =0$, respectively, when compared to the LQCD results. However, the consistency improves at $\q2_{max}$, particularly for $F_{0}^{B_cB}(\q2_{max})$ in Type-II formulation, where the difference reduces to $\sim 3\%$ with LQCD results. Furthermore, the LQCD results also show an increasing trend with respect to the $\q2$ variation likewise observed in Type-II$^{(}$*${}^{)}$ results. The characteristic feature of bottom-conserving transitions, which has been reported in our previous work~\cite{Hazra:2023zno}, is that these form factors in the small available $\q2$ range show near flat behavior.
   
        \item[(iii)] Similar to $B_c \to P$ transitions, we calculate the form factors for bottom-conserving $B_c \to V$ transitions for both Type-I and Type-II correspondences, as shown in Table~\ref{t3}. It should be noted that in $B_c \to V$ transitions, $V(\q2)$ and $A_2(\q2)$ form factors remain unaffected by the spurious contributions associated with the $B^{(i)}_{j}$ functions. Consequently, the results obtained in CLF approach for $\lambda=0$ and $\lambda=\pm$ polarization states of vector mesons are in agreement with each other, regardless of whether Type-I or Type-II correspondence schemes are employed. However, in Type-I scheme, these zero-mode contributions lead to inconsistency in $A_0(\q2)$ and $A_1(\q2)$ form factors for $B_c \to V$ transitions. As described in the methodology in Sec.~\ref{S2}, the Type-II scheme effectively resolves the issues corresponding to self-consistency and covariance of the matrix elements~\cite{Chang:2019mmh}. Therefore, in Type-II scheme, zero-mode contributions associated with the $B_1^{(2)}$ and $B_3^{(3)}$ functions vanish in the form factors $A_0(\q2)$ and $A_1(\q2)$, as explained in Sec.~\ref{S2}. Therefore, the form factors $A_0(\q2)$ and $A_1(\q2)$ corresponding to longitudinal ($\lambda=0$) and the transverse ($\lambda= \pm$) polarization states are numerically equal. Furthermore, we plot all the bottom-conserving $B_c \to B_{(s)}^{*}$ transition form factors to observe their variation with respect to $\q2$, as shown in Figure~\ref{f3}. The form factors $A_0(\q2)$, $A_1(\q2)$, and $A_2(\q2)$ display nearly a flat behavior with respect to $\q2$ likewise, $B_c \to P$ form factors. In addition, the form factor $V(\q2)$ shows a reasonable variation in magnitude corresponding to the available $\q2$. Although the variations in $V(\q2)$ form factors seem to be significant in Figure~\ref{f3} (due to their higher numerical values), however, are only roughly $20\%$ larger with respect to $\q2=0$. It should be noted that $B_c \to B^*_{(s)}$ form factors are more sensitive to the uncertainties in constituent quark masses and $\beta$ values, leading to larger collective uncertainties of the order of $\sim 30\%$ and $\sim 40\%$ for $V(0)$ and $A_2(0)$ form factors, respectively. Such significant uncertainties were anticipated, given the incorporation of a broad range of $\beta$ parameter values alongside variations in quark masses. Furthermore, the degree of sensitivity to quark mass and $\beta$ uncertainties varies among different form factors. On the other hand, the uncertainties are substantially large specifically for $A_0(0)$, \textit{i.e.}, up to $\sim 60\%$, in Type-I scheme. At the same time, the slope parameters also show larger uncertainties.
		
        \item[(iv)] In general, the transition form factors essentially involve the overlap integral of the initial and final state meson wave functions, which depend upon the internal degrees of freedom, mainly transverse momentum distributions and constituent quark masses. Furthermore, in CLFQM the actual magnitude of these transitions has contributions originating from vertex functions and current operators. Therefore, first we plot the overlap\footnote{The normalization of Gaussian-type radial wave function of meson is described by Choi \textit{et al.}~\cite{Choi:2021mni, Choi:2009ai}.} of initial and final wave functions at $\q2 \approx 0$, where we have integrated out $k_{\perp}^{2}$ using Eq.~\eqref{e5}, as shown in Figure~\ref{f6} with corresponding overlap factor. The larger wave function overlap can be explained by the internal momentum distribution peaks at $x_1 \sim 0$ for $\psi_{B_{(s)}^{(*)}}(x_1)$ and $x_1 \sim 0.25$ for $\psi_{B_c}(x_1)$, as per Eq.~\eqref{e5}. The location and width of the peak are governed by constituent quark masses, where heavier quark takes a larger fraction of momentum~\cite{Hwang:2010hw, Cheung:1996qt, Choi:2009ai}. This results in a large overlap between the initial and final states. The overlap factor inside the total integrand, therefore, leads to decisive change in magnitude of the total form factor. For further analysis, we also plot the total integrand defined by Eq.~\eqref{e27} with respect to the momentum fraction $x_1$ for the $B_c$ meson to $P(V)$ transition form factors at $\q2 \approx 0$, as shown in Figures~\ref{f8} and~\ref{f9}. To obtain these plots, we included the mass factors (given by Eq.~\eqref{e17}) into Eq.~\eqref{e27} and integrated out $\bm{k}_{\perp}$. It should be noted that the total integrand of transition form factors, \textit{e.g.}, $B_c \to B_{(s)}^{(*)}$ follow exactly same overlap region which is governed by the initial and final wave functions. The bottom-conserving transition form factors have larger amplitudes than the bottom-changing form factors (as seen in Figures~\ref{f8} and~\ref{f9}). The area under the curves gives the magnitude of the form factor for the respective transitions and we observe constructive interference for most of the transition form factors, except for $A_2^{B_cB^*}$. We observe that the overlap integrand of $A_2^{B_cB^*}$ traverses both positive and negative regions with respect to changes in $x_1$. The positive and negative peaks are due to the constructive and destructive interference of their corresponding wave functions, and therefore, should be added with their respective signs to give the total magnitude of the overlap integrand. It is worth noting that among the ${B_c\rightarrow B^*}$ transition form factors, the area under the peak corresponding to the $V(x_1)$ integrand is larger, which leads to the larger magnitude of the form factor $V^{B_c B^*}(0)$, as listed in the Table~\ref{t3}. Similar conclusions can be made for other transition form factors. Thus, the overlap integrand plots represent the true behavior of form factors at $\q2 = 0$. Furthermore, the magnitude of the overlap is expected to increase with respect to $\q2$ to reach a maximum at $\q2_{max}$. Since the available $\q2$ range is small, the overlap at $\q2_{max}$ is expected to be roughly the same as that at $\q2 = 0$. Therefore, a flat behavior of form factor is expected, as seen in Figures~\ref{f2} and~\ref{f3}.
		
        \item[(v)] Aforementioned, the choice of $q^+ = 0$ frame of reference restricts the calculation of the form factors only in the space-like region for momentum transfer $\q2 \leq 0$. To understand the physical decay process, we need to know the form factors in time-like region, \textit{i.e.}, $\q2 > 0$. This can be achieved by extrapolating the form factors as appropriate functions of $\q2$ (given by Eqs.~\eqref{e37} and~\eqref{e38}), for which the knowledge of form factors at $\q2=0$ (see Figures~\ref{f8} and~\ref{f9}) is crucial. While the two methods provide independent descriptions of the form factors in space-like and time-like regions, they are nonetheless complementary in nature~\cite{Jaus:1996np}. Thus, provide a complete description of the decay dynamics of the transition process for the full $\q2$ range. In our work, to determine the form factors over the entire range, we utilize parameterization in Eq.~\eqref{e37} that accommodates the contributions of meson resonances of relevant spin and parity for the entire $\q2$-channel. Similarly, the parameterization in Eq.~\eqref{e38} isolates meson resonances below the transition threshold for the corresponding meson poles given in Table~\ref{t2}. In the case of $B_c$ to $B_{(s)}^{(*)}$, we use resonances $D_{(s)}^{**}$ as pole masses to analyze $\q2$ behavior throughout the available range. This can be explained through the confining interaction between $c$ and $\bar{d}(\bar{s})$ to produce $D_{(s)}$ meson resonances that fluctuate into $W$-boson. In the physical region the form factors at $\q2 =0$ are larger than the values for bottom-changing transitions. This can be understood as follows, for $B_c \to B_{(s)}^{(*)}$ transition, the energy released to the final state is much smaller than $m_b$ (because $m_c \ll m_b$, and $M_0^\p \sim m_b$), therefore, the $b$ quark remains almost unaffected. This is reflected in the larger amplitude of the overlap integrand between the initial and final states. The pole at $M_{D_{(s)}^{**}}^2$ lies far from the $\q2_{max}~(\lesssim 1\text{~GeV}^2$), which is less than $\sim 25\% $ as compared to $M_{D_{(s)}^{**}}^2$ (square of the pole mass). Therefore, the effect of pole contribution in the $\q2$ variation of bottom-conserving $B_c \to B_{(s)}^{(*)}$ form factors are smaller. Furthermore, the form factors $F_1(\q2)$, $V(\q2)$, and $A_0(\q2)$ involving $M_{1^-}$ and $M_{0^-}$ poles are affected by roughly $(22-25)\%$ for $B_c \to B^{(*)}$ transitions, while $B_c \to B_{s}^{(*)}$ transitions are less affected, \textit{i.e.}, by $(5-7)\%$. Thus, these form factors show very small variations in the $0 \leq \q2 \leq \q2_{max}$. Similarly, $F_0(\q2)$, $A_1(\q2)$, and $A_2(\q2)$ are affected by $M_{0^+}$ and $M_{1^+}$ poles, which lie farther away from $\q2_{max}$, show least variation with $\q2$, and therefore, show near flat behavior. In addition, the variation between the numerical values of the form factors at $\q2 =0$ and $\q2_{max}$ for $B_c \to B_{(s)}^{(*)}$ form factors in Type-II* are slightly larger as compared to Type-II. This numerical variation between Type-II and Type-II* is less than $5\%$ corresponding to the parameterizations given by Eqs.~\eqref{e37} and~\eqref{e38}. Therefore, we expect that the variation in the form factors over a small $\q2$ range in bottom-conserving transitions can be reliably estimated by a simple VMD-type pole behavior. However, the parameterizations described by Eqs.~\eqref{e37} and~\eqref{e38} are necessary for the accuracy of the numerical evaluation of the form factors. Moreover, such extension beyond the available $\q2$ range is important for the understanding of semileptonic decays. This is due to the distinct feature of the semileptonic decays in which resonances are not only observed within the kinematic range of meson decay, however, also extend beyond the available $\q2$ region~\cite{Melikhov:2000yu, Hill:2005ju}.
		
        \item[(vi)] For $B_c \to B_{(s)}^{*}$ transitions, the slope parameters $a$ and $b$ are numerically closer to unity in magnitude and are positive, except for the form factors $A_2^{B_cB^*}$ in Type-II correspondence for Eq.~\eqref{e37}. Interestingly, the magnitude of the parameter $a$ is very small for $A_2^{B_cB_{s}^*}$ and is negative for $A_2^{B_cB^*}$ which explains the flat behavior, as shown in Figure~\ref{f3}. We found that the numerical values of all the form factors for the Type-I scheme (using Eq.~\eqref{e40} and parent pole mass) are less than one, except for $V(0)$, the same can be observed for Type-II and Type-II*. Although the numerical values of $V(0)$ between Type-I and Type-II$^{(}$*${}^{)}$ differ roughly by an average of $15\%$, the slope parameters are substantially different. Interestingly, the slope parameter $a$ is negative and greater than one for most of the form factors, except for $A_1^{B_cB_{(s)}^*}$, and the parameter $b$ has very large values ranging roughly from $130 - 1000$ (for $A_0(\q2)$ in Type-I) with positive sign. It may be noted that both slope parameters are exceptionally large for the form factor $A_0^{B_cB_{(s)}^*}$. Similar observations can be made for the remaining form factors, where the slope parameters $a$ and $b$ are typically large for the Type-I scheme. As observed in $B_c \to P$ bottom-conserving transitions, for $B_c \to B_{(s)}^{*}$ form factors, we observe smaller numerical values for the form factors along with a decreasing trend in the Type-I scheme, as compared to Type-II correspondence. In addition, we observe that the form factors $A_0(\q2)$ and $A_1(\q2)$ affected by the zero-mode contributions show a substantial decrease in the numerical values with respect to the Type-I scheme. Furthermore, the $A_1(0)$ form factors change by $\sim 23\%$ for both Type-II and Type-II* in addition to the $a$ and $b$ parameters. We want to emphasize that the numerical values of the form factors in Type-II$^{(}$*${}^{)}$ exhibit a significant variation in the magnitude of $A_0(0)$ form factors ranging from $(70-90)\%$ compared to Type-I scheme. The impact of the spectator quark mass on the numerical values of $B_c\rightarrow B_{(s)}^*$ transition form factors over the available $\q2$ is negligible, which has also been recently observed by LQCD calculations~\cite{Cooper:2020wnj}.
		
        \item[(vii)] Furthermore, we analyze the $z$-series parameterization of the form factors at maximum recoil point ($a'_0$), as given in Eq.~\eqref{e38}. The numerical results obtained from $z$-series parameterization are surprisingly consistent with those obtained from the $\q2$ dependence used in Eq.~\eqref{e37}. In addition, the free parameters $a'_1$ and $a'_2$ take very large values, as shown in Table~\ref{t3}. However, the sign for $a'_1$ parameter is consistently negative, and $a'_2$ parameter is positive, except for $A_2^{B_cB^{*}}(\q2)$. Further, the magnitude of $a'_2$ parameter is exceedingly larger than $a'_1$ parameter due to the fact that the coefficients take large values for smaller $\pm |z|_{max}$ (\textit{i.e.}, $ \approx \pm 0.0008$ for $B_c$ to $B_{(s)}^{(*)}$ transition). In addition, the uncertainties corresponding to the quark masses and $\beta$ values in $a'_2$ are larger than that of the $a'_1$ parameter. As already pointed out, the $\q2$ behavior of power series expansion, as shown by Type-II* in Figures~\ref{f2} and~\ref{f3}, is consistent with the $\q2$ behavior corresponding to Eq.~\eqref{e37}. However, it shows relatively larger variation towards the maximum $\q2$, particularly for $V(\q2)$ form factors. Therefore, we reemphasize that both $\q2$ formulations for Type-II$^{(}$*${}^{)}$ appear to be consistent with each other within very small numerical variations.		
	\end{itemize}
	
    \subsubsection{\textbf{Bottom-changing transition form factors}}
	\begin{itemize}
        \item[(i)] The bottom-changing transitions typically exhibit a wider range of $\q2$ compared to bottom-conserving transitions. In the case of $B_c \to D^{(*)}$ transition form factors, it is expected that the $\q2$ range will be considerably broader with respect to $B_c \to \eta_c{(J/\psi)}$ form factors, spanning from $0 \leq \q2 \lesssim 20$ GeV${}^2$. This extended range offers an opportunity to examine how the form factors are influenced by the dependence on $\q2$ and to highlight the importance of the resonance pole contribution below the threshold. We plot the bottom-changing $B_c$ to $P$ and $V$ transition form factors to observe their variation with respect to $\q2$, as shown in Figures~\ref{f4} and~\ref{f5}, respectively. The form factors remain the same at $\q2=0$ for all the bottom-changing transitions in both Type-II and Type-II*. For bottom-changing transitions, both the slope parameters of $B_c \to P$ and $B_c \to V$ form factors are positive and in the range of $a,b \subset (0,~2)$ and $a,b \subset (0,~3)$, respectively, as given in Table~\ref{t4}. 
		
        \item[(ii)] Similar to the bottom-conserving case, to understand the dynamics of the $B_c \to D^{(*)}$ transitions, we plot the wave function overlap between the initial $\psi_{B_c}(x_1)$ and final $\psi_{D^{(*)}}(x_1)$ wave functions at $\q2 = 0$, as shown in Figure~\ref{f7a}. Due to the limited overlap near $\q2=0$, the numerical values of the form factors are expected to be smaller as compared to $B_c \to B_{(s)}^{(*)}$ and $B_c \to \eta_{c}(J/\psi)$ transitions. Since the fraction of momentum carried by the spectator $c$ quark is of the order of the decaying $b$ quark, $u$ quark takes minimal momentum. Consequently, $\psi_{D^{(*)}}(x_1)$ exhibits its maximum near $x_1 \sim 1/4$ with a larger width, while the peak for $\psi_{B_c}(x_1)$ lies at $x_1 \sim 3/4$. The available $\q2$ for $B_c$ to $D^{(*)}$ transitions is significantly large ($0 \leq \q2 \lesssim 20$ GeV${}^2$), hence these $b \to u$ transitions involve $B^{**}$ poles fluctuating in the weak current $b\ov{u}$. Moreover, the $\q2_{max}$ is around $65\%$ of the $M_{B^{**}}^2 \lesssim 34~\text{GeV}^2$, which is not far away from the $\q2_{max}$, in contrast to $B_c \to B_{(s)}^{(*)}$ transitions. Thus, we expect reasonable contributions from the resonance poles in the available $\q2$ range, as shown in Figures~\ref{f4} and~\ref{f5}. As a result, the form factors will have larger numerical values at $\q2_{max}$, as can be seen from Table~\ref{t4}. Similar to $B_c \to B_{(s)}^{(*)}$ form factors, we also plot the total integrand for bottom-changing transition form factors, as shown in Figures~\ref{f8} and~\ref{f9}. Note that the total integrand of bottom-changing transition form factors shows a substantial decrease in magnitude, as compared to bottom-conserving transition form factors. Additionally, we observe constructive interference for all the bottom-changing transition form factors. Among bottom-conserving and bottom-changing transitions, $B_c \to \eta_c (J/\psi)$ form factors have intermediate amplitude due to the largest wave function overlap factor. 
  
        \item[(iii)] As stated before, we wish to emphasize that our numerical values of form factors $F_{0}^{B_cD_{(s)}}(\q2)$ and $F_{1}^{B_cD_{(s)}}(\q2)$ are in excellent agreement with the LQCD predictions~\cite{Cooper:2021bkt}. The form factors for LQCD results for both at $\q2 = 0$ and $\q2_{max}$ are as follows~\cite{Cooper:2021bkt}: $F_{0[1]}^{B_cD}(0) = 0.186 \pm 0.023~[0.186 \pm 0.023]$, $F_{0[1]}^{B_cD}(\q2_{max}) = 0.668 \pm 0.020~[1.50 \pm 0.18]$; $F_{0[1]}^{B_cD_s}(0) = 0.217 \pm 0.018~[0.217 \pm 0.018]$, $F_{0[1]}^{B_cD_s}(\q2_{max}) = 0.736 \pm 0.011~[1.45 \pm 0.12]$. The numerical values of $B_c \to D$ form factors differ by $\sim 9\%$ at $\q2 =0$ compared to the LQCD results. However, the agreement substantially improves at $\q2_{max}$ for $F_{0}^{B_cD}(\q2_{max})$ and $F_{1}^{B_cD}(\q2_{max})$ for Type-II and Type-II*, respectively. Furthermore, for $B_c \to D_s$ form factors, our results are in good agreement in comparison to the LQCD results, where the results match within $\sim 15\%$. It is interesting to note that the $\q2$ variation of Type-II* form factors in Figure~\ref{f4} show a behavior similar to that observed in LQCD~\cite{Cooper:2021bkt}. The numerical values of the form factors in Type-II* vary more sharply near the maximum $\q2$ than the form factors in Type-II. It is significant to note that the pole at $M_{B_{(s)}^{**}}^2$ lie away from $\q2_{max}$, \textit{i.e.}, $\sim (50-70)\%$ of $M_{B_{(s)}^{**}}^2$ for $B_c \to D_{(s)}^{(*)}$ transitions. Furthermore, the form factors $F_0(\q2)$ and $F_1(\q2)$ receive pole contributions from $M_{0^+}$ and $M_{1^-}$, respectively, which result in visibly different behavior corresponding to the squared mass of resonances. We observe similar $\q2$ behavior for $B_c \to D_{(s)}^{*}$ form factors. In addition, the form factors $V(\q2)$ and $A_0(\q2)$ which receive pole contributions from $M_{1^-}$ and $M_{0^-}$ poles show expected behavior. Whereas, the form factors $A_1(\q2)$ and $A_2(\q2)$ that receive contributions from $M_{1^+}$ poles vary less sharply, as expected. Furthermore, we notice that the effect of the variation in the quark masses and $\beta$ parameters lead to larger uncertainties in the $B_c \to D_{(s)}$ form factors as large as $\sim 40\%$, this has not been previously analysed and reported in the literature. In contrast, the uncertainties in $B_c \to \eta_c$ form factors are as small as $\sim 3\%$. Above stated observations highlight the importance of quantitative perspective of this analysis.
		
        \item[(iv)] One of the most peculiar aspects of bottom-changing transition form factors, especially for $B_c \to V$, is that they have larger values of $a$ and $b$ parameters due to the smaller magnitude of form factors as compared to bottom-conserving ones. It is worth mentioning that even though the numerical values of all the bottom-changing transition form factors at $\q2 =0$ are similar between Type-I and Type-II schemes (except $A_0(\q2)$ and $A_1(\q2)$), their respective slope parameters as well as values at $\q2_{max}$ differ significantly with larger magnitudes, observed exclusively for parameter $b$. This shows that the form factors with $\q2$ dependence given by Eq.~\eqref{e40} vary more sharply. It should be emphasized that, likewise bottom-conserving transition form factors in the Type-I scheme, we observe significant numerical variation in the magnitudes of the form factor $A_0(0)$ ($A_1(0)$), \textit{i.e.}, $\sim 30\%~(10\%)$ as compared to both Type-II and Type-II* for $B_c \to D_{(s)}^*$ transitions. Therefore, the effect of self-consistency cannot simply be determined from the numerical values of the affected form factors at $\q2 = 0$. Particularly, in the $z$-series parameterization (Type-II*), the form factors at the maximum recoil point ($a'_0$) have comparable values with the Type-II form factor at $\q2 = 0$; however, they differ significantly at $\q2_{max}$. For $B_c \to D^*$ transition, the numerical values of $A_0(0)$ between Type-I and Type-II* differ by $\sim 30\%$. On the other hand, the free parameters $a'_1$ and $a'_2$ have large values and follow the same pattern in all bottom-changing transitions. Among these, $B_c$ to $D_{(s)}^{(*)}$ transitions have smaller values of $a'_1$ and $a'_2$ as compared to transitions involving charmonia due to the larger value of $\pm |z|_{max}=\pm 0.039$ for $D_{(s)}^{(*)}$ mesons. 
  
        \item[(v)] In the case of $B_c \to D^*_{(s)}$ bottom-changing transitions, the form factors show increased sensitivity to uncertainties in constituent quark masses and $\beta$ values, resulting in more substantial collective uncertainties. For instance, we observe a maximum uncertainty of approximately $86\%$ for the $A_2^{B_cD^*}(0)$ form factor. As stated earlier, the quantitative analysis of $B_c \to D^*_{(s)}$ transition form factors highlights the critical role of uncertainties propagating through the form factors via the input parameters. We believe that these uncertainties are crucial for the accurate assessment of both semileptonic and nonleptonic decay processes. In addition, the slope parameters associated with these transitions also demonstrate increased uncertainties. It should be noted that the uncertainties corresponding to the quark mass are smaller than those of $\beta$ values for the form factors $A_0(0)$ and $A_1(0)$, while the remaining form factors show comparable variations. As previously said, we note the maximum collective uncertainties of approximately $60\%$ in the case of $V(0)$ and $A_2(0)$ for both Type-II and Type-II* scenarios, exhibiting similar behavior, \textit{i.e.}, demonstrating roughly comparable sensitivity to $\beta$ parameters and quark masses. In general, a comparison of the numerical values of the form factors between Type-I and Type-II correspondence reveals that the effect of self-consistency and covariance leads to significant changes in the numerical values of $A_{0}(\q2)$ and $A_{1}(\q2)$. The Type-I scheme exhibits a similar sensitivity to quark masses and $\beta$ values in the $B_c \to D_{(s)}^*$ transition form factors. In addition, as observed in bottom-conserving transitions, the $A_0(\q2)$ form factor shows a decreasing trend, contrasting with the behavior of the $A_1(\q2)$ form factor as $\q2$ is varied within the Type-I scheme. Such deviations between the two schemes are expected to be decisive for the study of weak semileptonic and nonleptonic decays. We also observe that the effects of self-consistency on bottom-changing transition form factors are smaller than those of bottom-conserving transition form factors.
		
        \item[(vi)] Among the bottom-changing transitions, we observe that $B_c$ decaying to charmonium states have larger numerical values of the form factors. This is due to the fact that in $B_c \to c\ov{c}$ meson transitions, the fractional momentum of the charm quark in the final state is of the order of the spectator $c$ quark. Therefore, $\psi_{\eta_c(J/\psi)}(x_1)$ have a peak near $x \sim 1/2$ which shows a larger overlap with $\psi_{B_c}(x_1)$ at $ x \sim 3/4$ as compared to the overlap between $\psi_{B_c}(x_1)$ and $\psi_{D^{(*)}_{(s)}}(x_1)$ (see Figure~\ref{f7c}); in fact, the overlap is even larger than bottom-conserving transitions. Thus, the overlap plots (as shown in Figures~\ref{f8} and~\ref{f9}) for the total integrand show the importance of the vertex functions and other factors including masses. This results in an intermediate integrand amplitude of the $B_c \to \eta_c(J/\psi)$ form factors that lies between $B_c \to B_{(s)}^{(*)}$ and $B_c \to D_{(s)}^{(*)}$. A similar trend can be observed for Type-II* and Type-I results using the $\q2$ dependence given by Eqs.~\eqref{e38} and~\eqref{e40}, respectively. It may be noted that for $B_c \to \eta_{c}(J/\psi)$, the resonance poles $M_{B_c^{**}}^2$ lies much farther as compared to $B_c \to D_{(s)}^{*}$ transitions, which is $\sim (21-26)\%$ of $M_{B_{c}^{**}}^2$. Furthermore, we observed that similar to other bottom-changing transition form factors, $B_c \to \eta_c(J/\psi)$ form factors show an increasing behavior toward the maximum $\q2$ though less sharply, as shown in Figures~\ref{f4} and~\ref{f5}. Both Type-II and Type-II* $\q2$ formulations show roughly similar behavior. In addition, it is interesting to note that the effects of self-consistency on bottom-changing $B_c \to J/\psi$ transition form factors are minimal as compared to both bottom-conserving and other bottom-changing transition form factors. Interestingly, we note that $B_c \to J/\psi$ form factors are least affected by the quark mass and $\beta$ uncertainties (for both Type-I and Type-II schemes), \textit{i.e.}, the maximum uncertainty of $\sim 14\%$ for $A_0(0)$ form factor, while rest of the form factors have even smaller uncertainties.
	\end{itemize}
    We have employed the Type-II correspondence to vector meson emitting transitions for both bottom-conserving and bottom-changing decays. Moreover, we confirm that on the application of Type-II correspondence, the $B_c$ to $V$ transition form factors are self-consistent, \textit{i.e.}, zero-mode contributions vanish numerically. We now proceed to calculate the branching ratios of semileptonic $B_c \to Pl\nu_{l}$ and $B_c \to Vl\nu_{l}$ decays involving $B_c \to P$ and $V$ transition form factors, respectively.

    \subsection{Semileptonic decays} \label{S3_B}
    In this subsection, we study the branching ratios of the semileptonic $B_c$ meson decays obtained by using the transition form factors given in Tables~\ref{t3} and~\ref{t4}. We list our predictions of the branching ratios of $B_c^+ \to Vl^+ \nu_{l}$ in Type-II correspondence as shown in Table~\ref{t5}. We have also computed these branching ratios using CLF form factors by employing Type-I correspondence, and the results are presented in column $3$ of Table~\ref{t6}. As already discussed in the form factors, we also employ the $\q2$ formulation given by Eq.~\eqref{e38} referred to as Type-II*, to obtain the semileptonic branching ratios of $B_c$ decays. We also computed the uncertainties in the branching ratios propagating through form factor uncertainties. The uncertainties corresponding to the quark masses and $\beta$ values are treated independently. Furthermore, we compare these results with other theoretical predictions from Refs.~\cite{Li:2023wgq, Ivanov:2006ni, Faustov:2022ybm} and~\cite{Wang:2008xt}, as given in Table~\ref{t6}. In addition, we list relative decay widths, the average values of other observables for the $B_c$ transitions, including the FB asymmetry ($\langle A_{\rm FB} \rangle$), convexity parameter ($\langle C_F^l \rangle$), longitudinal (transverse) ($\langle P_{L(T)}^l \rangle$) polarization of the charged lepton, and asymmetry parameter ($\alpha^*$) in Table~\ref{t5}. Furthermore, we plotted the $\q2$ variation of the differential decay rates and $A_{\rm FB}(\q2)$ of $B_c^+ \to Vl^+ \nu_l$ decays in Figures~\ref{f10} and~\ref{f11}, respectively.
    \subsubsection{\textbf{Bottom-conserving decays}}
    The bottom-conserving CKM-enhanced $(\Delta b = 0, \Delta C =-1, \Delta S = -1)$ and CKM-suppressed $(\Delta b = 0, \Delta C =-1, \Delta S = 0)$ semileptonic decay modes of $B_c$ mesons undergo kinematic suppression due to the large mass of the $B_{(s)}^*$ meson in the final states. These semileptonic decay processes provide an excellent opportunity to observe the effects of form factors on the branching ratios and, therefore, to test the theoretical models. In addition to form factors, kinematic and CKM factors play an important role in determining their magnitude. We analyzed $B_c^+ \to B_{(s)}^{*0}l^+\nu_{l}$ decays using the self-consistent CLFQM. We have observed the following.
	\begin{itemize}
        \item[(i)] We observe that the branching ratios of bottom-conserving decays are of $\mathcal{O}(10^{-2})$ to $\mathcal{O}(10^{-3})$ despite the kinematic suppression. Among these decays, the CKM-enhanced modes have dominant branching ratios, \textit{i.e.}, $\mathcal{B}(B_c^+ \to B_{s}^{*0} e^+\nu_e) =(3.53^{+0.15+0.49}_{-0.24-0.81}) \t 10^{-2}$ and $\mathcal{B}(B_c^+ \to B_{s}^{*0} \mu^+\nu_\mu) =(3.30^{+0.14+0.46}_{-0.22-0.77}) \t 10^{-2}$, as listed in Table~\ref{t5}. This is due to the fact that the kinematic suppression is predominated by the CKM factor $(V_{cs})$. On the other hand, the branching ratios of $B_c \to B^*l\nu_l$ decays involving $c \to d$ transition (governed by $V_{cd}$) are smaller by an order of magnitude. In general, the branching ratios of $P \to V$ semileptonic decays are expected to be larger than $P \to P$ decays, which can also be observed from our results. We found that our results are in good agreement with recent LQCD predictions within the uncertainties~\cite{Cooper:2020wnj}. Although we focused on $P \to V$ semileptonic decays of the $B_c$ meson, we also list $B_c^+ \to Pl^+\nu_l$ decays in CLFQM for Type-II, Type-II*, and Type-I, as shown in columns 2, 3, and 4 of Table~\ref{ta1}, respectively, in Appendix~\ref{aB}. The bottom-conserving branching ratios for lattice results are as follows~\cite{Cooper:2020wnj}: $\mathcal{B}(B_c^+ \to B^{0}l^+\nu_{l}) = (8.47 \pm 0.31 \pm 0.43 \pm 0.24) \t 10^{-4}$ and $\mathcal{B}(B_c^+ \to B_s^{0}l^+\nu_{l}) = (1.348 \pm 0.046 \pm 0.033 \pm 0.043) \t 10^{-2}$. In an effort to ensure the reliability of the CLF approach, we compare the decay width ratios of our results with LQCD expectations:
        \[~~~~~~~~~~~~~~~~~~~~~~~~~~~~~~~~~~~~\text{Type-II}~~~~~~~~~~~~~~~~\text{~Type-II*}~~~~~~~~~~~~~~~~~\text{LQCD~\cite{Cooper:2020wnj}}\]
		\[ \frac{\Gamma(B_c^+ \to B_{s}^{0} e^+\nu_e)|V_{cd}|^2}{\Gamma(B_c^+ \to B^{0} e^+\nu_e)|V_{cs}|^2} = 0.88^{+0.20+0.21}_{-0.20-0.10}~~~~~~~~~~~0.82^{+0.19+0.02}_{-0.20-0.00} ~~~~~~~~~~~~ 0.759\pm 0.044; \]
		\[ \frac{\Gamma(B_c^+ \to B_{s}^{0} \mu^+\nu_{\mu})|V_{cd}|^2}{\Gamma(B_c^+ \to B^{0} \mu^+\nu_{\mu})|V_{cs}|^2} = 0.87^{+0.20+0.20}_{-0.20-0.09}~~~~~~~~~~0.81^{+0.19+0.02}_{-0.20-0.00}~~~~~~~~~~~~ 0.759\pm 0.044.\]
        Our results are in good agreement with LQCD ratios for Type-II* $\q2$ formulation; however, are slightly larger for Type-II $\q2$ formulation\footnote{Note that the uncertainties in the ratios of the branching fractions are bound to increase because of their additive nature. As mentioned before, we have ignored the uncertainties of the CKM factors in our analysis.}. Moreover, the semileptonic branching ratios of bottom-conserving modes for Type-II* are larger by $\sim (22 - 28) \% $ as compared to Type-II results. It may be noted that the uncertainties in our branching ratios for Type-II$^{(}$*${}^{)}$, stemming individually from both quark masses and $\beta$ parameters, are generally modest, with maximum deviations of the $\mathcal{O}(20\%)$, as shown in Table~\ref{ta1}. The form factors, $F_0(\q2)$ and $F_1(\q2)$, are not subject to self-consistency issues within CLFQM. Consequently, the numerical discrepancies observed in the Type-II correspondence scheme for decays involving $F_0(\q2)$ and $F_1(\q2)$ form factors can be attributed to variations arising from different $\q2$ formulations. Similarly, for $B_c \to B_{(s)}^*l\nu_l$ decays, we predict 
        \[~~~~~~~~~~~~~~~~~~~~~~~~~~~~~~~~~~~~~\text{Type-II}~~~~~~~~~~~~~~~~~\text{~Type-II*}\]
		\[ \frac{\Gamma(B_c^+ \to B_{s}^{*0} e^+\nu_e)|V_{cd}|^2}{\Gamma(B_c^+ \to B^{*0} e^+\nu_e)|V_{cs}|^2} = 0.87^{+0.12+0.35}_{-0.09-0.20}~~~~~~~~~~~0.83^{+0.09+0.19}_{-0.06-0.17}; \]
		\[ \frac{\Gamma(B_c^+ \to B_{s}^{*0} \mu^+\nu_{\mu})|V_{cd}|^2}{\Gamma(B_c^+ \to B^{*0} \mu^+\nu_{\mu})|V_{cs}|^2} = 0.86^{+0.11+0.35}_{-0.09-0.20}~~~~~~~~~~~0.82^{+0.09+0.18}_{-0.06-0.16}.\]

        \item[(ii)] Due to larger uncertainties in the form factors corresponding to $\beta$ parameters than constituent quark masses, the semileptonic branching ratios show greater sensitivity to variation of $\beta$ parameters leading to enhanced uncertainties. The uncertainties in $\beta$ (quark mass) result in a maximum change in branching ratios of the order of $\sim 33\%$ ($\sim 12\%$) for $B_c^+ \to B^{*0} l^+ \nu_l$ decays. On the other hand, the uncertainties for $B_c^+ \to B_s^{*0} l^+ \nu_l$ are relatively smaller, with a maximum deviation of approximately $18 \%$ inclusive of the uncertainties from both quark mass and $\beta$ values. Notably, for $B_c^+ \to B_{(s)}^{*0} l^+ \nu_l$ decays, uncertainties range from $\sim (18 -45)\%$ collectively. Such an expanded range of uncertainties would provide a reasonable scope for experimental investigations.
		
        \item[(iii)] The $B_c \to Vl\nu_{l}$ branching ratios are influenced mainly by the form factors $V(\q2),~ A_1(\q2),$ and $A_2(\q2)$. However, it is worth mentioning that the contribution of the form factor $A_0(\q2)$ to these branching ratios can be considered insignificant (see Eq.~\eqref{e41}). It is well known that in the semileptonic $P \to V$ weak decays, the contribution from the form factor $A_2(\q2)$ can be ignored due to the negligible coefficient in the decay rates~\cite{Neubert:1991we, Wang:2008xt}. Furthermore, the branching ratios of the semileptonic decays depend upon the magnitude and signs of the form factors. We want to emphasize that the numerical values of the form factors, especially $A_0(\q2)$ and $A_1(\q2)$ have changed significantly in Type-II correspondence. Therefore, to quantify the effect of self-consistency on the branching ratios of the semileptonic decay modes, we compare our results with those of Type-I correspondence. We found that the numerical results for Type-II scheme (using Eq.~\eqref{e37}) are enhanced by $\sim (50-60)\%$ as compared to the branching ratios in Type-I scheme. Similar observations can be made for the comparison between Type-II* and Type-I results because the results between Type-II* and Type-II differ by less than $\sim 10\%$ for bottom-conserving modes. As expected, the differences between the results for Type-I and Type-II correspondences (inclusive of Type-II*) are sufficiently large and hence cannot be ignored. It may be emphasized that the uncertainties in the Type-I scheme results, arising from variations in form factors, are considerably larger than those in the Type-II scheme results, in some cases differing by an order of magnitude. In addition, we also compare our results with other works~\cite{Li:2023wgq, Faustov:2022ybm, Wang:2008xt, Ivanov:2006ni}, as listed in columns $4-7$ of Table~\ref{t6}. We found that our results for bottom-conserving semileptonic decays are of the same order as compared to predictions from other theoretical models, except for $B_c^+ \to B^{*0} l^+\nu_l$ by Li \textit{et al.}~\cite{Li:2023wgq} using the CLFQM framework within Type-I scheme. 	
        \item[(iv)] The mass difference between the electrons and muons has minimal impact ($\sim 6\%$) on the branching ratios and other physical observables of bottom-conserving semileptonic $B_c \to B_{(s)}^{*}$ decays. Additionally, the comparative variation of bottom-conserving semileptonic differential decay rates for $e$ and $\mu$ lepton modes with respect to $\q2$ are plotted in Figures~\ref{f10a} and~\ref{f10b}. It should be noted that in semileptonic decay processes, the physical observables depend on the mass of the final lepton, with $\q2_{min} = m^2_{l}$ (assuming mass of neutrino is negligible). The differential decay rate plots show distinct peaks corresponding to the lepton mass for the available $\q2$ range, with the same end-points at $\q2_{max}$ as expected. We also have calculated relative longitudinal and transverse decay widths, and their ratios for bottom-conserving $B_c^+ \to Vl^+\nu_l$ decays, as shown in columns $4$, $5$, and $6$ of Table~\ref{t5}, respectively. It is noteworthy to mention that the longitudinal component of the decay widths dominates the transverse component. The magnitude of this difference is relatively modest, with the longitudinal component exceeding the transverse by approximately $(4-6)\%$. The longitudinal decay widths of $B_c \to B_{(s)}^{*}l\nu_{l}$ decays decrease with increasing lepton mass but marginally. 
		
        \item[(v)] We also calculated the expectation values of FB asymmetry, $\langle A_{FB} \rangle$, using Eq.~\eqref{e50}, as shown in column $7$ of Table~\ref{t5}. It is noteworthy that all the $A_{FB}(B_c^+ \to Vl^+\nu_l)$ values are negative. The negative values of $A_{FB}$ in bottom-conserving semileptonic decays reflect the dominance of the parity-violating helicity structure-function, $\mathcal{H}_P$, particularly with a larger contribution from $H_{-}$ amplitude. The $\mathcal{H}_{SL}$ contributions are negligible for electron decay modes, and their values increase with increasing lepton mass. We notice that the magnitude of $\mathcal{H}_{SL}$ in $A_{FB}(B_c^+ \to B_{(s)}^{*} \mu^+\nu_{\mu})$ is larger by $(16-19)\%$ leading to more negative value. Addressing the minimal discrepancy between Type-II and Type-II* in form factors and semileptonic branching ratios, a similar trend can be observed for $A_{FB}$ and other observables. Note that the calculations for observables beyond branching ratios have not been reported in the existing literature in CLFQM approach. Consequently, we exclusively present the physical observables obtained from Type-II formulation. Furthermore, we plot the $\q2$ variation of the FB asymmetry of $B_c^+ \to B_{(s)}^{*}l^+ \nu_l$ decays in Figures~\ref{f11a} and~\ref{f11b}. Notably, $A_{FB}$ shows a distinct rise near $\q2_{min}$, particularly in the electron decay mode. The high precision calculation shows that $A_{FB} \to 0$ as $\q2 \to 0$. Such behavior can also be seen in other works~\cite{Sun:2023iis, Pandya:2023ldv, Colangelo:2021dnv}.  
        \item[(vi)] Furthermore, we calculated the mean values of $\langle C_F^l \rangle$, $\langle P_L^l \rangle$, and  $\langle P_T^l \rangle$, as shown in columns $8$, $9$, and $10$ of Table~\ref{t5}, respectively. It is noteworthy that the bottom-conserving semileptonic decay modes have a negative lepton-side convexity parameter, $C_F^l$, which is due to the predominance of the longitudinal helicity component, $\mathcal{H}_L$. Additionally, the transverse polarization parameter, $P_T^l$ is very small for $e$ decay modes, \textit{i.e.}, $\mathcal{O}(10^{-3})$. Furthermore, we also computed the asymmetry parameter, $\alpha^*$ by using Eq.~\eqref{e54}, as illustrated in column $11$ of Table~\ref{t5}. The asymmetry parameter, $\alpha^*$, is consistently negative for all $B_c$ to $V$ semileptonic decays, indicating the dominance of the longitudinal helicity components, $\mathcal{H}_L$ and $\mathcal{H}_S$. Notably, parameters such as $C_F^l$, $P_L^l$, and $\alpha^*$ show a decrease in magnitude with increases in lepton mass. 
	\end{itemize}
    \subsubsection{\textbf{Bottom-changing decays}}
    In this subsection, we focus on the bottom-changing CKM-enhanced $(\Delta b = -1, \Delta C =-1, \Delta S = 0)$ and CKM-suppressed $(\Delta b = -1, \Delta C =0, \Delta S = 0)$ semileptonic decay modes of $B_c$ mesons, which involve the charm mesons in the final states. One notable aspect of bottom-changing semileptonic decays is that they include $\tau^+ \nu_{\tau}$, alongside $e^+ \nu_e$ and $\mu^+ \nu_{\mu}$ lepton pairs in the final state. We have analyzed and listed our major findings on $B_c \to D^*(J/\psi)l\nu_l$ decays as follows:
	\begin{itemize}
        \item[(i)] The branching ratios of bottom-changing $B_c^+ \to Vl^+\nu_l$ decays range from $\mathcal{O}(10^{-2})$ to $\mathcal{O}(10^{-5})$, as given in Table~\ref{t5}. Among these decays, $B_c^+ \to J/\psi e^+\nu_e$ and $B_c^+ \to J/\psi \mu^+\nu_\mu$ are most dominant with branching ratios $(2.35^{+0.36+0.04}_{-0.33-0.12}) \t 10^{-2}$ and $(2.34^{+0.36+0.04}_{-0.33-0.12}) \t 10^{-2}$, respectively, since $B_c^+ \to J/\psi l^+\nu_l$ decays are both CKM- and kinematically enhanced. On the other hand, the CKM-suppressed $B_c^+ \to D^{*0} l^+\nu_{l}$ decays, involving $b \to u$ transition, have smaller branching ratios, \textit{i.e.}, $\mathcal{O}(10^{-5})$. Similar to bottom-conserving decays, the semileptonic $B_c \to V$ branching ratios of bottom-changing decays generally show greater sensitivity to variations in the $\beta$ parameter than to uncertainties in the constituent quark masses, with the exception of $B_c^+ \to J/\psi l^+ \nu_l$. The branching ratios of $B_c^+ \to D^{*0}l^+\nu_l$ demonstrate a maximum variation of approximately $78\% (39\%)$, while $B_c^+ \to J/\psi l^+ \nu_l$ shows variations of the order of $5\% (15\%)$ corresponding to uncertainties in $\beta$ (quark masses). As previously mentioned, the $B_c \to Vl\nu_{l}$ branching ratios are predominantly determined by $V(\q2)$ and $A_1(\q2)$, while $A_2(\q2)$ and $A_0(\q2)$ have minimal impact. Therefore, the larger uncertainties in these branching ratios can be primarily attributed to the collective influence of form factors $V(\q2)$ and $A_1(\q2)$. Furthermore, the uncertainties in $B_c^+ \to D^{*0}l^+\nu_l$ decays resulting from quark mass variations exhibit a more symmetric distribution compared to those observed in bottom-conserving decays.
        
        \item[(ii)] Similar to bottom-conserving $B_c \to P$ semileptonic decays, the branching ratios of bottom-changing $B_c^+ \to D^0l^+\nu_l$ decays (Table~\ref{t1}) are consistent with the recent LQCD results~\cite{Cooper:2021bkt}. The bottom-changing branching ratios for lattice results are as follows~\cite{Cooper:2021bkt}: $\mathcal{B}(B_c^+ \to D^{0}e^+\nu_{e}) = (3.37 \pm 0.48 \pm 0.08 \pm 0.42) \t 10^{-5}$ and $\mathcal{B}(B_c^+ \to D^{0}\tau^+\nu_{\tau}) = (2.29 \pm 0.23 \pm 0.06 \pm 0.29) \t 10^{-5}$. Interestingly, unlike bottom-conserving $B_c^+ \to P l^+\nu_l$ decays, the semileptonic branching ratios of bottom-changing $B_c^+ \to D^0l^+\nu_l$ decays for Type-II* formulation are smaller by $\sim (39 - 48) \% $ when compared to Type-II predictions. It should be noted that the $B_c \to D$ branching ratios exhibit a maximum uncertainty of $\sim 67\% ~(50\%)$ in Type-II (Type-II*) correspondence due to variations in quark masses affecting the form factors. Additionally, uncertainties arising from variations in the $\beta$ parameter contribute to a maximum deviation of $\sim 40\% ~(36\%)$. These variations in the branching ratio arise from differences in $F_0(q^2)$ and $F_1(q^2)$ corresponding to the $q^2$ formulation, despite being self-consistent.

        \item[(iii)] In bottom-changing semileptonic $B_c$ transitions, the phase space is usually larger compared to bottom-conserving transitions. Specifically, the semileptonic branching ratios involving $B_c \to D^*$ and $B_c \to J/\psi$ have $\sim 44\%$ and $75\%$ difference between the $e$ (or $\mu$) and $\tau$ semileptonic decays, respectively. It is worth noting that the mass difference between the electron and muon does not significantly affect $b \to u(c)$ semileptonic decays. As observed before, the branching ratios decrease with increasing lepton mass, \textit{i.e.}, the branching ratios of $B_c \to J/\psi e(\mu){\nu}_{e(\mu)}$ are larger roughly by a factor of $4$ to that of $\mathcal{B}(B_c \to J/\psi \tau \nu_{\tau})$. Similarly, for $B_c \to D^* e(\mu){\nu}_{e(\mu)}$ decays, the branching ratios of corresponding decay modes are $\sim 2$ times larger than $\mathcal{B}(B_c \to D^* \tau \nu_{\tau})$. In the case of $B_c \to D^*e(\mu){\nu}_{e(\mu)}$ decays, the relative longitudinal and transverse decay widths are equal. Conversely, for $B_c \to J/\psi e(\mu){\nu}_{e(\mu)}$ decays, the relative longitudinal decay widths exhibit a larger magnitude compared to the transverse decay widths. Notably, in all bottom-changing semileptonic decays involving a $\tau$ lepton in the final state, the relative transverse decay widths exceed the longitudinal decay widths. The LQCD prediction for the branching ratio $B_c^+ \to J/\psi \mu^+\nu_{\mu}$ is $(1.50 \pm 0.11 \pm 0.10 \pm 0.03) \%$~\cite{Harrison:2020gvo}. It is interesting to note that their result exhibits significant deviation from the majority of theoretical models~\cite{Ivanov:2006ni, Colangelo:1999zn, Patnaik:2019jho, Huang:2007kb, Qiao:2012vt, Rui:2016opu}, including our Type-II$^{(*)}$ predictions, which span a range of $(2.07 - 6.70)\%$. The exceptions to this trend are the results reported in Refs.~\cite{Wang:2008xt} and~\cite{Faustov:2022ybm}, as listed in Table~\ref{t6}. The discrepancies observed among these results can be attributed to variations in input parameters—primarily quark and pole masses—as well as the diverse $q^2$ parameterizations employed in form factor calculations. In addition, we calculate the LFU ratios between $\tau$ and $e(\mu)$ leptons for both Type-II and Type-II* results as follows:
		\[~~~~~~~~~~~~~~~~~~~~~~~~~~~~~~~~~~~~~~~~~~~~~~~~\text{Type-II}~~~~~~~~~~~~~\text{~Type-II*}~~~~~~~~~~~~~~~\text{LQCD~\cite{Harrison:2020nrv}}\]
		\[ ~\mathcal{R}_{D^{*}} = \frac{\mathcal{B}(B_c^+ \to {D}^{*0}\tau^+{\nu}_{\tau})}{\mathcal{B}(B_c^+ \to {D}^{*0}e(\mu)^+{\nu}_{e(\mu)})} =~0.56^{+0.30+0.60}_{-0.30-0.00}~~~~~~~~~ 0.60^{+0.20+0.50}_{-0.16-0.00}~~~~~~~~~~~~~~~~~~~~~-~~~~~~~~~; \]
		\[ \mathcal{R}_{J/\psi} = \frac{\mathcal{B}(B_c^+ \to J/\psi\tau^+{\nu}_{\tau})}{\mathcal{B}(B_c^+ \to J/\psi e(\mu)^+{\nu}_{e(\mu)})} =~ 0.25^{+0.05+0.02}_{-0.05-0.00}~~~~~~~ 0.25^{+0.05+0.02}_{-0.06-0.00}~~~~~~~~0.2582 \pm 0.0038.\]
		Note that the experimental measurements for the LFU ratios involving $b \to c\tau \nu_{\tau}$ for $J/\psi$ in the final state is $\mathcal{R}_{J/\psi} = 0.71 \pm 0.18 \pm 0.17$~\cite{LHCb:2017vlu}, which is much larger than the theoretical estimates. We want to emphasize that the current SM predictions for these ratios fall within a range of $0.25 - 0.28$~\cite{Sun:2023iis, Faustov:2022ybm, Anisimov:1998uk, Kiselev:2002vz, Hernandez:2006gt}, including ours. It is worth mentioning that the difference between multiple approaches is very small, which also agrees with the LQCD observation~\cite{Harrison:2020nrv}. Furthermore, the experimental observation is substantially larger than the theoretical expectations, even though the cumulative uncertainties in the experimental value are of the order of $50\%$. Thus, more experimental observations would result in a clear picture to establish the scope of new physics beyond the SM in these decays. Similarly, for bottom-conserving $B_c \to B_{(s)}^*$ semileptonic decays, we found
		\[~~~~~~~~~~~~~~~~~~~~~~~~~~~~~~~~~~~~~~~~~~~~\text{Type-II}~~~~~~~~~~~~~~~~\text{~Type-II*}~~~~\]
		\[ \mathcal{R}_{B^{*}} = \frac{\mathcal{B}(B_c^+ \to {B}^{*0}\mu^+{\nu}_{\mu})}{\mathcal{B}(B_c^+ \to {B}^{*0}e^+{\nu}_{e})} =~ 0.95^{+0.16+0.40}_{-0.13-0.25}~~~~~~~~~~~~~~ 0.95^{+0.13+0.24}_{-0.10-0.22}; \]
		\[\mathcal{R}_{B_s^{*}} = \frac{\mathcal{B}(B_c^+ \to {B}_s^{*0}\mu^+{\nu}_{\mu})}{\mathcal{B}(B_c^+ \to {B}_s^{*0}e^+{\nu}_{e})} =~ 0.93^{+0.09+0.31}_{-0.06-0.18}~~~~~~~~~~~~~~0.94^{+0.07+0.17}_{-0.04-0.15},\]
		which is in good agreement with Ref.~\cite{Faustov:2022ybm}.
		
        \item[(iv)] As previously noted, the self-consistency effects are expected to be significant in semileptonic $B_c\to D^*$ decays. The branching ratios of $B_c^+\to D^{*0} l^+\nu_{l}$ decays in the Type-II$^{(}$*${}^{)}$ show a variation ranging from $\sim (57-78)\%$ compared to those in the Type-I scheme. However, self-consistency has a minimal effect on the branching ratios of semileptonic decays of $B_c$ to $J/\psi$ states, with variation of $\sim 20\%$ across Type-I results, when compared to Type-II$^{(}$*${}^{)}$ results. The uncertainty in the branching fractions for the semileptonic $B_c\to D^*$ decays is substantial in the Type-I scheme, reaching an even larger value of $\sim 200\%$ for the $B_c \to D^*\tau \nu_{\tau}$ decay mode. In contrast, the uncertainties associated with the $B_c \to J/\psi l \nu_{l}$ decays are significantly smaller. It may be noted that for $B_c^+ \to D^{*0}l^+\nu_l$ decays for Type-II*, the branching ratios are larger than those of Type-II by $\sim (24 - 29) \% $, this behavior is opposite to the observation made for $B_c^+ \to P l^+\nu_l$ decays. However, the $B_c^+ \to J/\psi l^+\nu_l$ decays differ by less than $\sim 1\%$ on comparison between Type-II and Type-II*. Additionally, to compare our results with other works, we have included the branching ratios from literature~\cite{Li:2023wgq, Faustov:2022ybm, Ivanov:2006ni, Wang:2008xt}, as presented in Table~\ref{t6}. Interestingly, a similar order of discrepancy can be observed in Type-I correspondence scheme results from other works~\cite{Li:2023wgq, Wang:2008xt} as compared to that of Type-II correspondence predictions from our work. For $B_c^+ \to J/\psi l^+ \nu_l$, numerical results of the branching ratios are consistent with other literature; in fact, all the models yield branching ratios of the same order, as mentioned earlier. In general, we observe substantial differences in the numerical values of branching ratios for bottom-changing semileptonic decays from different models that range up to $\sim 60\%$. Particularly, the discrepancy among $\mathcal{B}(B_c^+ \to D^{*0} e(\mu)^+ \nu_{e(\mu)})$ results in the Type-I scheme from other works~\cite{Li:2023wgq, Wang:2008xt} and Type-II scheme in our work also range from $\sim (20-57)\%$. We have also plotted the $\q2$ variation of the differential decay rates of $B_c^+ \to D^{*0}l^+\nu_l$ and $B_c^+ \to J/\psi l^+\nu_l$ in Figures~\ref{f10c} and~\ref{f10d}, respectively. 
		
        \item[(v)] Furthermore, we calculated the $A_{FB}$ for bottom-changing semileptonic decays, and listed in column $7$ of Table~\ref{t5}. The $A_{FB}$ for bottom-changing decays are consistently negative in numerical values due to the dominant contributions from $\mathcal{H}_{SL}$, with a larger magnitude for the $H_0$ helicity amplitude. The exception is evident in the decays that involve an electron in the final state, where $\mathcal{H}_{P}$ is predominantly large because of the larger magnitude of $H_-$ helicity amplitude. However, the contributions from $\mathcal{H}_{P}$ decrease with the lepton mass. Further, as the mass of the lepton increases, the $A_{FB}({B_c\to D^*\tau \nu_{\tau}})$ increases by approximately $20\%$ as compared to $A_{FB}({B_c\to D^*e \nu_{e}})$; however, $A_{FB}(B_c\to J/\psi\tau \nu_{\tau})$ increase up to $30\%$ from $A_{FB}(B_c\to J/\psi e \nu_{e})$. 
		
        \item [(vi)] We observe a behavior similar to that of $A_{FB}$ for observables such as $C_F^l$ and $P_L^l$ with respect to the lepton mass. In this case, the numerical values of channels involving $e$ and $\mu$ are almost identical, while the decays involving $\tau$ show some significant change. Furthermore, the $\alpha^*$ value lead to an observable difference in the case of ${B_c\to J/\psi\tau \nu_{\tau}}$ decay, which have $34\%$ variation with respect to ${B_c\to J/\psi e (\mu)\nu_{e (\mu)}}$ decays. This distinction arises from the influence of the lepton's mass on the decay process. It should be noted that for $B_c\to J/\psi l \nu_{l}$, the uncertainty corresponding to the $\beta$ values is negligible for the physical observables like $A_{FB}$, $C_F^l$, and $\alpha^*$, as shown in Table~\ref{t5}.
        \end{itemize}
	
    \subsection{Nonleptonic decays} \label{S3_C}
    In this subsection, we discuss our predictions for the branching ratios of nonleptonic $B_c \to PV$ decays. Aforementioned, the nonleptonic decays of the $B_c$ meson consist of CKM-enhanced ($\Delta b=0,\Delta C=-1,\Delta S=-1$; $\Delta b=-1,\Delta C=-1,\Delta S=0$; and $\Delta b=-1,\Delta C=0,\Delta S=1$), CKM-suppressed ($\Delta b=0,\Delta C=-1,\Delta S=0$; $\Delta b=-1,\Delta C=-1,\Delta S=1$; and $\Delta b=-1,\Delta C=0,\Delta S=0$), and CKM-doubly-suppressed ($\Delta b=0,\Delta C=-1,\Delta S=1$; $\Delta b=-1,\Delta C=1,\Delta S=1$; and $\Delta b=-1,\Delta C=1,\Delta S=0$) bottom-conserving and bottom-changing decay modes. We calculated the decay amplitude using the decay constants listed in Table~\ref{t7}. Among the form factors listed in Tables~\ref{t3} and~\ref{t4} for $B_c \to P$ and $B_c \to V$ transitions, only the form factors $F_1(\q2)$ and $A_0(\q2)$ are relevant for the numerical evaluation of the branching ratios of $B_c \to PV$ decays. Since the $A_0(\q2)$ form factor is affected by self-consistency issues related to the $B^{(i)}_{j}$ functions, the study of nonleptonic $B_c \to PV$ decays provides an excellent opportunity to investigate such effects between Type-I and Type-II correspondence. We determine the branching ratios of nonleptonic $B_c$ decays involving color-favored diagram (Class-I), color-suppressed diagram (Class-II), and their interference (Class-III) for both large $N_c$ limit and $N_c = 3$, as given by Eqs.~\eqref{e60} and~\eqref{e61} in Sec.~\ref{S2_C}. We list all the possible bottom-conserving $B_c \to PV$ decays in Table~\ref{t8}. The Tables~\ref{t9},~\ref{t10}, and~\ref{t11} show our predictions for bottom-changing decays. Likewise semileptonic decays, we also calculate the uncertainties in branching ratios originating from the uncertainties in the form factors. Furthermore, we compare our results with other theoretical models, namely RIQM~\cite{Naimuddin:2012dy, Nayak:2022qaq}, RCQM~\cite{Ivanov:2006ni}, RQM~\cite{Ebert:2003wc}, QCDF~\cite{Sun:2015exa}, pQCD~\cite{Rui:2012qq, Rui:2014tpa}, and CLFQM (Type-I)~\cite{Zhang:2023ypl}, as given in Tables~\ref{t12}$-$\ref{t14}. We list our key findings as follows.
	\begin{itemize}
        \item[(i)] For bottom-conserving decay modes, the branching ratios of $B_c$ meson decays into $B^{(*)}$ and $B_s^{(*)}$ mesons in the final state ranges from $\mathcal{O}(10^{-2}) \text{~to~ }\mathcal{O}(10^{-6})$ for the Type-II formulation and up to $\mathcal{O}(10^{-5})$ for Type-II*, as shown in Table~\ref{t8}. It is well known for the case of CKM-favored decays that the CKM-enhancement dominates the kinematic suppression, resulting in branching ratios of $\mathcal{O}(10^{-2}) \sim \mathcal{O}(10^{-3})$ for $N_c = \infty$. Among them, the most dominant CKM and color-favored (Class I) decays are $B_c^+ \to \pi^+ B_s^{*0}$ and $B_c^+ \to B_s^{0} \rho^+$, which have branching ratios of $(4.86^{+0.01+0.73}_{-0.14-1.08}) \t 10^{-2}$ and $(3.46^{+0.53+0.25}_{-0.54-0.63}) \t 10^{-2}$, respectively. It is worth noting that for $B_c \to B_{(s)}^{(*)}$ transition, the mass of the spectator $b$ quark is significantly larger than that of the decaying $c$ quark, and the whole momentum is carried by the $b$ quark. Therefore, the transition form factors at $\q2=0$ in such case differ up to $\sim 28\%$ from those at maximum momentum transfer between the initial and final states. This increase in the form factor at $\q2_{max}$ leads to an enhancement of up to $\sim 40\%$ in the branching ratio of $B_c^+ \to \pi^+ B_s^{*0}/\ov{K}^0B^{*+}$ decays (involving $A_0(\q2)$ form factors) for both Type-II and Type-II* formulations. However, the decays involving the form factor $F_1(\q2)$ are affected by less than $14\%$ at $\q2_{max}$. Furthermore, we observe that the CKM-favored nonleptonic bottom-conserving decays exhibit uncertainties typically ranging from $\sim (15-50)\%$, correlating with the uncertainties in their respective form factors. These uncertainties are notably enhanced in color-suppressed channels characterized by lower branching ratios. Moreover, the branching ratios of these decays generally demonstrate increased sensitivity to variations in the $\beta$ values, with a few exceptions to this trend.
		
        \item [(ii)] We want to emphasize that $A_0(\q2)$ transition form factors are affected by the self-consistency problems, and their contribution to semileptonic decays involving vector meson in the final state are suppressed in general. However, nonleptonic $B_c \to PV$ decays that explicitly involve $A_0(\q2)$ form factors would give quantitative measure of self-consistency effects between Type-II and Type-I correspondence schemes. Therefore, we compare our predictions in Type-II and Type-II* with the results in Type-I correspondence, as listed in columns $2,~4$, and $6$ of Table~\ref{t8}. It may be noted that the results in the tables follow the order in which decays involving $A_0(\q2)$ are listed first, decays involving $F_1(\q2)$ are listed thereafter, and Class-III decays involving both (if allowed) are given last for each CKM mode. We found that the results of the Type-I scheme for CKM-favored bottom-conserving modes are significantly smaller for $B_c^+ \to \pi^+B_s^{*0}/\ov{K}^0B^{*+}$ decays. The branching ratio of color-favored $B_c^+ \to \pi^+ B_s^{*0}$ decay in the Type-I scheme is $\sim 90\%$ smaller than that of the Type-II scheme. However, the branching ratio of color-suppressed $B_c^+ \to \ov{K}^0 B^{*+}$ decay changes by $\mathcal{O}(10^{-2})$ in Type-II scheme as compared to Type-I predictions. In addition, as previously noted, the uncertainties in the branching ratios arising from variations in the form factors are substantially larger (ranging from $(70-180)\%$) for the Type-I scheme compared to the Type-II scheme involving $A_0(\q2)$ form factors, as evident from Table~\ref{t8}. Moreover, to accurately assess the magnitude of self-consistency effects, we compare the numerical results of Type-I and Type-II schemes utilizing an identical $\q2$ formulation for both\footnote{Note that the numerical results of Type-II correspondence scheme for Eq.~\eqref{e40} (with parent pole mass): $A_0^{B_cB^{*}}(0)=0.50,~a=-9.92,~b=356.83$ and $A_0^{B_cB_s^{*}}(0)=0.62,~a=-4.25,~b=285.25$ are used.}, \textit{i.e.}, for Eq.~\eqref{e40}, we found that $\mathcal{B}(B_c^+ \to \pi^+ B_s^{*0})$ decay decrease by $\sim 88\%$, while $\mathcal{B}(B_c^+ \to \ov{K}^0 B^{*+})$ decay decreases by $\mathcal{O}(10^{-2})$. Consequently, these substantial discrepancies between the Type-I and Type-II scheme predictions indicate that the effects of self-consistency on such decays are significant and cannot be ignored. In addition, we observe that the difference between the Type-II and Type-II* formulations yields larger variations in the branching ratios for decays involving $F_1(\q2)$ form factor than those involving $A_0(\q2)$ form factor. However, the maximum differences among Type-II and Type-II* is $19\%$ and $27\%$ for decays involving $A_0(\q2)$ and $F_1(\q2)$ form factors, respectively, where the Type-II* formulation predicts larger branching ratios. We reiterate that the form factor $F_1(\q2)$ does not exhibit any self-consistency issues. Therefore, the observed changes in the numerical results of the Type-II correspondence scheme for the decays involving only $F_1(\q2)$ form factor can be attributed to variations arising from the $\q2$ formulations.
		
        \item[(iii)] In the bottom-conserving CKM-suppressed $(\Delta b=0,\Delta C=-1,\Delta S=0)$ modes, the branching ratios for the dominant decays are $\mathcal{B}(B_c^+ \to B^0 \rho^+) = (2.77^{+0.58+0.32}_{-0.49-0.68}) \t 10^{-3}$, $\mathcal{B}(B_c^+ \to K^+ B_s^{*0}) = (2.45^{+0.04+0.37}_{-0.10-0.64}) \t 10^{-3}$, and $\mathcal{B}(B_c^+ \to \pi^+ B^{*0}) = (2.27^{+0.08+0.50}_{-0.16-0.67}) \t 10^{-3}$. All the above stated decays involve color-favored (Class-I) processes. The next order branching ratios are of $\mathcal{O}(10^{-4})$, which correspond to the color-suppressed process, as shown in Table~\ref{t8}. It is interesting to note that the branching ratios of CKM-doubly-suppressed decays are of $\mathcal{O}(10^{-4}) \sim \mathcal{O}(10^{-6})$ with dominant branching ratio, $\mathcal{B}(B_c^+ \to K^+ B^{*0}) = (1.29^{+0.07+0.30}_{-0.11-0.45}) \t 10^{-4}$ for color-favored decay. As observed in CKM-enhanced decays, apart from the variation due to different $\q2$ formulations, the branching ratios of the decays (in Type-II scheme) involving $A_0(\q2)$ form factors change substantially as compared to those of the Type-I scheme. We wish to emphasize that the branching ratios of the decays involving $A_0(\q2)$ form factors and color-favored processes in CKM-suppressed and -doubly-suppressed modes are more seriously affected by self-consistency. The branching ratios of these decays change roughly by $\mathcal{O}(10^{-2}) \sim \mathcal{O}(10^{-3})$ for Type-I scheme as compared to Type-II predictions. Likewise, for CKM-enhanced decays within the Type-II correspondence, the Type-II* branching ratios are larger $\sim (10-26)\%$ as compared Type-II predictions for both CKM-suppressed and -doubly-suppressed modes. In addition, the Type-I decays involving $A_0(q^2)$ form factors are subject to substantial uncertainties, reaching up to a maximum of $\sim 180\%$ (for both CKM-suppressed and -doubly-suppressed modes). It is intriguing to note that despite the nearly symmetric uncertainties in the form factors, the uncertainties in the nonleptonic branching ratios are more asymmetric. The substantial discrepancies observed for nonleptonic bottom-conserving weak decays in Type-I scheme (when compared to self-consistent Type-II results) within the CLFQM framework highlight the inherent inconsistencies in Type-I schemes. These deviations cast doubt on the reliability of results obtained through the Type-I scheme. Furthermore, the uncertainties in Type-II* formulation are in general smaller than those of Type-II formulation with a few exceptions.
  
        \item[(iv)] In addition to the large $N_c$ limit, we also predict branching ratios at $N_c=3$, as shown in columns $3,~5$, and $7$ of Table~\ref{t8} for Type-II, Type-II*, and Type-I, respectively. Aforementioned, we have considered tree-dominated $B_c$ decays and have neglected the small nonfactorizable and penguin contributions within our formalism. As previously mentioned, the number of color degrees of freedom ($N_c$) is usually treated as a phenomenological parameter in weak meson decays to account for nonfactorizable contributions. In the present case, we have used the $N_c=3$ based on the model-independent analysis of $B$ decays, which suggests that $a_2$ has a smaller magnitude~\cite{Browder:1995gi}. We get $a_1=1.09$ and $a_2=-0.09$ (from Eq.~\eqref{e61}) at $N_c=3$ for bottom-conserving $B_c$ decays. Since the bottom-conserving weak decays do not involve any Class-III decays, we expect an overall decrease in the branching ratios of these decays corresponding to smaller values of $a_1$ and $a_2$ at $N_c=3$. We observe that the numerical values of color-suppressed decays at $N_c=3$ are more seriously affected on account of substantial reduction in the magnitude of coefficient $a_2$. Given that we performed calculations for both $N_c = 3$ and at the large $N_c$ limit, we disregarded the uncertainties in the parameters $a_1$ and $a_2$. Consequently, these predictions can be interpreted as representing a reasonable range of numerical results within the current formalism.
		
        \item[(v)] In the case of bottom-changing $B_c$ decays to $D^{(*)}$, $D_s^{(*)}$, and $\eta_c (J/\psi)$ mesons in the final state, we enlist the branching ratio predictions in Tables~\ref{t9},~\ref{t10}, and~\ref{t11}. The most dominant CKM-enhanced decay modes, $B_c^+ \to \eta_c \rho^+$, $B_c^+ \to D_s^+ J/\psi$, $B_c^+ \to \eta_c D_s^{*+}$, and $B_c^+ \to \pi^+ J/\psi$ have branching ratios $(3.91^{+0.13+0.11}_{-0.12-0.24}) \t 10^{-3}$, $(2.44^{+0.99+0.00}_{-0.81-0.07}) \t 10^{-3}$, $(1.69^{+0.05+0.29}_{-0.00-0.26}) \t 10^{-3}$, and $(1.65^{+0.31+0.10}_{-0.29-0.15}) \t 10^{-3}$, respectively, at large $N_c$ limit. Among these, $B_c^+ \to \eta_c \rho^+$ and $B_c^+ \to \pi^+ J/\psi$ decays are color-favored (Class-I) decays, while $B_c^+ \to D_s^+ J/\psi$ and $B_c^+ \to \eta_c D_s^{*+}$ are Class III type decays. We wish to emphasize that the $B_c^+ \to D_s^+ J/\psi$ and $B_c^+ \to \eta_c D_s^{*+}$ decays receive contributions from both color-favored and -suppressed diagrams and interfere destructively at large $N_c$ limit. However, for $N_c=3$, color-favored and -suppressed contributions for both of these decays interfere constructively, yielding larger branching ratios due to the positive values of $a_1$ and $a_2$ (as shown in Eq.~\eqref{e61}). In the CKM-enhanced ($\Delta C=-1,\Delta S=0$) mode, the branching ratios of $B_c^+ \to \ov{D}^0 D^{*+}$ and $B_c^+ \to D^+ \ov{D}^{*0}$ decays are of $\mathcal{O}(10^{-5})$, which falls within the experimental upper limits~\cite{Workman:2022ynf}. In contrast, for ($\Delta C=0,\Delta S=1$) mode, the next order branching ratios for the CKM-favored decays, \textit{e.g.}, $B_c^+ \to \pi^0 D_s^{*+}$, $B_c^+ \to K^+ {D}^{*0}$, $B_c^+ \to {D}^{0} K^{*+}$, $B_c^+ \to D_s^{+} \rho^0$, etc., remain highly suppressed. The branching ratios of these decays range from $\mathcal{O}(10^{-7}) \sim \mathcal{O}(10^{-10})$ as they occur through suppressed $b \to u$ weak transitions. We observe that the uncertainties in the branching ratios of CKM-favored ($\Delta C=-1,\Delta S=0$) and color-suppressed decays are larger (up to $\sim 90\%$). Conversely, the uncertainties for color-favored decays involving $B_c \to \eta_c(J/\psi)$ transitions are roughly below $25\%$. Interestingly, the Class-III decays in ($\Delta C=0,\Delta S=1$) mode have intermediate uncertainties of approximately $40\%$ or less. Furthermore, the dominant branching ratios of bottom-changing decays are smaller as compared to those of bottom-conserving decays. As expected, due to the smaller values of $a_1$ and $a_2$ at $N_c=3$, the branching ratios of all the decays show a decreasing trend, except for Class-III decays\footnote{Note that the reduction in the values of $a_1$ and $a_2$ at $N_c=3$ leads to a proportional decrease in uncertainties across all decay modes, including Class-III decays. This comprehensive uncertainty reduction occurs despite the additive nature of uncertainties, as both color-favored and color-suppressed contributions experience a decrease in magnitude.}. 
		
        \item[(vi)] In the CKM-suppressed ($\Delta C = -1, \Delta S = 1$) decay mode, the dominant $B_c^+ \to \eta_c K^{*+}$ and $B_c^+ \to K^{+} J/\psi$ decays have branching ratios of $\mathcal{O}(10^{-4})$, and the branching ratios for the rest of the decays are of $\mathcal{O}(10^{-6})$. For ($\Delta C = 0, \Delta S = 0$) mode, the branching ratios are of $\mathcal{O}(10^{-4}) \sim \mathcal{O}(10^{-9})$, where the dominant modes $B_c^+ \to \eta_c D^{*+}$ and $B_c^+ \to D^{+} J/\psi$ belong to Class III decays. These decays arise from destructive interference between color-favored and color-suppressed processes, and have the branching ratios of $\mathcal{O}(10^{-4})$ and $\mathcal{O}(10^{-5})$, respectively. Aforementioned, at $N_c = 3$, both coefficients $a_1$ and $a_2$ become positive, which enhances their branching ratios as compared to the values at $N_c = \infty$. Furthermore, $B_c$ meson decaying to ${D}^0 \rho^+$ and $\pi^+ {D}^{*0}$ in the final states are the only decays that involve the color-favored diagram and have branching ratios of $\mathcal{O}(10^{-6})$. In addition, $\mathcal{B}(B_c^+ \to D_{s}^+ \ov{D}^{*0})=(4.08^{+1.36+0.51}_{-1.19-0.51}) \t 10^{-6}$ and $\mathcal{B}(B_c^+ \to \ov{D}^0 D_{s}^{*+})=(2.92^{+0.21+1.97}_{-0.46-1.57}) \t 10^{-6}$ at large $N_c$ limit, which are within the experimental upper limit~\cite{Workman:2022ynf}. As previously observed, decays involving $B_c \to \eta_c(J/\psi)$ transition form factors show varying degrees of uncertainty. For the CKM-favored and -suppressed Class-III modes, these uncertainties range from $\sim (15-45)\%$, whereas Class-I decays demonstrate a more moderate variation of $\sim (5-25)\%$, as given in Tables~\ref{t9} and~\ref{t10}. 

        \item[(vii)] Since we have focused on the discrepancies arising because of the self-consistency problem in form factors and consequently on the decays of $B_c$ meson, we compare our results of the Type-II scheme with those of the Type-I bottom-changing decays. We found that CKM and color-favored bottom-changing decays involving $A_0(\q2)$ form factors suffer a change in branching ratios between $(25-58)\%$. However, the branching ratios of dominant Class-III decays, which involve $F_1(\q2)$ and $A_0(\q2)$ form factors, change by $\sim (20-56)\%$. In the Type-I scheme, we observe that the branching ratios for bottom-changing CKM-suppressed Class-I decays, influenced by the $A_0(q^2)$ form factor (subject to self-consistency issues), decrease by approximately an order of magnitude, with associated uncertainties exceeding $150\%$. Moreover, as previously noted for bottom-conserving decays, the uncertainties in bottom-changing CKM-favored decays affected by self-consistency issues are markedly more pronounced in the Type-I scheme, reaching over $200\%$. It may be noted that, in the above mentioned changes corresponding to self-consistency, we have only considered the branching ratios up to $\mathcal{O}(10^{-6})$. We infer that, similar to bottom-conserving decays, bottom-changing decays are significantly impacted by self-consistency issues, particularly for color-favored decays. The substantial discrepancies between Type-I and Type-II scheme predictions underscore that the effects of self-consistency on such decays are significant and warrant careful consideration. 
		
        \item[(viii)] It is worth noticing that all of the CKM-doubly-suppressed $B_c$ decays belong to the Class-III category. The color-favored and color-suppressed amplitudes interfere destructively to give the branching ratios $\mathcal{O}(10^{-6}) \sim \mathcal{O}(10^{-7})$ for these decays. As intended, the branching ratios of these modes are enhanced at $N_c =3$. However, the effects of self-consistency on the branching ratios of these decays are roughly $(20-90)\%$. In addition, the uncertainties in the branching ratios of Type-II decays, corresponding to variations in quark mass and $\beta$ values, range from $\sim (10-70)\%$ and $\sim(20-90)\%$, respectively. On the other hand, in the Type-I scheme, the uncertainties become exceptionally large, making the results questionable. Furthermore, all the bottom-changing CKM-doubly-suppressed $B_c \to PV$ decays such as, $B_c^+ \to {D}^0D_{(s)}^{*+}$ and $B_c^+ \to {D}_{(s)}^+D^{*0}$ are within the observed experimental upper limit~\cite{Workman:2022ynf}. In the case of bottom-changing decays, both CKM-favored and -suppressed, the difference in branching ratios between Type-II and Type-II* predictions typically remains below $\sim 10\%$. Notable exceptions include $\mathcal{B}(B_c^+ \to \ov{D}^0 D^{*+})$, $\mathcal{B}(B_c^+ \to D^+ \ov{D}^{*0})$, and $\mathcal{B}(B_c^+ \to D_s^+ \ov{D}^{0})$, where differences of up to $\sim 20\%$ are observed. For CKM-doubly suppressed decays, the differences are more substantial, ranging from $\sim (14-32)\%$. Consistent with previous observations, Type-II* branching ratios are in general larger than those of Type-II. However, for all the $B_c$ decays to two charmed mesons in the final state (including Class-III decays), the branching ratios are lower than those predicted by Type-II.
		
        \item[(ix)] It should be noted that the recent experimental observations provide the ratios of branching fractions of nonleptonic $B_c$ decays involving a $J/\psi$ meson in the final state. Therefore, we compared our results with the experimental values reported by LHCb and ATLAS~\cite{ATLAS:2022aiy, LHCb:2016vni, ATLAS:2015jep, HFLAV:2022esi}. The ratios of the branching fractions determined theoretically are expressed as follows:
		\[~~~~~~~~~~~~~~~~~~~~~~~~~~~~~\text{Type-II}~~~~~~~~~~~\text{~Type-II*}~~~~~~~~~~~~~\text{Experimental value}\]
		\[ 
        \frac{\mathcal{B}(B_c^+ \to J/\psi D_s^+)}{\mathcal{B}(B_c^+ \to J/\psi \pi^+)} =
        \begin{array}{ll}
          ~3.35^{+0.71+0.37}_{-0.78-0.25}   &  ~~~2.82^{+0.63+0.33}_{-0.68-0.23}\\
           (1.48^{+0.56+0.14}_{-0.66-0.09})  & ~~(1.45^{+0.49+0.16}_{-0.57-0.09})
        \end{array} 
        ~~~~2.76 \pm 0.33 \pm 0.33~\text{\cite{ATLAS:2022aiy}};~~~~ \]
		\[ 
        \frac{\mathcal{B}(B_c^+ \to J/\psi K^+)}{\mathcal{B}(B_c^+ \to J/\psi \pi^+)} =
        \begin{array}{ll}
          ~0.08^{+0.02+0.01}_{-0.02-0.01}   & ~~~0.08^{+0.02+0.01}_{-0.02-0.01} \\
          (0.07^{+0.02+0.01}_{-0.02-0.01})   & ~~(0.08^{+0.02+0.01}_{-0.02-0.01})
        \end{array}
        ~~~~0.079 \pm 0.007\pm 0.003~\text{\cite{LHCb:2016vni}},\]
        where the values in the parentheses are obtained for large $N_c$ limit. We wish to point out that our results for $N_c = 3$ match well with the experimental values within the uncertainties. Similarly, we compare the ratio of the branching fractions for the nonleptonic $B_c^+ \to J/\psi \pi^+$ decay to the semileptonic $B_c^+ \to J/\psi \mu^+ \nu_{\mu}$ decay with the experiment, as given below,
		\[~~~~~~~~~~~~~~~~~~~~~~~~~~~~~~~\text{Type-II}~~~~~~~~~~~\text{~Type-II*}~~~~~~~~~~~~~~~\text{Experimental value}\]
        \[ 
        \frac{\mathcal{B}(B_c^+ \to J/\psi \pi^+)}{\mathcal{B}(B_c^+ \to J/\psi \mu^+ \nu_{\mu})} =
        \begin{array}{ll}
        ~0.06^{+0.01+0.01}_{-0.02-0.00}  & ~~~0.06^{+0.01+0.01}_{-0.01-0.00} \\
        (0.07^{+0.02+0.01}_{-0.02-0.00})  & ~~(0.07^{+0.02+0.01}_{-0.02-0.00})
        \end{array}~~0.0469 \pm 0.0028 \pm 0.0046~\text{\cite{LHCb:2014rck}}. \]
        We note that our results though larger in magnitude are very close to experimental observation including the errors.
	\end{itemize}
    
    Finally, we compare our numerical results of the branching ratios with those of other theoretical models, such as RIQM~\cite{Naimuddin:2012dy, Nayak:2022qaq}, RCQM~\cite{Ivanov:2006ni}, RQM~\cite{Ebert:2003wc}, QCDF~\cite{Sun:2015exa}, pQCD~\cite{Rui:2012qq, Rui:2014tpa}, and CLFQM (Type-I)~\cite{Zhang:2023ypl}, as shown in Tables~\ref{t12}$-$\ref{t14}. All branching ratio predictions from different models are of the same order, with a few exceptions. Among them, our numerical results for the bottom-conserving branching ratios of $B_c$ decays involving a $B$ meson in the final state match well with the QCDF~\cite{Sun:2015exa} results. We observe that our Type-II predictions for the most dominant bottom-changing CKM-favored $B_c$ decays, \textit{i.e.}, involving $\eta_c \rho^+$, $D_{s}^{+} J/\psi$, $\eta_c D_{s}^{*+}$, and $\pi^+ J/\psi$ in the final state, match very well with the predictions of RCQM~\cite{Ivanov:2006ni}, except $B_c^+ \to \eta_c D_{s}^{*+}$ decay. Notably, for these decays, the predictions from other theoretical models are larger as compared to our results. We also compared our Type-I results with CLFQM (Type-I)~\cite{Zhang:2023ypl} and observed that their values are of the same order but larger than ours by roughly $(30-70)\%$, due to the different input parameters and the exponential $\q2$ formulation used in their work.
	
    \section{Summary and Conclusions} \label{S4}
    In this work, we provide a comprehensive analysis of weak transition form factors, semileptonic decays, and nonleptonic decays of the $B_c$ meson involving $P$ and $V$ mesons in CLFQM. We employed Type-II correspondence in the CLF approach to resolve the self-consistency issues due to the presence of residual $\omega$-dependencies associated with the $B^{(i)}_{j}$ functions, which remain independent of zero-mode contributions. It may be noted that the issues of inconsistency and violation of covariance in Type-I correspondence, which affect the $A_{0}(\q2)$ and $A_1(\q2)$ form factors, can be simultaneously resolved by $M^{\p(\p\p)} \to M_0^{\p(\p\p)}$ considered in Type-II correspondence~\cite{Chang:2018zjq}. However, the quantitative measure of these effects in Type-II correspondence has never been studied in semileptonic as well as nonleptonic decays of the $B_c$ meson. In this analysis, the effects of self-consistency originating from transition form factors on weak decays are quantitatively established. Furthermore, comprehensive investigations into the impacts of self-consistency and covariance on bottom-conserving and bottom-changing semileptonic and nonleptonic decays within the CLFQM framework are conducted. Two primary objectives are pursued: (i) the impact of self-consistency on weak semileptonic and nonleptonic decays is examined using modified form factors within a CLFQM approach, and (ii) self-consistency in bottom-conserving transition form factors, previously unexplored, is established and its effects on bottom-conserving weak decays are quantified. Furthermore, ambiguities related to the $\q2$ parameterization are addressed in the analysis to provide a more robust understanding of these decay processes. The self-consistency affects the numerical results of the form factors $A_{0}(\q2)$ and $A_1(\q2)$, which in turn appear in the semileptonic and nonleptonic decays of the $B_c$ meson. It is well known that the coefficient of $A_0(\q2)$ form factor is suppressed in the semileptonic decay rates; therefore, semileptonic decays only provide a comprehensive picture that corresponds to the effects originating from $A_1(\q2)$ form factor. Thus, to observe the effect of $A_0(\q2)$ form factor, we calculated the $B_c \to PV$ decays which involve $F_1(\q2)$ and $A_0(\q2)$ form factors. Therefore, we calculated the transition form factors in CLFQM formalism in Tables~\ref{t3} and~\ref{t4}. In the current work, we thoroughly examined the appropriate $\q2$ formulations, especially for bottom-conserving transitions involving $B_c \to V(P)$ form factors. Therefore, we have analyzed two different $\q2$ formulations in Type-II correspondence referred to as Type-II and Type-II* by using Eqs.~\eqref{e37} and~\eqref{e38}, respectively. We also compared our results with Type-I correspondence for the $\q2$ formulation in Eq.~\eqref{e40} to assess the effects of self-consistency quantitatively. In addition, we have incorporated the uncertainties in form factors originating from quark masses and $\beta$ parameters in our analysis. Consequently, we observed their implications on semileptonic and nonleptonic weak decays of $B_c$ meson. In addition, we calculated the experimentally significant physical observables, namely, the FB asymmetry, lepton-side convexity parameter, longitudinal (transverse) polarization of the charged lepton, and asymmetry parameter. We list our major conclusions as follows.
	\begin{itemize}
        \item We reconfirmed that the form factors $A_{0}(\q2)$ and $A_1(\q2)$ in CLFQM Type-I correspondence scheme acquire zero-mode contributions through $B^{(i)}_{j}$ functions, which results in different numerical values for the longitudinal and transverse polarization states. These issues are resolved within Type-II correspondence, which ensures self-consistency and covariance of matrix elements. It may be emphasized that the zero-mode contributions in Type-II correspondence vanish numerically, though exist formally in the analytical relations of the form factors. For bottom-conserving transitions, the numerical results of the Type-II$^{(}$*${}^{)}$ form factors, $A_{0}(0)$ and $A_1(0)$, show a significant change of $(70-90)\%$ and $\sim 23\%$, respectively, as compared to those of the Type-I scheme. Similarly, for bottom-changing transitions, we observed that the numerical values of the form factor $A_{0}(0)(A_1(0))$ in Type-II correspondence, for both Eqs.~\eqref{e37} and~\eqref{e38}, roughly vary by $\sim 30\%~(10\%)$ as compared to Type-I for $B_c \to D_{(s)}^*$ transitions. We also observe that these form factors are sensitive to $\q2$ formulations, resulting in significantly different slope parameters (coefficients). Therefore, we conclude that the improvement in the numerical results of Type-II correspondence cannot simply be determined from the variation of form factors at $\q2 =0$; the modification in the numerical values of slope parameters also plays a significant role in the quantitative evaluation of these effects. Furthermore, the Type-II correspondence influences $B_c \to J/\psi$ transition form factors minimally, as compared to both bottom-conserving and other bottom-changing transition form factors.
		
        \item We also found that the $M^{\p(\p\p)} \to M_0^{\p(\p\p)}$ transformation, in general, affects the numerical values of all the transition form factors irrespective of the spin-parity of the final state meson. Therefore, the numerical values of the form factors which do not suffer from self-consistency issues have also been modified. We found that the numerical results for the Type-II$^{(}$*${}^{)}$ form factors $F^{B_cB_{(s)}}(\q2)$ are in very good agreement with the LQCD observations for both at $\q2=0$ and $\q2_{max}$. On the other hand, the numerical values of the form factors $F^{B_cD_{s}}(\q2)~(F^{B_cD}(\q2))$ are in good agreement with the LQCD predictions within $\sim 15\%$($\sim 9\%$). 
		
        \item We found that $\mathcal{B}(B_c \to B_s^{(*)}l\nu_l)$ and $\mathcal{B}(B_c \to J/\psi(\eta_c) l\nu_l)$ are the most dominant among the $B_c \to V(P)l\nu_l$ semileptonic decays. Our results for $\mathcal{B}(B_c^+ \to B_{(s)}^{0}l^+\nu_{l})$ are in good agreement with the recent LQCD predictions. In addition, the decay width ratios of bottom-conserving semileptonic decays involving pseudoscalar meson ($B_{s}^{0}$ and $ B^{0}$) in final state for Type-II* match well with LQCD expectations. Furthermore, the decays involving the $\tau$ lepton have the lowest branching ratios among all the decays because of the significantly larger mass of the $\tau$ lepton. We quantified the effect of self-consistency on the branching ratios of the semileptonic decay modes by comparing our results with those of Type-I correspondence. We found that the numerical results for the Type-II scheme are larger by $(50-60)\%$, $(57-78)\%$, and around $20\%$ as compared to the branching ratios in the Type-I scheme involving $B_c \to B_{(s)}^{*}$, $B_c \to D^{*}$, and $B_c \to J/\psi$ semileptonic decays, respectively. Furthermore, we found that our LFU ratio involving $b \to c\tau \nu_{\tau}$ for $J/\psi$ in the final state match well with LQCD and other theoretical models; however, are smaller than the experimental measurement.
		
        \item For the nonleptonic $B_c$ decays, branching ratios will be affected by the self-consistency issues for decays involving $A_0(\q2)$ transition form factors. These decays presented an excellent opportunity to observe these effects in a quantitative manner. Interestingly, the branching ratios of CKM and color-favored bottom-conserving $B_c \to P V$ decays are observed to be affected by approximately $\sim 90\%$, while those of bottom-changing decays are impacted by $\sim (25-57)\%$. However, the color-favored CKM-suppressed and doubly-suppressed modes are more seriously affected, where some of the branching ratios are changing by $\sim 100\%$. Therefore, we conclude that the self-consistency effects are predominant in $B_c \to P V$ decays. Furthermore, it is observed that the impact of uncertainties associated with quark mass and $\beta$ parameters is more pronounced in bottom-changing transitions (expect for $B_c \to J/\psi$) and decays. Notably, the substantial uncertainties in the slope parameters of $\q2$ formulations do not significantly affect the branching ratio values in semileptonic and nonleptonic decays. 
        
        \item Finally, we conclude that both bottom-conserving and bottom-changing decays are significantly affected by self-consistency issues arising through the form factors. These impacts can influence the branching ratios by up to two orders of magnitude, with certain decay channels exhibiting particularly large uncertainties in Type-I scheme. Consequently, the substantial variation in predictions, coupled with uncertainties of greater magnitude, casts doubt on the validity of the results obtained through Type-I scheme. Furthermore, the observed discrepancies between Type-I and Type-II scheme predictions highlight the crucial role of self-consistency considerations. These findings emphasize the critical importance of thoroughly evaluating self-consistency effects in future studies of such decays.
 	\end{itemize}
    
    Thus, the agreement between our predictions in Type-II correspondence scheme and the LQCD results assures the reliability of our numerical results for $B_c$ meson decays. We wish to remark that the we have ignored nonfactorizable processes, for example, W-exchange, W-annihilation, penguin processes in our analysis of nonleptonic $B_c$ weak decays. However, the study of nonfactorizable contributions and CP-symmetries can more reliably be carried out in model-independent manner that requires huge amount of experimental data. We hope that the experimental observation of these $B_c$ weak decays can help to shed some light on the underlying physics of the $B_c$ meson.
	
	\section*{Acknowledgment}
	The authors are pleased to express their thanks to H. Y. Cheng, C. K. Chua, H. M. Choi, C. R. Ji, and Gautam Bhattacharyya for their helpful comments and discussions. The author (RD) gratefully acknowledge the financial support by the Department of Science and Technology (SERB:CRG/2018/002796), New Delhi.
 
 \newpage
	\appendix
	\section*{Appendix}
        \section{Branching ratios of $B_c^+ \to Pl^{+}\nu_l$ decays }
	\label{aB}
        We list the numerical values of $B_c^+ \to Pl^{+}\nu_l$ semileptonic decays using the form factors given in Tables~\ref{t3} and~\ref{t4}, and the numerical inputs are discussed in Sec.~\ref{S3}.
	
	\begin{table}[ht]
        \caption{Branching ratios of $B_c^+ \to Pl^{+}\nu_l$ decays. For the definitions of Type-II, Type-II*, and Type-I, refer to the caption of Table~\ref{t3}.}
		\label{ta1}
\begin{tabular}{c | c  c  c}
			\hline \hline
			{Decay}                      &  Type-II             &  Type-II* & Type-I \\ \hline
			{$B_c^+ \to B^{0} e^+ \nu_e$} 	 		& $(9.12^{+1.76+0.82}_{-1.55-1.72}) \t 10^{-4}$ & $(1.27^{+0.25+0.00}_{-0.23-0.02}) \t 10^{-3}$ & $(5.00^{+0.74+1.84}_{-0.68-2.04}) \t 10^{-4}$ \\
			{$B_c^+ \to B^{0} \mu^+ \nu_{\mu}$}  	& $(8.79^{+1.71+0.76}_{-1.50-1.63}) \t 10^{-4}$ & $(1.21^{+0.24+0.00}_{-0.21-0.02}) \t 10^{-3}$ & $(4.79^{+0.69+1.73}_{-0.64-1.91}) \t 10^{-4}$ \\	\hline
			{$B_c^+ \to B_s^{0} e^+ \nu_e$ }     	& $(1.56^{+0.20+0.10}_{-0.23-0.22}) \t 10^{-2}$ & $(2.03^{+0.29+0.00}_{-0.31-0.04}) \t 10^{-2}$ & $(1.06^{+0.10+0.24}_{-0.12-0.33}) \t 10^{-3}$ \\
			{$B_c^+ \to B_s^{0} \mu^+ \nu_\mu$ } 	& $(1.48^{+0.20+0.09}_{-0.22-0.20}) \t 10^{-2}$ & $(1.90^{+0.28+0.00}_{-0.30-0.04}) \t 10^{-2}$ & $(1.01^{+0.10+0.22}_{-0.12-0.30}) \t 10^{-3}$ \\ \hline
			{$B_c^+ \to D^{0}e^+ \nu_e$} 			& $(2.89^{+1.95+1.17}_{-1.12-1.15}) \t 10^{-5}$ & $(1.77^{+0.91+0.64}_{-0.57-0.55}) \t 10^{-5}$ & $(1.36^{+1.02+0.80}_{-0.55-0.60}) \t 10^{-5}$ \\
			{$B_c^+ \to D^{0}\mu^+ {\nu}_\mu$} 		& $(2.89^{+1.94+1.17}_{-1.12-1.15}) \t 10^{-5}$ & $(1.77^{+0.91+0.64}_{-0.56-0.55}) \t 10^{-5}$ & $(1.36^{+1.01+0.80}_{-0.54-0.60}) \t 10^{-5}$ \\
			{$B_c^+ \to D^{0}\tau^+ {\nu}_\tau$ }	& $(2.18^{+1.47+0.77}_{-0.86-0.83}) \t 10^{-5}$ & $(1.13^{+0.58+0.41}_{-0.36-0.35}) \t 10^{-5}$ & $(7.83^{+6.93+5.36}_{-3.52-3.93}) \t 10^{-6}$ \\
\hline \hline
		\end{tabular}
	\end{table}
	

	\newpage
	\begin{table}[ht]
		\caption{\label{t1} Constituent quark masses and the Gaussian parameters $\beta$ for P and V mesons.\footnote[1]{Note that here $q$ denotes either $u$ or $d$ quark.}}
		\begin{tabular}{c  c  c  c  c   c}
			\hline \hline
			\multicolumn{6}{c}{Constituent quark masses (in GeV)} \\ \hline
			\multicolumn{2}{c}{$m_{u} = m_{d} = 0.26 \pm 0.04;$}   & $m_{s} =0.45 \pm 0.05;$ &  \multicolumn{2}{c}{$m_{c} =1.45 \pm 0.20;$} &  $m_{b} =4.64 \pm 0.20$  \\ \hline \hline
			\multicolumn{6}{c}{Gaussian parameters $\beta$ (in GeV)} \\ \hline 
			$^{2S+1} L_J$		& $^1S_0$	& $^3S_1$ 	& $^{2S+1} L_J$		& $^1S_0$	& $^3S_1$	\\ \hline
			$\beta_{c\bar q}$	& $0.4656^{+0.0217}_{-0.0212}$	& $0.4255 \pm 0.0426$	& $\beta_{b\bar q}$	& $0.5547^{+0.0260}_{-0.0261}$	& $0.5183 \pm 0.0518$  \\
			$\beta_{c\bar s}$	& $0.5358^{+0.0137}_{-0.0135}$	& $0.4484 \pm 0.0448$	& $\beta_{b\bar s}$	& $0.6103^{+0.0330}_{-0.0331}$	& $0.5589 \pm 0.0559$	\\
			$\beta_{c\bar c}$	& $0.7690 \pm 0.0049$	& $0.6492 \pm 0.0069$	& $\beta_{b\bar c}$	& $0.9207 \pm 0.0921$	& $-$	\\ \hline \hline
		\end{tabular}
	\end{table}
	
 \begin{table}[ht]
		\caption{Decay constants for P and V mesons (in MeV)\footnote[2]{Available experimental values are listed. The numerical values in the parentheses are from LQCD. Note that we only listed the central values (uncertainties are ignored).}.}
		\label{t7} 
		\begin{tabular}{c c  c c}
			\hline  \hline 
			\multicolumn{4}{c}{Decay Constants }	\\ \hline
			$f_{\pi}$	 	& $130.56$~\cite{Workman:2022ynf} &
			$f_{\rho}$	 	& $210$~\cite{Bharucha:2015bzk}  \\
			$f_{K}$	     	& $155.7$~\cite{Workman:2022ynf} &
			$f_{K^{*}}$	 	& $204$~\cite{Bharucha:2015bzk}	\\
			$f_{\eta}$ 		& $(181.14)$~\cite{Dowdall:2013rya} &
			$f_{\phi}$	 	& $(228.5)$~\cite{Chakraborty:2017hry}	    \\
			$f_{D}$      	& $203.8$~\cite{Workman:2022ynf}   &
			$f_{D^{*}}$	 	& $(223.5)$~\cite{Lubicz:2017asp}	\\
			$f_{D_s}$    	& $250.1$~\cite{Workman:2022ynf}   &
			$f_{{D_s}^{*}}$ & $213$~\cite{BESIII:2023zjq}	\\
			$f_{\eta_c}$ 	& $335$~\cite{Workman:2022ynf}      &
			$f_{J/\psi}$	& $416$~\cite{Workman:2022ynf}	\\
			\hline \hline
		\end{tabular}
	\end{table}
	\begin{table}[ht]
		\caption{Transition pole masses for $B_c \to P$ and $V$ form factors (in GeV). }
		\label{t2}
		\begin{tabular}{c | c | c | c | c}
			\hline 	\hline
			\multirow{2}{*}{Quark transition}& $F_1(\q2), V(\q2)$ & $F_0(\q2)$	         & $A_0(\q2)$		  & $A_1(\q2), A_2(\q2)$ \\ \cline{2-5}
			& $J^P = 1^-$   & $0^+$	& $0^-$	   & $1^+$ \\ \hline
			\multicolumn{5}{c}{Bottom-conserving transitions} \\ \hline
			$c \to d$	    & $2.010$ & $2.308$ &  $1.870$ & $2.422$ \\
			$c \to s$	    & $2.112$ & $2.318$ & $1.968$  & $2.460$ \\  \hline 
			\multicolumn{5}{c}{Bottom-changing transitions} \\ \hline
			$b \to u$	    & $5.325$ & $5.670$ &  $5.279$ & $5.726$ \\
			$b \to s$	    & $5.415$ & $5.762$ &  $5.367$ & $5.829$ \\
			$b \to c$	    & $6.473$ & $6.836$ & $6.274$  & $6.866$ \\ \hline \hline
		\end{tabular}
	\end{table}

	\begin{table}[ht]
		\caption{Form factors of bottom-conserving transitions. Note that Type-II (Type-II*) results represent the transition form factors calculated using $\q2$ dependence given in Eq.~\eqref{e37} (Eq.~\eqref{e38}) with transition pole masses as listed in Table~\ref{t1}. Type-I results represent the transition form factors calculated using $\q2$ dependence given in Eq.~\eqref{e40} with parent pole mass ($M_{B_c}$).}
		\label{t3}
		\scalebox{0.70}{
		\begin{tabular}{c|  c  c  c  c  |c  c  c  c} \hline 	\hline
			\multirow{3}{*}{Form Factor} & \multicolumn{4}{c|}{Type-II (Type-II*)} & \multicolumn{4}{c}{Type-I} \\ 
			\cline{2-9}
			& $F(0)$ & \multirow{2}{*}{$F(\q2_{max})$} & $a$ & $b$ & \multirow{2}{*}{$F(0)$} & \multirow{2}{*}{$F(\q2_{max})$} & \multirow{2}{*}{$a$} & \multirow{2}{*}{$b$} \\  
			& $(a'_0)$ &  & $(a'_1)$ & $(a'_2)$ &  &  &  &  \\  \hline
			\multicolumn{9}{c}{$B_c \to P$ transitions} \\ \hline
			\multirow{2}{*}{$F_0^{B_cB}$} 
			& $0.60^{+0.05+0.01}_{-0.05-0.03}$ & $0.79^{+0.10+0.03}_{-0.09-0.08}$ & $0.39^{+0.21+0.04}_{-0.16-0.20}$ & $0.74^{+0.08+0.61}_{-0.07-0.31}$ 
			& \multirow{2}{*}{$0.53^{+0.04+0.03}_{-0.04-0.05}$} & \multirow{2}{*}{$0.48^{+0.02+0.09}_{-0.02-0.12}$} & \multirow{2}{*}{$-2.77^{+0.81+4.16}_{-0.84-8.05}$} & \multirow{2}{*}{$193.24^{+23.70+90.89}_{-20.63-54.55}$}  \\
			& ($0.62^{+0.05+0.01}_{-0.05-0.02}$)    & ($0.86^{+0.09+0.00}_{-0.09-0.01}$)    & ($-41.53^{+10.69+6.16}_{-12.72-6.95}$)  & ($816.36^{+539.60+299.35}_{-418.37-227.35}$) 
			&  &  &  &  \\
			\multirow{2}{*}{$F_1^{B_cB}$} 
			& $0.60^{+0.05+0.02}_{-0.05-0.04}$ & $0.82^{+0.08+0.07}_{-0.08-0.15}$ & $0.31^{+0.11+0.09}_{-0.08-0.39}$ & $0.79^{+0.06+0.68}_{-0.05-0.34}$	
			& \multirow{2}{*}{$0.53^{+0.04+0.05}_{-0.04-0.08}$} & \multirow{2}{*}{$0.36^{+0.02+0.14}_{-0.02-0.14}$} & \multirow{2}{*}{$-9.40^{+0.64+9.63}_{-0.46-18.11}$} & \multirow{2}{*}{$336.00^{+28.16+179.78}_{-25.74-114.81}$} \\ 
			& ($0.62^{+0.05+0.01}_{-0.05-0.02}$)    & ($0.97^{+0.09+0.00}_{-0.09-0.01}$)    & ($-57.32^{+9.42+8.68}_{-10.46-8.32}$)  & ($1570.88^{+443.22+425.90}_{-373.30-402.62}$) 
			&  &  &  &  \\
\hline \hline
		\end{tabular}}
	\end{table}
	
	\begin{table}[ht!]
		\caption{Form factors of bottom-changing transitions. For the definitions of Type-II, Type-II*, and Type-I, refer to the caption of Table~\ref{t3}.}
		\label{t4}
		\scalebox{0.75}{
			\begin{tabular}{c | c  c c c  |c  c  c  c} \hline 	\hline
				\multirow{3}{*}{Form Factor} & \multicolumn{4}{c|}{Type-II (Type-II*)} & \multicolumn{4}{c}{Type-I} \\ 
				\cline{2-9}
				& $F(0)$ & \multirow{2}{*}{$F(\q2_{max})$} & $a$ & $b$ & \multirow{2}{*}{$F(0)$} & \multirow{2}{*}{$F(\q2_{max})$} & \multirow{2}{*}{$a$} & \multirow{2}{*}{$b$} \\  
				& $(a'_0)$ &  & $(a'_1)$ & $(a'_2)$ &  &  &  &  \\  \hline
				\multicolumn{9}{c}{$B_c \to P$ transitions} \\ \hline
				\multirow{2}{*}{$F_0^{B_cD}$} 
				& $0.17^{+0.04+0.03}_{-0.03-0.03}$ & $0.67^{+0.22+0.08}_{-0.19-0.15}$ & $1.09^{+0.09+0.27}_{-0.10-0.23}$ & $0.79^{+0.33+0.58}_{-0.29-0.32}$
				& \multirow{2}{*}{$0.17^{+0.04+0.03}_{-0.03-0.03}$}    & \multirow{2}{*}{$0.22^{+0.14+0.12}_{-0.08-0.10}$}      & \multirow{2}{*}{$2.36^{+0.08+0.21}_{-0.11-0.22}$}      & \multirow{2}{*}{$3.82^{+1.13+1.97}_{-1.01-1.22}$}  \\
				& ($0.17^{+0.04+0.03}_{-0.03-0.03}$)    & ($0.85^{+0.13+0.01}_{-0.12-0.02}$)    & ($-1.56^{+0.18+0.15}_{-0.14-0.09}$)   & ($4.45^{+0.15+1.81}_{-0.23-1.77}$) 
				&  &  &  &  \\
				\multirow{2}{*}{$F_1^{B_cD}$} 
				& $0.17^{+0.04+0.03}_{-0.03-0.03}$ & $0.74^{+0.42+0.30}_{-0.27-0.31}$ & $1.51^{+0.05+0.19}_{-0.07-0.20}$ & $1.59^{+0.48+0.93}_{-0.43-0.55}$
				& \multirow{2}{*}{$0.17^{+0.04+0.03}_{-0.03-0.03}$}    & \multirow{2}{*}{$0.10^{+0.06+0.07}_{-0.04-0.05}$}      & \multirow{2}{*}{$3.01^{+0.03+0.00}_{-0.07-0.11}$}      & \multirow{2}{*}{$9.02^{+1.92+3.90}_{-1.78-2.49}$} \\ 
				& ($0.17^{+0.04+0.03}_{-0.03-0.03}$)    & ($1.50^{+0.25+0.05}_{-0.22-0.08}$)    & ($-2.45^{+0.35+0.06}_{-0.34-0.00}$)   & ($12.11^{+1.11+1.25}_{-1.33-1.64}$) 
				&  &  &  &   \\
\hline \hline
		\end{tabular}}
	\end{table}

	\begin{turnpage}
	\begin{table}[ht]
		\caption{Branching ratios, relative decay widths and physical observables of $B_c^+ \to Vl^{+}\nu_l$ decays in Type-II CLFQM. }
		\label{t5}
		\begin{tabular}{c  c  c  c  c  c  c  c  c  c  c}
			\hline 	\hline
			Quark      & \multirow{2}{*}{Decay}  & \multirow{2}{*}{$\mathcal{B}$}   & \multirow{2}{*}{$\frac{\Gamma_{L}}{\Gamma}$} & \multirow{2}{*}{$\frac{\Gamma_{T}}{\Gamma}$}   & \multirow{2}{*}{$\frac{\Gamma_{L}}{\Gamma_{T}}$} & \multirow{2}{*}{$\langle A_{FB}\rangle$} & \multirow{2}{*}{$\langle C_{F}^{l}\rangle$}& \multirow{2}{*}{$\langle P_{L}^{l}\rangle$} & \multirow{2}{*}{$\langle P_{T}^{l}\rangle$} & \multirow{2}{*}{$\alpha^*$}\\ 
			Transition &  &  &  &  &  &  & & & & \\ \hline
			\multirow{2}{*}{$c \to d$} &$B_c^+ \to B^{*0} e^+ \nu_e$   & $(2.09^{+0.20+0.38}_{-0.25-0.69}) \t 10^{-3}$ & $0.53$ & $0.47$ & $1.13$ & $-0.17^{+0.02+0.01}_{-0.03-0.00}$ & $-0.44^{+0.02+0.00}_{-0.02-0.03}$ & $1.00$ & $-1.03 \t 10^{-3}$ & $-0.39^{+0.02+0.00}_{-0.01-0.02}$  \\
			& $B_c^+ \to B^{*0} \mu^+ \nu_{\mu}$                       & $(1.98^{+0.19+0.37}_{-0.23-0.66}) \t 10^{-3}$ & $0.52$ & $0.48$ & $1.11$ & $-0.20^{+0.02+0.01}_{-0.03-0.00}$ & $-0.35^{+0.02+0.00}_{-0.01-0.02}$ & $0.91$ & $-0.13$            & $-0.38^{+0.02+0.00}_{-0.01-0.02}$  \\ \hline
			\multirow{2}{*}{$c \to s$} &$B_c^+ \to B_s^{*0} e^+ \nu_e$ & $(3.53^{+0.15+0.49}_{-0.24-0.81}) \t 10^{-2}$ & $0.53$ & $0.47$ & $1.14$ & $-0.14^{+0.02+0.01}_{-0.02-0.00}$ & $-0.45^{+0.02+0.00}_{-0.01-0.01}$ & $1.00$ & $-1.20 \t 10^{-3}$ & $-0.39^{+0.02+0.00}_{-0.01-0.01}$  \\
			& $B_c^+ \to B_s^{*0} \mu^+ \nu_\mu$	                   & $(3.30^{+0.14+0.46}_{-0.22-0.77}) \t 10^{-2}$ & $0.53$ & $0.47$ & $1.11$ & $-0.18^{+0.02+0.01}_{-0.02-0.00}$ & $-0.35^{+0.02+0.00}_{-0.01-0.01}$ & $0.90$ & $-0.15$            & $-0.38^{+0.01+0.00}_{-0.01-0.01}$  \\ \hline
			\multirow{3}{*}{$b \to u$} & $B_c^+ \to D^{*0}e^+ \nu_e$   & $(1.04^{+0.38+0.00}_{-0.39-0.81}) \t 10^{-4}$ & $0.50$ & $0.50$ & $1.00$ & $-0.25^{+0.05+0.24}_{-0.07-0.00}$ & $-0.37^{+0.10+0.47}_{-0.08-0.06}$ & $1.00$ & $-1.49 \t 10^{-4}$ & $-0.33^{+0.09+0.43}_{-0.06-0.05}$  \\
\end{tabular}
	\end{table}
\end{turnpage}
	
	\begin{table}[ht]
		\caption{Branching ratios of $B_c^+ \to Vl^{+}\nu_l$ decays. For the definitions of Type-II, Type-II*, and Type-I, refer to the caption of Table~\ref{t3}.}
		\label{t6}
		\scalebox{0.92}{
		\begin{tabular}{c | c  c | c  c  c  c}
			\hline \hline
			
			\multirow{2}{*}{Decay}                      & \multicolumn{2}{c|}{Ours}             & \multirow{2}{*}{\cite{Li:2023wgq}} & \multirow{2}{*}{\cite{Faustov:2022ybm}} & \multirow{2}{*}{\cite{Wang:2008xt}} & \multirow{2}{*}{\cite{Ivanov:2006ni}} \\
			\cline{2-3}
			& Type-II (Type-II*) & Type-I &  &  &   &   \\\hline
			\multirow{2}{*}{$B_c^+ \to B^{*0} e^+ \nu_e$}         
			& $(2.09^{+0.20+0.38}_{-0.25-0.69}) \t 10^{-3}$ & \multirow{2}{*}{$(9.42^{+0.63+3.28}_{-0.95-4.57}) \t 10^{-4}$} & \multirow{2}{*}{$7.77 \t 10^{-4}$} & \multirow{2}{*}{$1.79 \t 10^{-3}$} & \multirow{2}{*}{$1.41 \t 10^{-3}$} & \multirow{2}{*}{$6.30 \t 10^{-3}$} \\
			& ($(2.32^{+0.19+0.31}_{-0.26-0.37}) \t 10^{-3}$) &  &  &  &  &   \\
			\multirow{2}{*}{$B_c^+ \to B^{*0} \mu^+ \nu_{\mu}$}   
			& $(1.98^{+0.19+0.37}_{-0.23-0.66}) \t 10^{-3}$ & \multirow{2}{*}{$(8.57^{+0.59+9.89}_{-0.87-4.25}) \t 10^{-4}$} & \multirow{2}{*}{$7.36\t 10^{-4}$ } & \multirow{2}{*}{$1.72 \t 10^{-3}$} & \multirow{2}{*}{$1.34 \t 10^{-3}$} & \multirow{2}{*}{-                } \\
			& ($(2.20^{+0.18+0.29}_{-0.24-0.35}) \t 10^{-3}$) &  &  &  &  &   \\\hline
			\multirow{2}{*}{$B_c^+ \to B_s^{*0} e^+ \nu_e$ }      
			& $(3.53^{+0.15+0.49}_{-0.24-0.81}) \t 10^{-2}$ & \multirow{2}{*}{$(1.80^{+0.02+0.35}_{-0.08-0.65}) \t 10^{-2}$} & \multirow{2}{*}{$1.42 \t 10^{-2}$} & \multirow{2}{*}{$2.30 \t 10^{-2}$} & \multirow{2}{*}{$1.96 \t 10^{-2}$} & \multirow{2}{*}{$2.37 \t 10^{-2}$} \\
			& ($(3.76^{+0.13+0.32}_{-0.23-0.41}) \t 10^{-2}$) &  &  &  &  &   \\
\hline \hline
		\end{tabular}}
	\end{table}
	
	
	\begin{turnpage}
	\begin{table}[ht]
		\caption{Branching ratios of bottom-conserving $B_c \to PV$ decays for CKM-favored, -suppressed and -doubly-suppressed modes. For the definitions of Type-II, Type-II*, and Type-I, refer to the caption of Table~\ref{t3}.}
		\label{t8}
  \scalebox{0.92}{
		\begin{tabular}{ c | c  c | c  c  |c c}
			\hline 	\hline
			\multirow{2}{*}{Decay}        & \multicolumn{2}{c|}{Type-II} & \multicolumn{2}{c|}{Type-II*} & \multicolumn{2}{c}{Type-I} \\ 
			\cline{2-7}
			& $N_c = \infty$ & $N_c = 3$ & $N_c = \infty$	& $N_c = 3$	& $N_c = \infty$ &  $N_c = 3$	 \\ \hline
			\multicolumn{7}{c}{CKM-favored ($\Delta b=0,~\Delta C=-1,~\Delta S=-1$)}   \\ \hline
			$B_c^+ \to \pi^+ B_s^{*0}$	  & $(4.86^{+0.01+0.73}_{-0.14-1.08}) \t 10^{-2}$ & $(3.64^{+0.01+0.55}_{-0.11-0.81}) \t 10^{-2}$ & $(5.21^{+0.03+0.55}_{-0.17-0.66}) \t 10^{-2}$ & $(3.90^{+0.02+0.41}_{-0.13-0.50}) \t 10^{-2}$   & $(4.48^{+0.00+4.03}_{-0.36-2.95}) \t 10^{-3}$ & $(3.35^{+0.00+3.02}_{-0.27-2.21}) \t 10^{-3}$ \\ 
			$B_c^+ \to \ov{K}^0 B^{*+}$   & $(7.74^{+0.44+1.81}_{-0.63-2.71}) \t 10^{-3}$ & $(2.41^{+0.14+0.56}_{-0.20-0.84}) \t 10^{-4}$ & $(9.46^{+0.54+1.22}_{-0.79-1.47}) \t 10^{-3}$ & $(2.95^{+0.17+0.38}_{-0.25-0.46}) \t 10^{-4}$	& $(1.38^{+0.06+2.43}_{-0.04-1.17}) \t 10^{-5}$ & $(4.29^{+0.18+7.58}_{-0.12-3.64}) \t 10^{-7}$ \\ 
			$B_c^+ \to B^+ \ov{K}^{*0}$   & $(3.09^{+0.69+0.43}_{-0.57-0.90}) \t 10^{-3}$ & $(9.61^{+2.14+1.35}_{-1.76-2.79}) \t 10^{-5}$ & $(4.19^{+0.84+0.00}_{-0.74-0.07}) \t 10^{-3}$ & $(1.30^{+0.26+0.00}_{-0.23-0.02}) \t 10^{-4}$	& $(8.20^{+0.99+6.52}_{-0.93-4.87}) \t 10^{-4}$ & $(2.55^{+0.31+2.03}_{-0.29-1.52}) \t 10^{-5}$ \\ 
			$B_c^+ \to B_s^0 \rho^+$      & $(3.46^{+0.53+0.25}_{-0.54-0.63}) \t 10^{-2}$ & $(2.59^{+0.39+0.19}_{-0.40-0.47}) \t 10^{-2}$ & $(4.12^{+0.58+0.00}_{-0.62-0.06}) \t 10^{-2}$ & $(3.08^{+0.44+0.00}_{-0.47-0.05}) \t 10^{-2}$	& $(1.67^{+0.15+0.68}_{-0.18-0.76}) \t 10^{-2}$ & $(1.25^{+0.11+0.51}_{-0.14-0.57}) \t 10^{-2}$ \\   \hline
			\multicolumn{7}{c}{CKM-suppressed ($\Delta b=0,~\Delta C=-1,~\Delta S=0$)}   \\
\hline \hline 
		\end{tabular}}
	\end{table}

	\begin{table}[ht]
		\caption{Branching ratios of bottom-changing $B_c \to PV$ decays for CKM-favored modes. For the definitions of Type-II, Type-II*, and Type-I, refer to the caption of Table~\ref{t3}.}
		\label{t9}
  \scalebox{0.92}{
		\begin{tabular}{ c | c  c | c  c | c c}
			\hline 	\hline
			\multirow{2}{*}{Decay}        & \multicolumn{2}{c|}{Type-II} & \multicolumn{2}{c|}{Type-II*} & \multicolumn{2}{c}{Type-I}  \\ 
			\cline{2-7}
			& $N_c = \infty$ & $N_c = 3$ & $N_c = \infty$	& $N_c = 3$	& $N_c = \infty$ & $N_c = 3$ \\ \hline
			\multicolumn{7}{c}{$\Delta b=-1,~\Delta C=-1,~\Delta S=0$}   \\ \hline  
			$B_c^+ \to \pi^+ J/\psi$   		 & $(1.65^{+0.31+0.10}_{-0.29-0.15}) \t 10^{-3}$ & $(1.40^{+0.27+0.09}_{-0.24-0.12}) \t 10^{-3}$ & $(1.66^{+0.32+0.10}_{-0.29-0.15}) \t 10^{-3}$ & $(1.41^{+0.27+0.09}_{-0.25-0.12}) \t 10^{-3}$ & $(1.23^{+0.09+0.04}_{-0.08-0.08}) \t 10^{-3}$ & $(1.05^{+0.07+0.04}_{-0.07-0.07}) \t 10^{-3}$ \\  
			$B_c^+ \to \ov{D}^0 D^{*+}$	   	 & $(4.10^{+0.83+2.94}_{-0.78-2.31}) \t 10^{-5}$ & $(7.79^{+1.57+5.59}_{-1.47-4.40}) \t 10^{-6}$ & $(3.54^{+0.66+2.67}_{-0.63-1.96}) \t 10^{-5}$ & $(6.73^{+1.26+5.07}_{-1.20-3.73}) \t 10^{-6}$ & $(2.44^{+0.58+4.81}_{-1.79-1.91}) \t 10^{-6}$ & $(4.64^{+1.11+9.13}_{-3.39-3.63}) \t 10^{-7}$ \\  
			$B_c^+ \to D^+ \ov{D}^{*0}$	   	 & $(3.94^{+2.05+1.29}_{-1.28-1.21}) \t 10^{-5}$ & $(7.48^{+3.90+2.45}_{-2.44-2.30}) \t 10^{-6}$ & $(3.27^{+1.64+1.11}_{-1.03-0.97}) \t 10^{-5}$ & $(6.22^{+3.12+2.11}_{-1.96-1.83}) \t 10^{-6}$ & $(3.13^{+1.92+1.37}_{-1.15-1.22}) \t 10^{-5}$ & $(5.95^{+3.65+2.61}_{-2.18-2.31}) \t 10^{-6}$ \\  
			$B_c^+ \to \eta_c \rho^+$        & $(3.91^{+0.13+0.11}_{-0.12-0.24}) \t 10^{-3}$ & $(3.33^{+0.11+0.10}_{-0.10-0.20}) \t 10^{-3}$ & $(3.96^{+0.12+0.11}_{-0.12-0.23}) \t 10^{-3}$ & $(3.37^{+0.10+0.09}_{-0.10-0.20}) \t 10^{-3}$ & $(3.90^{+0.13+0.12}_{-0.12-0.24}) \t 10^{-3}$ & $(3.32^{+0.11+0.10}_{-0.11-0.20}) \t 10^{-3}$ \\ \hline 
			\multicolumn{7}{c}{$\Delta b=-1,~\Delta C=0,~\Delta S=1$}   \\
\hline  \hline
		\end{tabular}}
	\end{table}

	\begin{table}[ht]
		\caption{Branching ratios of bottom-changing $B_c \to PV$ decays for CKM-suppressed modes. For the definitions of Type-II, Type-II*, and Type-I, refer to the caption of Table~\ref{t3}.}
		\label{t10}
  \scalebox{0.92}{
		\begin{tabular}{ c | c  c | c  c| c c}
			\hline 	\hline
			\multirow{2}{*}{Decay}        & \multicolumn{2}{c|}{Type-II} & \multicolumn{2}{c|}{Type-II*} & \multicolumn{2}{c}{Type-I} \\ 
			\cline{2-7}
			& $N_c = \infty$ & $N_c = 3$ & $N_c = \infty$	& $N_c = 3$	& $N_c = \infty$ & $N_c = 3$	 \\ \hline
			\multicolumn{7}{c}{$\Delta b=-1,~\Delta C=-1,~\Delta S=1$}   \\ \hline
			$B_c^+ \to K^+ J/\psi$     		 & $(1.23^{+0.23+0.07}_{-0.21-0.11}) \t 10^{-4}$ & $(1.05^{+0.20+0.06}_{-0.18-0.09}) \t 10^{-4}$ & $(1.25^{+0.24+0.07}_{-0.22-0.11}) \t 10^{-4}$ & $(1.06^{+0.20+0.06}_{-0.19-0.09}) \t 10^{-4}$ & $(9.19^{+0.65+0.32}_{-0.64-0.63}) \t 10^{-5}$ & $(7.82^{+0.56+0.27}_{-0.54-0.54}) \t 10^{-5}$ \\   
			$B_c^+ \to \ov{D}^0 D_s^{*+}$    & $(2.92^{+0.21+1.97}_{-0.46-1.57}) \t 10^{-6}$ & $(5.55^{+0.41+3.74}_{-0.87-2.98}) \t 10^{-7}$ & $(2.74^{+0.18+1.85}_{-0.39-1.40}) \t 10^{-6}$ & $(5.21^{+0.34+3.52}_{-0.73-2.67}) \t 10^{-7}$ & $(4.36^{+0.00+6.69}_{-1.70-3.23}) \t 10^{-7}$ & $(8.29^{+0.00+12.70}_{-3.24-6.13}) \t 10^{-8}$ \\   
			$B_c^+ \to D_s^+ \ov{D}^{*0}$    & $(4.08^{+1.36+0.51}_{-1.19-0.51}) \t 10^{-6}$ & $(7.76^{+2.58+0.96}_{-2.26-0.96}) \t 10^{-7}$ & $(3.43^{+1.14+0.48}_{-0.98-0.46})\t 10^{-6}$ & $(6.53^{+2.16+0.91}_{-1.86-0.87}) \t 10^{-7}$ & $(3.54^{+1.36+0.62}_{-1.13-0.63}) \t 10^{-6}$ & $(6.72^{+2.59+1.17}_{-2.14-1.20}) \t 10^{-7}$ \\
			$B_c^+ \to \eta_c K^{*+}$  		 & $(1.94^{+0.06+0.05}_{-0.06-0.11}) \t 10^{-4}$ & $(1.65^{+0.05+0.05}_{-0.05-0.10}) \t 10^{-4}$ & $(1.95^{+0.06+0.05}_{-0.06-0.11}) \t 10^{-4}$ & $(1.66^{+0.05+0.05}_{-0.05-0.10}) \t 10^{-4}$ & $(1.93^{+0.06+0.06}_{-0.06-0.12}) \t 10^{-4}$ & $(1.64^{+0.05+0.05}_{-0.05-0.10}) \t 10^{-4}$ \\ \hline
			\multicolumn{7}{c}{$\Delta b=-1,~\Delta C=0,~\Delta S=0$}   \\
\hline 
			\hline 
		\end{tabular}}
	\end{table}
	\begin{table}[ht]
		\caption{Branching ratios of bottom-changing $B_c \to PV$ decays for CKM-doubly-suppressed modes. For the definitions of Type-II, Type-II*, and Type-I, refer to the caption of Table~\ref{t3}.}
		\label{t11}
  \scalebox{0.92}{
		\begin{tabular}{ c | c  c | c  c| c c}
			\hline 	\hline
			\multirow{2}{*}{Decay}        & \multicolumn{2}{c|}{Type-II} & \multicolumn{2}{c|}{Type-II*} & \multicolumn{2}{c}{Type-I} \\ 
			\cline{2-7}
			& $N_c = \infty$ & $N_c = 3$ & $N_c = \infty$	& $N_c = 3$	& $N_c = \infty$ &  $N_c = 3$ \\ \hline
			\multicolumn{7}{c}{$\Delta b=-1,~\Delta C=1,~\Delta S=1$}   \\ \hline
			$B_c^+ \to {D}^0 D_s^{*+}$  & $(3.06^{+2.22+0.60}_{-1.20-0.63}) \t 10^{-6}$ & $(6.55^{+3.06+2.38}_{-2.02-2.22}) \t 10^{-6}$ & $(2.32^{+1.66+0.46}_{-0.90-0.43}) \t 10^{-6}$ & $(5.39^{+2.38+2.03}_{-1.59-1.74}) \t 10^{-6}$	& $(3.47^{+2.86+1.13}_{-1.43-1.19}) \t 10^{-6}$ & $(4.34^{+2.45+2.21}_{-1.57-1.85}) \t 10^{-6}$ \\  
			$B_c^+ \to D_s^+ {D}^{*0}$  & $(5.69^{+0.91+5.34}_{-0.88-3.88}) \t 10^{-6}$ & $(1.10^{+0.24+0.70}_{-0.22-0.58}) \t 10^{-5}$ & $(4.79^{+0.66+4.74}_{-0.67-3.18}) \t 10^{-6}$ & $(9.27^{+1.86+6.27}_{-1.76-4.76}) \t 10^{-6}$	& $(5.46^{+297.99+139.86}_{-0.00-0.00}) \t 10^{-9}$ & $(9.44^{+0.50+12.42}_{-4.32-5.89}) \t 10^{-7}$ \\   \hline
			\multicolumn{7}{c}{$\Delta b=-1,~\Delta C=1,~\Delta S=0$}   \\ \hline 
			$B_c^+ \to D^+ {D}^{*0}$    & $(2.06^{+0.24+1.73}_{-0.30-1.29}) \t 10^{-7}$ & $(3.60^{+0.83+2.43}_{-0.73-1.95}) \t 10^{-7}$ & $(1.80^{+0.18+1.58}_{-0.25-1.11}) \t 10^{-7}$ & $(3.09^{+0.66+2.20}_{-0.59-1.64}) \t 10^{-7}$	& $(5.52^{+54.35+100.42}_{-0.00-0.00}) \t 10^{-10}$ & $(3.36^{+0.16+4.98}_{-1.57-2.32}) \t 10^{-8}$ \\
			$B_c^+ \to {D}^0 D^{*+}$    & $(1.93^{+1.21+0.42}_{-0.71-0.42}) \t 10^{-7}$ & $(3.48^{+1.70+1.27}_{-1.09-1.17}) \t 10^{-7}$ & $(1.58^{+0.96+0.35}_{-0.56-0.32}) \t 10^{-7}$ & $(2.90^{+1.36+1.10}_{-0.88-0.94}) \t 10^{-7}$	& $(2.30^{+1.72+0.81}_{-0.92-0.82}) \t 10^{-7}$ & $(2.38^{+1.33+1.13}_{-0.84-0.96}) \t 10^{-7}$ \\ 	\hline \hline
		\end{tabular}}
	\end{table}

\end{turnpage}
	\begin{table}[ht]
		\caption{Branching ratios of bottom-conserving $B_c \to PV$ decays as predicted in other models. }
		\label{t12}
		\scalebox{0.92}{
			\begin{tabular}{ c | c c c c }
				\hline 	\hline
				{Decay}        & {RIQM~\cite{Naimuddin:2012dy}} & {RCQM\cite{Ivanov:2006ni}} & {RQM\cite{Ebert:2003wc}} & {QCDF~\cite{Sun:2015exa}} \\ \hline
				$B_c^+ \to \pi^+ B_s^{*0}$	  & $8.61 \t 10^{-2}$ & $2.1 \t 10^{-2}$ & $1.6 \t 10^{-2}$  & $-$ \\ 
				$B_c^+ \to \ov{K}^0 B^{*+}$   & $2.26 \t 10^{-2}$ & $8.8 \t 10^{-4}$ & $1.1 \t 10^{-3}$  & $-$ \\ 
				$B_c^+ \to B^+ \ov{K}^{*0}$   & $1.83 \t 10^{-2}$ & $1.1 \t 10^{-3}$ & $9.0 \t 10^{-4}$  & $3.72 \t 10^{-3}$ \\ 
				$B_c^+ \to B_s^0 \rho^+$      & $9.97 \t 10^{-2}$ & $2.3 \t 10^{-2}$ & $1.4 \t 10^{-2}$  & $4.44 \t 10^{-2}$ \\   \hline
				$B_c^+ \to K^+ B_s^{*0}$      & $4.99 \t 10^{-3}$ & $1.3 \t 10^{-3}$ & $1.1 \t 10^{-3}$  & $-$ \\
				$B_c^+ \to \pi^+ B^{*0}$      & $5.55 \t 10^{-3}$ & $5.7 \t 10^{-4}$ & $2.6 \t 10^{-4}$  & $-$ \\
				$B_c^+ \to \pi^0 B^{*+}$      & $4.70 \t 10^{-4}$ & $2.0 \t 10^{-5}$ & $1.0 \t 10^{-5}$  & $-$ \\
				$B_c^+ \to B^+ \rho^0$        & $1.06 \t 10^{-3}$ & $7.1 \t 10^{-5}$ & $5.0 \t 10^{-5}$  & $2.86 \t 10^{-4}$ \\
				$B_c^+ \to B^+ \omega$        & $-$ 			  & $-$ 			 & $-$ 				 & $2.05 \t 10^{-4}$ 	\\
				$B_c^+ \to B^0 \rho^+$        & $1.30 \t 10^{-2}$ & $2.0 \t 10^{-3}$ & $1.3 \t 10^{-3}$  & $5.32 \t 10^{-3}$ \\
				$B_c^+ \to B_s^0 K^{*+}$      & $4.00 \t 10^{-4}$ & $1.1 \t 10^{-4}$ & $3.0 \t 10^{-5}$  & $1.25 \t 10^{-4}$ \\ \hline
				$B_c^+ \to K^+ B^{*0}$	      & $3.90 \t 10^{-4}$ & $3.6 \t 10^{-5}$ & $4.0 \t 10^{-5}$  & $-$ \\ 
				$B_c^+ \to B^+ K^{*0}$	      & $-$ 			  & $-$ 			 & $-$ 				 & $1.07 \t 10^{-5}$ 	\\
				$B_c^+ \to B^0 K^{*+}$	      & $3.20 \t 10^{-5}$ & $4.8 \t 10^{-5}$ & $4.0 \t 10^{-5}$  & $1.06 \t 10^{-4}$ \\ 	\hline \hline  
		\end{tabular}}
	\end{table}
	\begin{table}[ht]
		\caption{Branching ratios of bottom-changing $B_c^+ \to D_{(s)}^{(*)+}\ov{D}^{(*)0}$ and $D_{(s)}^{(*)+}{D}^{(*)0}$ decays as predicted in other models. }
		\label{t13}
  \scalebox{0.92}{
		\begin{tabular}{ c |c c  c c }
			\hline 	\hline
			{Decay}        & {RIQM~\cite{Naimuddin:2012dy}} & {pQCD~\cite{Rui:2012qq}} &{RCQM\cite{Ivanov:2006ni}} & {PDG~\cite{Workman:2022ynf}} \\  \hline
			$B_c^+ \to {D}^0 D_s^{*+}$	& $4.18 \t 10^{-7}$ & $2.5 \t 10^{-6}$ & $9.3 \t 10^{-6}$ & $<9.0 \t 10^{-4}$ \\  
			$B_c^+ \to D_s^+ {D}^{*0}$  	& $2.25 \t 10^{-7}$ & $1.9 \t 10^{-6}$ & $1.3 \t 10^{-6}$ & $<6.6 \t 10^{-4}$ \\  
			$B_c^+ \to D^+ {D}^{*0}$    	& $8.00 \t 10^{-9}$ & $7.0 \t 10^{-8}$ & $5.2 \t 10^{-8}$ & $<3.7 \t 10^{-4}$ \\
			$B_c^+ \to {D}^0 D^{*+}$    	& $1.80 \t 10^{-8}$ & $9.0 \t 10^{-8}$ & $4.4 \t 10^{-7}$ & $<2.0 \t 10^{-4}$ \\ 
			$B_c^+ \to \ov{D}^0 D_s^{*+}$   		& $3.15 \t 10^{-7}$ & $7.0 \t 10^{-8}$ & $6.5 \t 10^{-7}$ & $<5.3 \t 10^{-4}$  \\   
			$B_c^+ \to D_s^+ \ov{D}^{*0}$   		& $4.91 \t 10^{-7}$ & $2.6 \t 10^{-7}$ & $2.4 \t 10^{-6}$ & $<4.6 \t 10^{-4}$ \\
			$B_c^+ \to \ov{D}^0 D^{*+}$	   		& $1.60 \t 10^{-6}$ & $1.2 \t 10^{-6}$ & $8.8 \t 10^{-6}$ & $<3.8 \t 10^{-4}$  \\  
			$B_c^+ \to D^+ \ov{D}^{*0}$	   		& $2.61 \t 10^{-5}$ & $3.4 \t 10^{-6}$ & $3.8 \t 10^{-5}$ & $<6.5 \t 10^{-4}$  \\  \hline \hline 
		\end{tabular}}
	\end{table}
	
	\begin{table}[ht]
		\caption{Branching ratios of bottom-changing $B_c \to PV$ decays involving one charmonium ($\eta_c(J/\psi$) state as predicted in other models. }
		\label{t14}
  \scalebox{0.92}{
		\begin{tabular}{ c | c c c c}
			\hline 	\hline
			{Decay}        &  {CLFQM (Type-I)\cite{Zhang:2023ypl}} & {RIQM~\cite{Nayak:2022qaq}} & {pQCD~\cite{Rui:2014tpa}} &{RCQM\cite{Ivanov:2006ni}} \\  \hline
			$B_c^+ \to D_s^+ J/\psi$   		& $6.09 \t 10^{-3}$ & $1.15 \t 10^{-3}$ & $8.05 \t 10^{-3}$ & $3.4 \t 10^{-3}$ \\
			$B_c^+ \to D^+ J/\psi$   		& $2.00 \t 10^{-4}$ & $3.69 \t 10^{-5}$ & $2.80 \t 10^{-4}$ & $1.5 \t 10^{-4}$ \\  
			$B_c^+ \to K^+ J/\psi$     		& $1.60 \t 10^{-4}$ & $3.00 \t 10^{-5}$ & $1.90 \t 10^{-4}$ & $1.3 \t 10^{-4}$ \\
			$B_c^+ \to \pi^+ J/\psi$   		& $1.97 \t 10^{-3}$ & $3.80 \t 10^{-4}$ & $2.33 \t 10^{-3}$ & $1.7 \t 10^{-3}$ \\   \hline
			$B_c^+ \to \eta_c D_s^{*+}$  	& $6.97 \t 10^{-3}$ & $2.16 \t 10^{-3}$ & $1.65 \t 10^{-2}$ & $3.7 \t 10^{-3}$ \\ 
			$B_c^+ \to \eta_c D^{*+}$	    & $3.10 \t 10^{-4}$ & $7.60 \t 10^{-5}$ & $5.80 \t 10^{-4}$ & $1.9 \t 10^{-4}$ \\
			$B_c^+ \to \eta_c K^{*+}$  		& $3.40 \t 10^{-4}$ & $6.30 \t 10^{-5}$ & $5.70 \t 10^{-4}$ & $2.5 \t 10^{-4}$ \\ 
			$B_c^+ \to \eta_c \rho^+$  		& $6.01 \t 10^{-3}$ & $1.20 \t 10^{-3}$ & $9.83 \t 10^{-3}$ & $4.5 \t 10^{-3}$ \\	\hline \hline 
		\end{tabular}}
	\end{table}
	
	\newpage
	\FloatBarrier
	\begin{figure}
		\centering
		\begin{subfigure}[b]{0.48\textwidth}
			\centering
			\includegraphics[width=\textwidth]{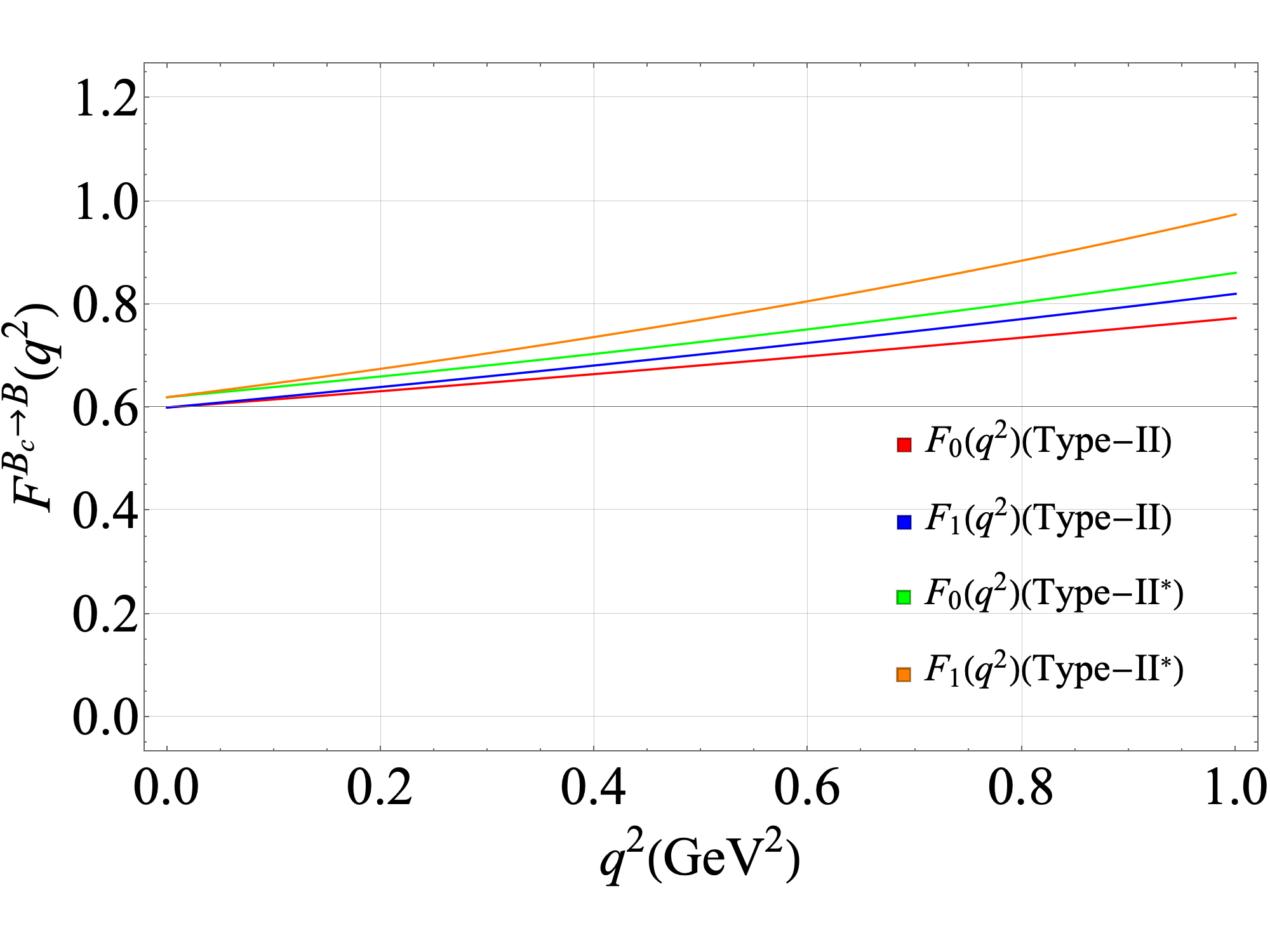}
			\caption{$B_c \to B$ transition}
			\label{f2a}
		\end{subfigure}
		\hfill
		\begin{subfigure}[b]{0.48\textwidth}
			\centering
			\includegraphics[width=\textwidth]{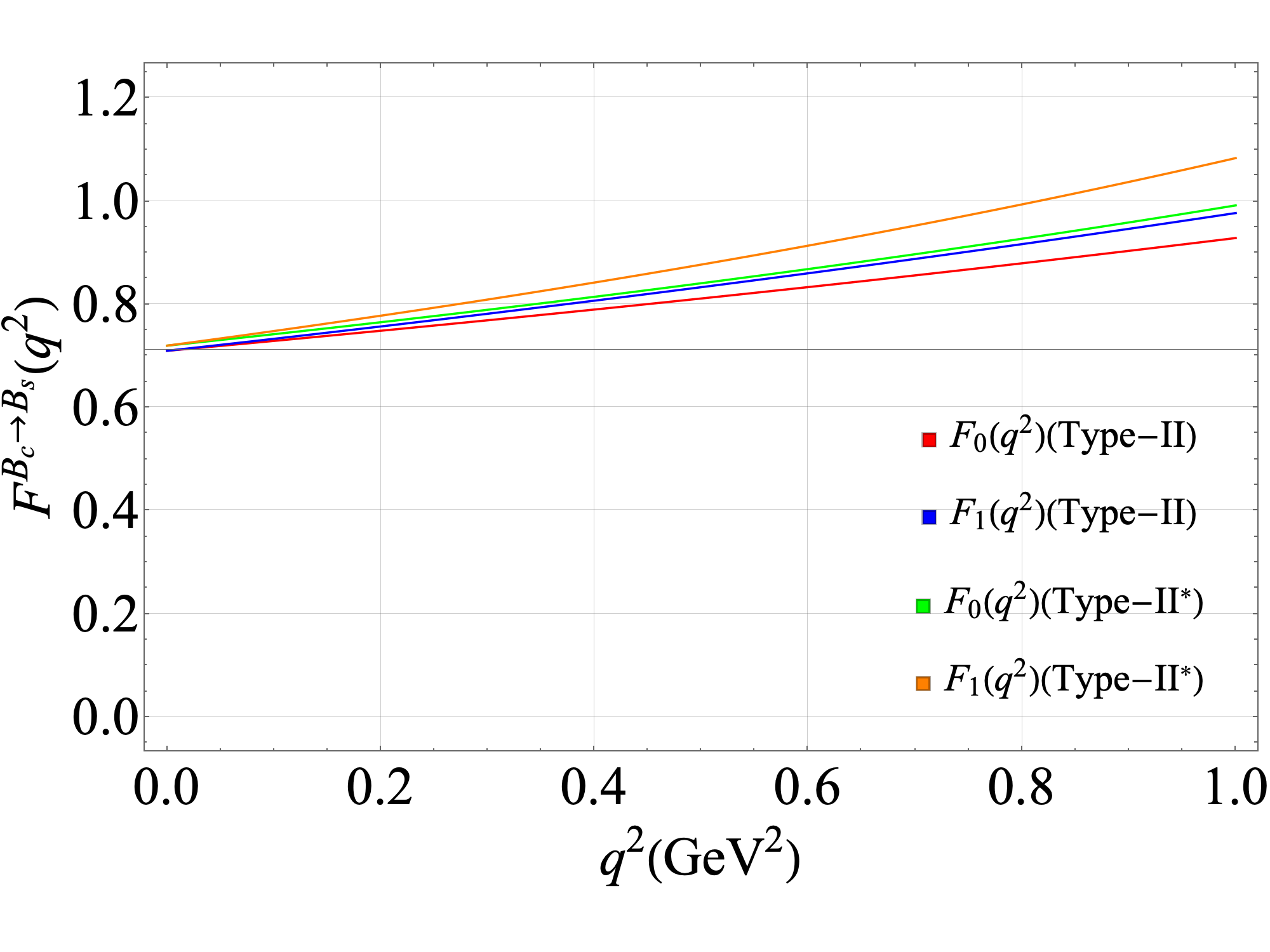}
			\caption{$B_c \to B_s$ transition}
			\label{f2b}
		\end{subfigure}
		\caption{$\q2$ dependence of bottom-conserving $B_c \to P$ form factors in Type-II (Type-II*) CLFQM using Eq.~\eqref{e37} (Eq.~\eqref{e38}).}
		\label{f2}
	\end{figure}
	
	\begin{figure}
		\centering
		\begin{subfigure}[b]{0.48\textwidth}
			\centering
			\includegraphics[width=\textwidth]{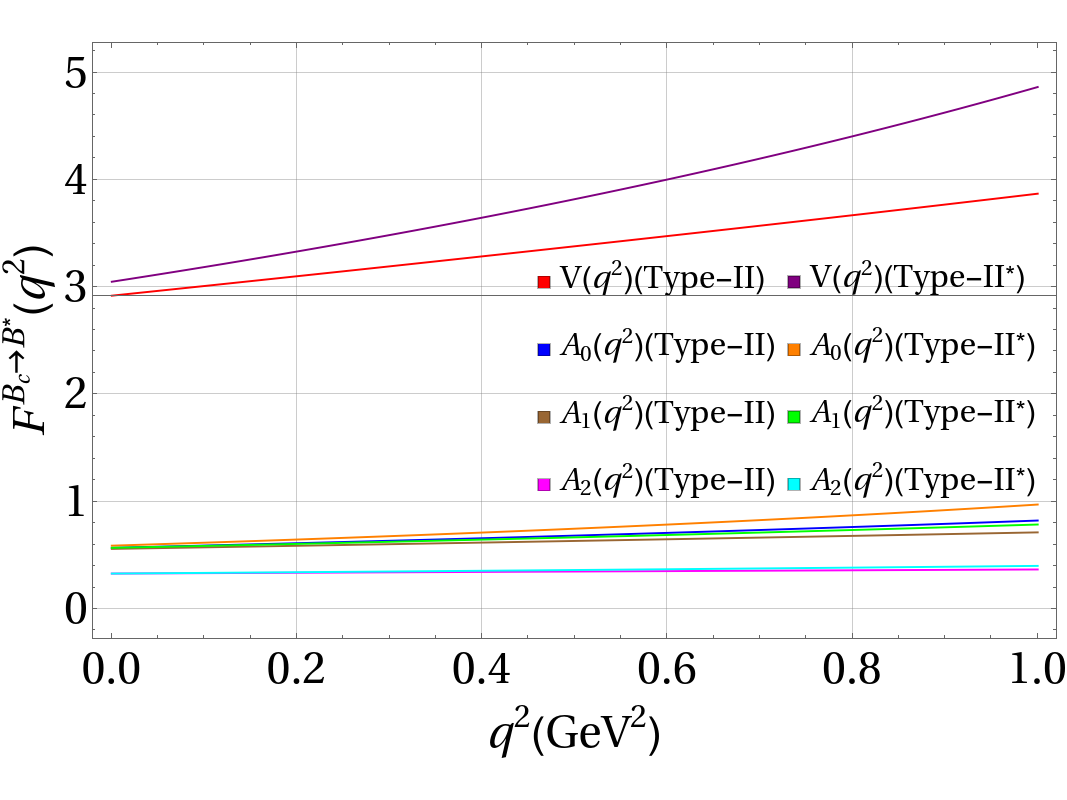}
			\caption{$B_c \to B^*$ transition}
			\label{f3a}
		\end{subfigure}
		\hfill
		\begin{subfigure}[b]{0.48\textwidth}
			\centering
			\includegraphics[width=\textwidth]{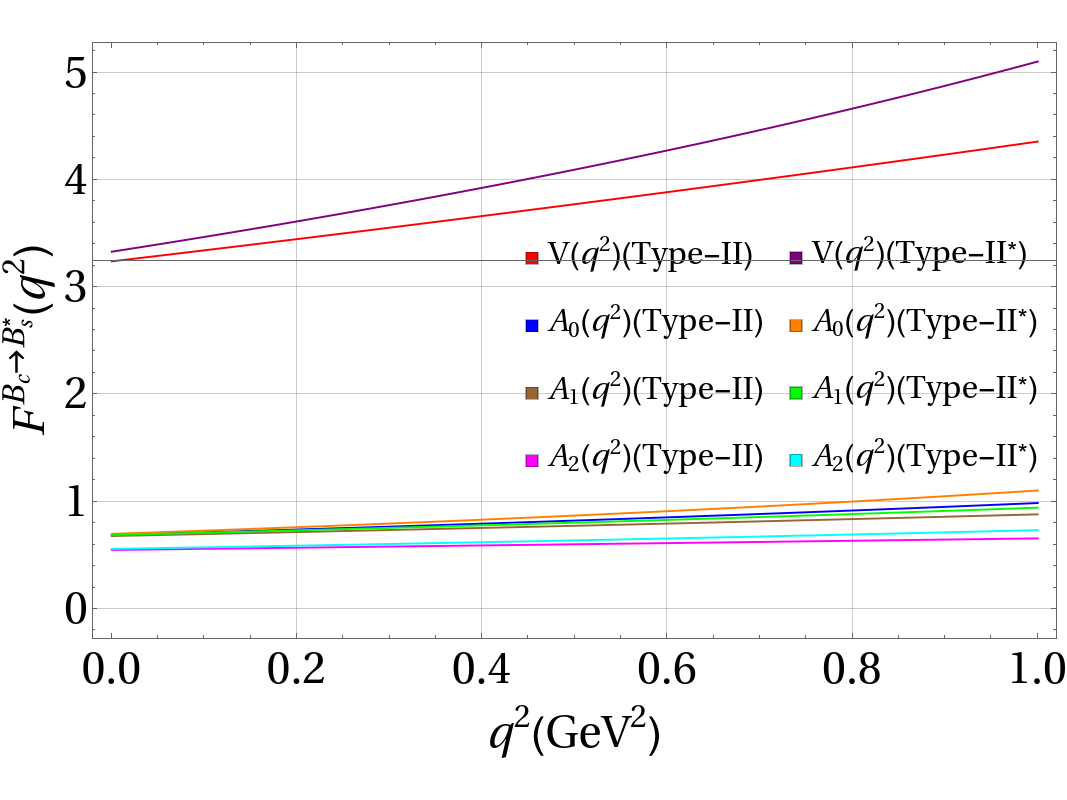}
			\caption{$B_c \to B_s^*$ transition}
			\label{f3b}
		\end{subfigure}
		\caption{$\q2$ dependence of bottom-conserving $B_c \to V$ form factors in Type-II (Type-II*) CLFQM using Eq.~\eqref{e37} (Eq.~\eqref{e38}).}
		\label{f3}
	\end{figure}

	\begin{figure}
		\centering
		\begin{subfigure}[b]{0.48\textwidth}
			\centering
			\includegraphics[width=\textwidth]{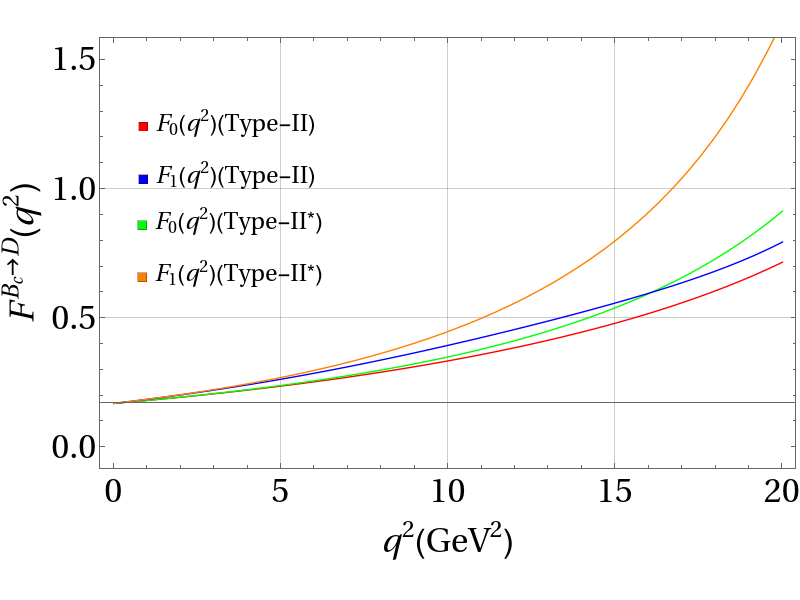}
			\caption{$B_c \to D$ transition}
			\label{f4a}
		\end{subfigure}
		\hfill
		\begin{subfigure}[b]{0.48\textwidth}
			\centering
			\includegraphics[width=\textwidth]{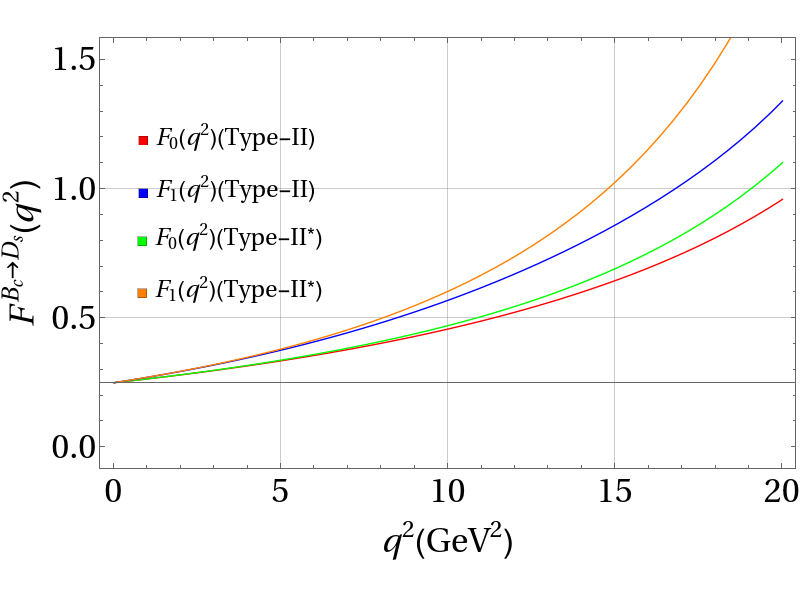}
			\caption{$B_c \to D_s$ transition}
			\label{f4b}
		\end{subfigure}
		\hfill
		\begin{subfigure}[b]{0.48\textwidth}
			\centering
			\includegraphics[width=\textwidth]{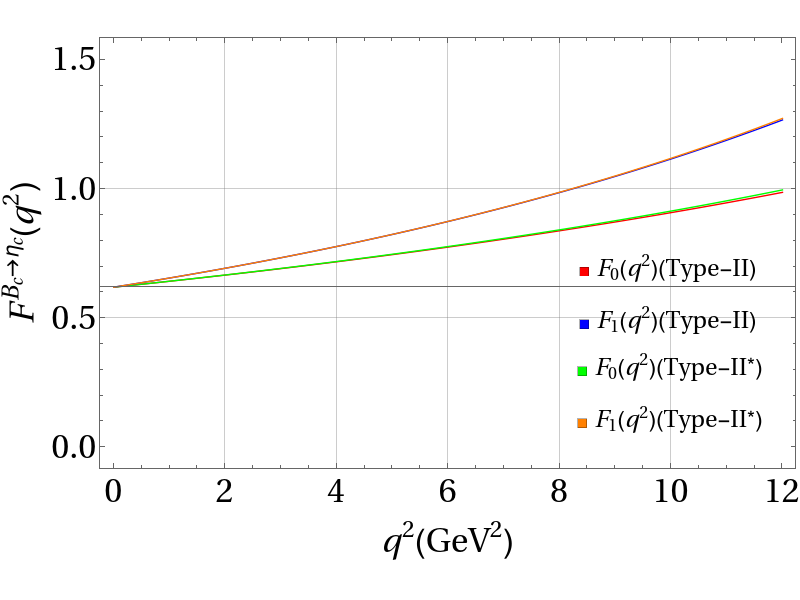}
			\caption{$B_c \to \eta_c$ transition}
			\label{f4c}
		\end{subfigure}
		\caption{$\q2$ dependence of bottom-changing $B_c \to P$ form factors in Type-II (Type-II*) CLFQM using Eq.~\eqref{e37} (Eq.~\eqref{e38}).}
		\label{f4}
	\end{figure}
	
	\begin{figure}
		\centering
		\begin{subfigure}[b]{0.48\textwidth}
			\centering
			\includegraphics[width=\textwidth]{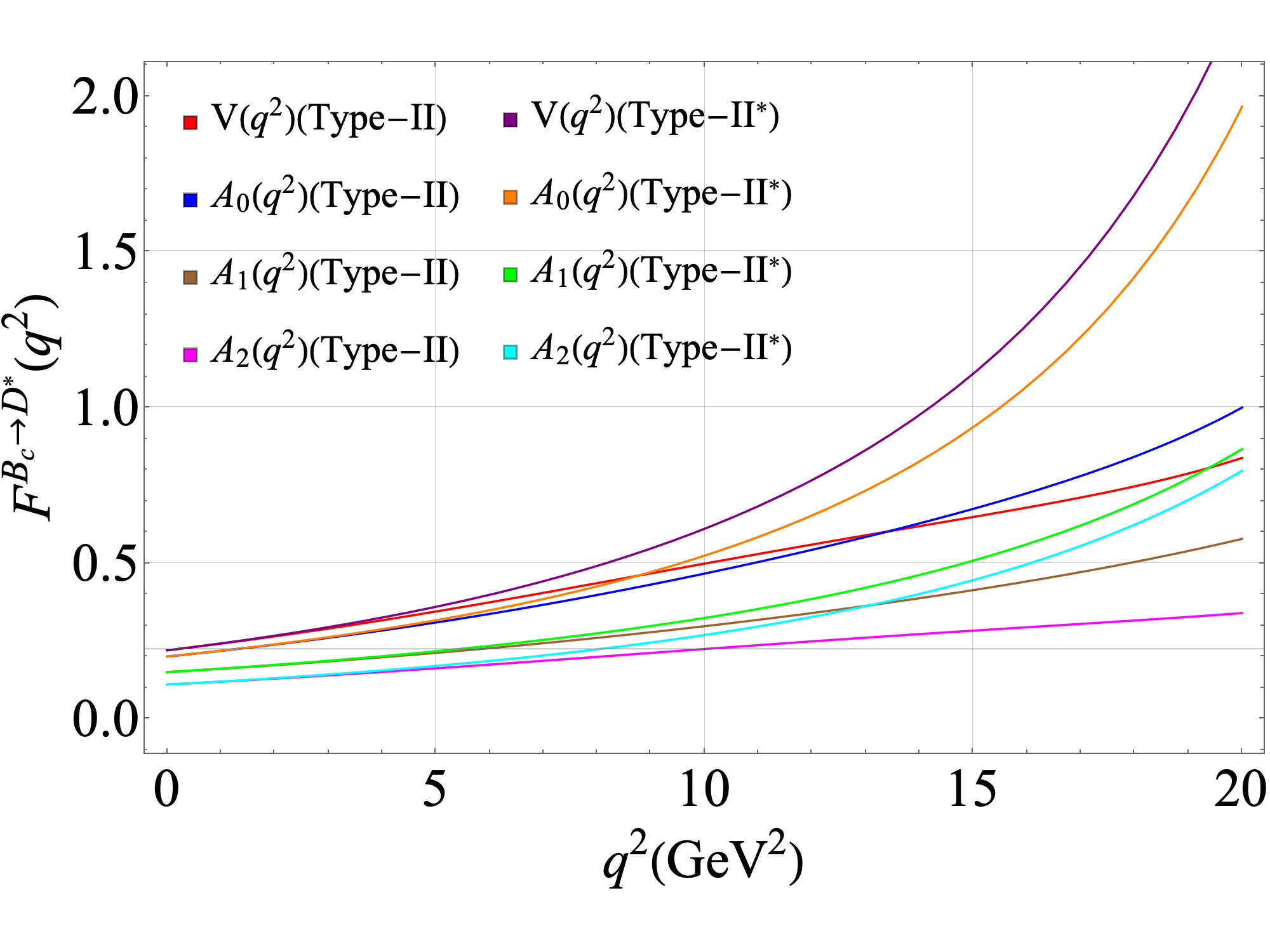}
			\caption{$B_c \to D^*$ transition}
			\label{f5a}
		\end{subfigure}
		\hfill
		\begin{subfigure}[b]{0.48\textwidth}
			\centering
			\includegraphics[width=\textwidth]{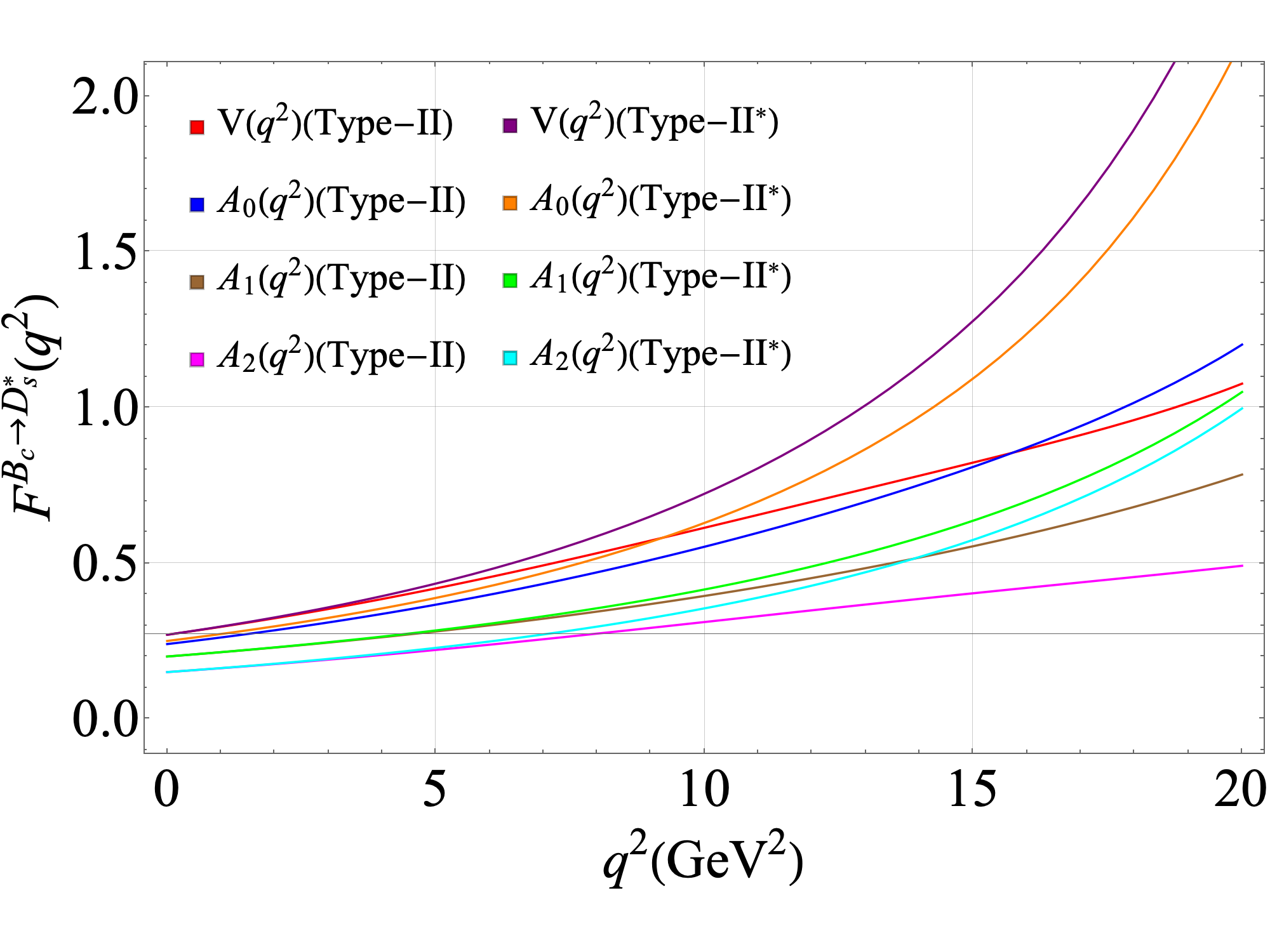}
			\caption{$B_c \to D_s^*$ transition}
			\label{f5b}
		\end{subfigure}
		\hfill
		\begin{subfigure}[b]{0.48\textwidth}
			\centering
			\includegraphics[width=\textwidth]{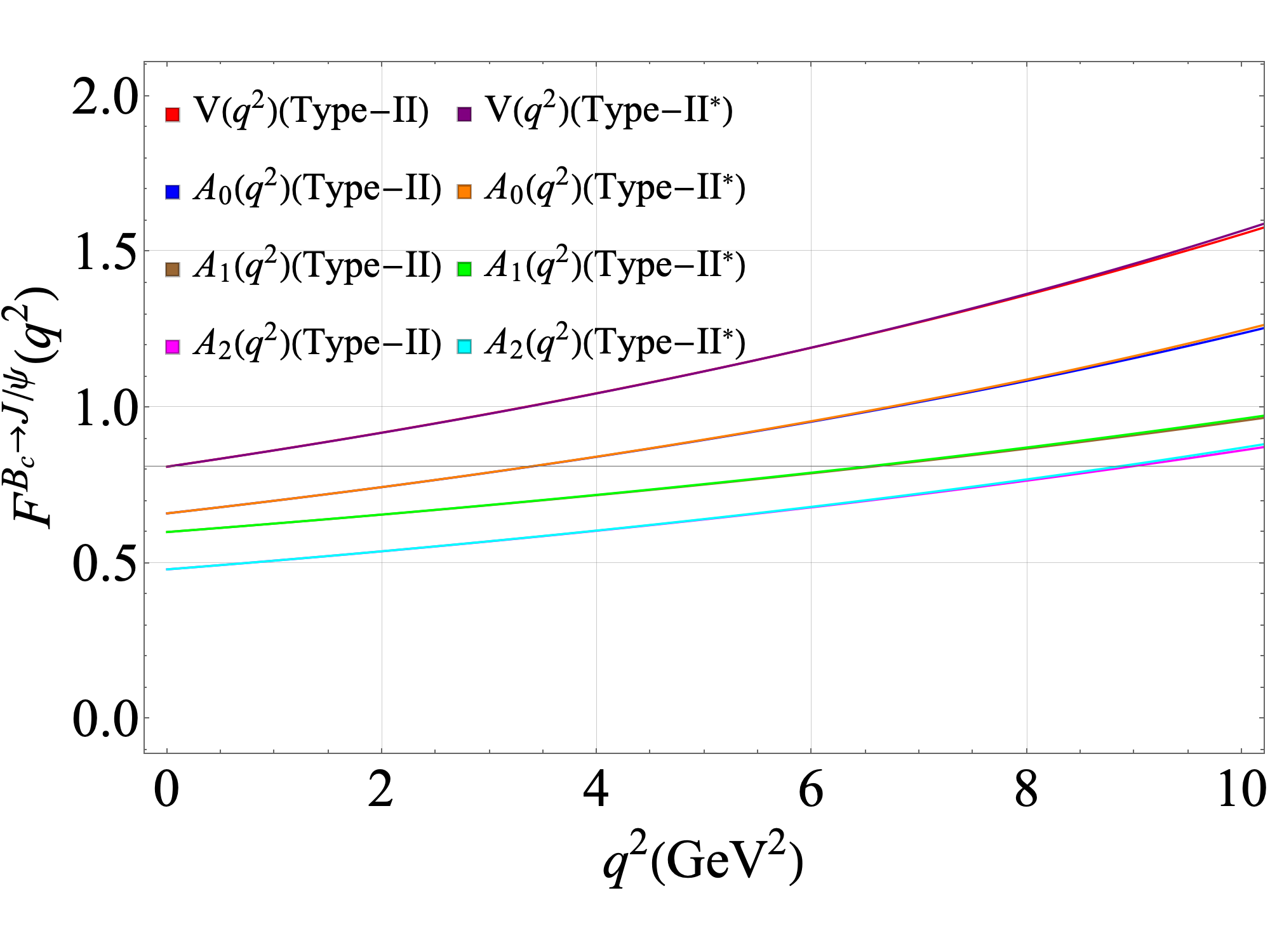}
			\caption{$B_c \to J/\psi$ transition}
			\label{f5c}
		\end{subfigure}
		\caption{$\q2$ dependence of bottom-changing $B_c \to V$ form factors in Type-II (Type-II*) CLFQM using Eq.~\eqref{e37} (Eq.~\eqref{e38}).}
		\label{f5}
	\end{figure}
	
	\begin{figure}
		\centering
		\begin{subfigure}[b]{0.47\textwidth}
			\centering
			\includegraphics[width=\textwidth]{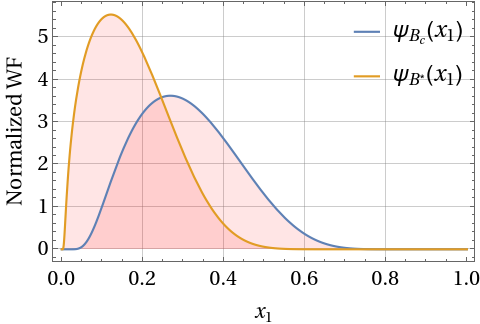}
			\caption{$B_c$ and $B^*$ (Overlap area $= 0.723$)}
			\label{f6a}
		\end{subfigure}
		\hfill
		\begin{subfigure}[b]{0.47\textwidth}
			\centering
			\includegraphics[width=\textwidth]{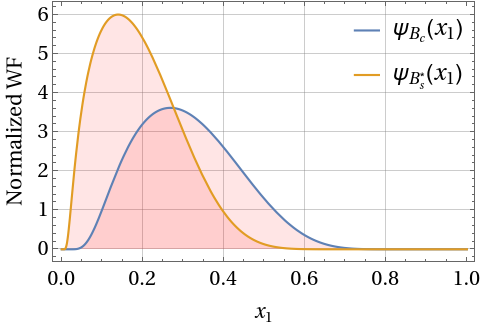}
			\caption{$B_c$ and $B_s^*$ (Overlap area $= 0.831$)}
			\label{f6b}
		\end{subfigure}
		\caption{Overlap plots of $B_c$ and $B^*,~B_s^*$ light-front wave function using Eq.~\eqref{e5}, in Type-II CLFQM. Note that overlap plots of $B_c$ and $B,~B_s$ wave function are similar to $B^*,~B_s^*$, with roughly $10\%$ increase in overlap area.}
		\label{f6}
	\end{figure}
	
	\begin{figure}
		\centering
		\begin{subfigure}[b]{0.47\textwidth}
			\centering
			\includegraphics[width=\textwidth]{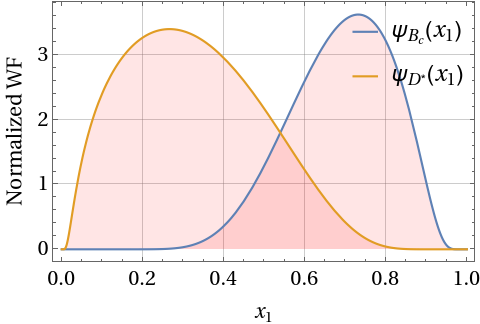}
			\caption{$B_c$ and $D^*$ (Overlap area $= 0.345$)}
			\label{f7a}
		\end{subfigure}
		\hfill
		\begin{subfigure}[b]{0.47\textwidth}
			\centering
			\includegraphics[width=\textwidth]{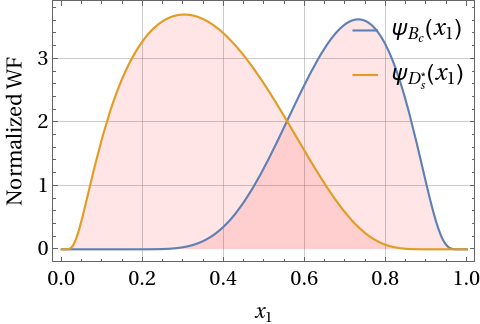}
			\caption{$B_c$ and $D_s^*$ (Overlap area $= 0.397$)}
			\label{f7b}
		\end{subfigure}
		\hfill
		\begin{subfigure}[b]{0.47\textwidth}
			\centering
			\includegraphics[width=\textwidth]{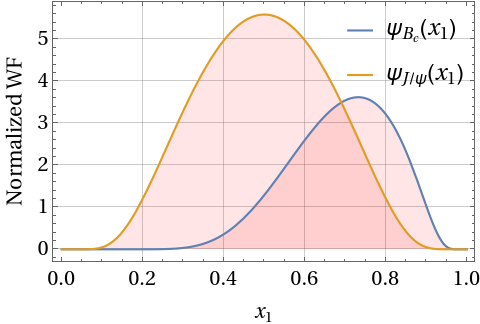}
			\caption{$B_c$ and $J/\psi$ (Overlap area $= 0.874$)}
			\label{f7c}
		\end{subfigure}
		\caption{Overlap plots of $B_c$ and $D^*,~D_s^*,~J/\psi$ light-front wave function using Eq.~\eqref{e5}, in Type-II CLFQM. Note that overlap plots of $B_c$ and $D,~D_s,~\eta_c$ wave function will be similar; however, we notice approximately $15\%$ and $26\%$ change between $B_c$ and $D,~\eta_c$, as well as between $B_c$ and $D_s$, respectively.}
		\label{f7}
	\end{figure}
	
	\begin{figure}
		\centering
		\begin{subfigure}[b]{0.48\textwidth}
			\centering
			\includegraphics[width=\textwidth]{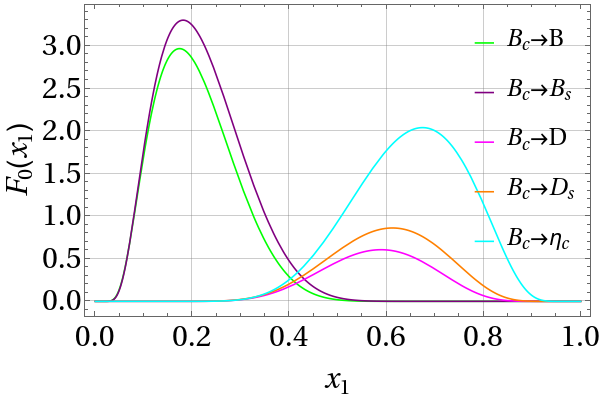}
			\caption{$F_0(x_1)$ versus $x_1$}
			\label{f8a}
		\end{subfigure}
		\hfill
		\begin{subfigure}[b]{0.48\textwidth}
			\centering
			\includegraphics[width=\textwidth]{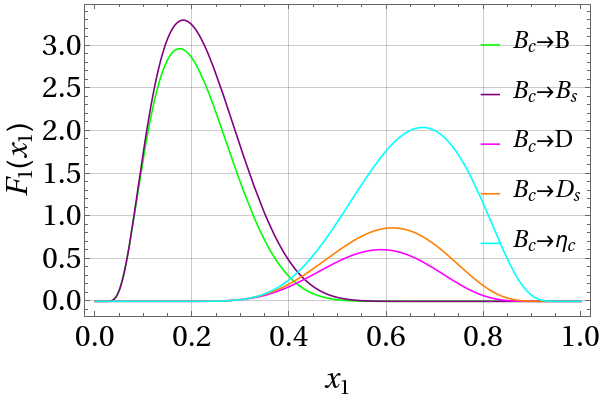}
			\caption{$F_1(x_1)$ versus $x_1$}
			\label{f8b}
		\end{subfigure}
		\caption{Dependence of form factor, $F(x_1)$ on $x_1$ for $B_c\to P$ transition at $\q2 \approx 0~\text{GeV}^2$, in Type-II CLFQM using Eq.~\eqref{e27}.}
		\label{f8}
	\end{figure}
	
	\begin{figure}
		\centering
		\begin{subfigure}[b]{0.48\textwidth}
			\centering
			\includegraphics[width=\textwidth]{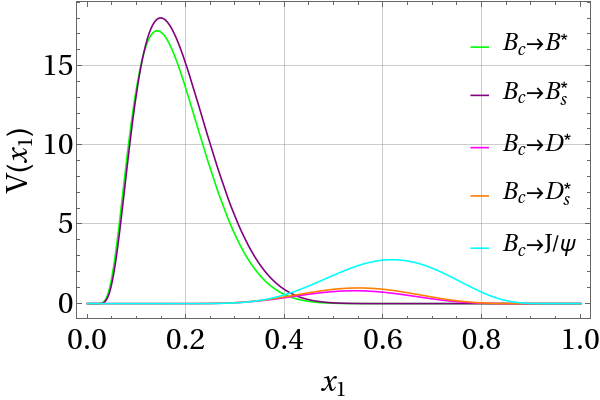}
			\caption{$V(x_1)$ versus $x_1$}
			\label{f9a}
		\end{subfigure}
		\hfill
		\begin{subfigure}[b]{0.48\textwidth}
			\centering
			\includegraphics[width=\textwidth]{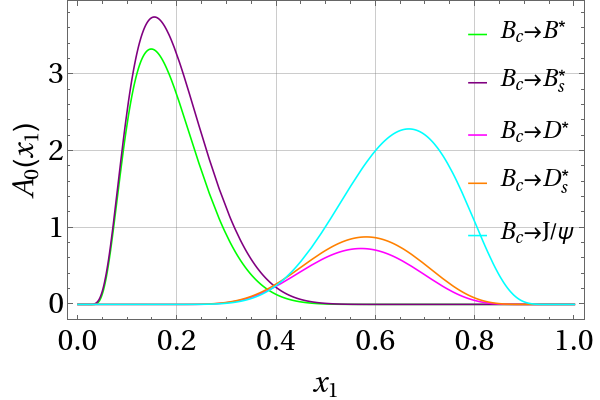}
			\caption{$A_0(x_1)$ versus $x_1$}
			\label{f9b}
		\end{subfigure}
		\hfill
		\begin{subfigure}[b]{0.48\textwidth}
			\centering
			\includegraphics[width=\textwidth]{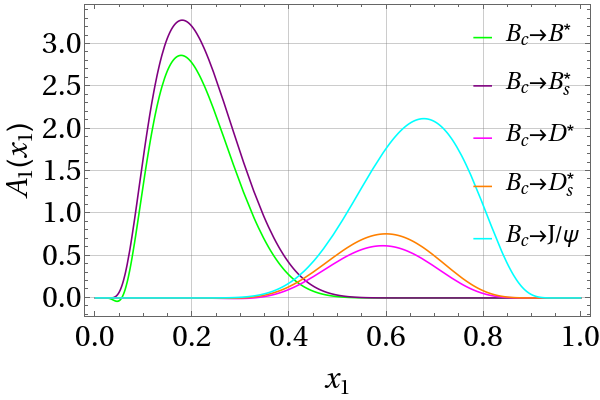}
			\caption{$A_1(x_1)$ versus $x_1$}
			\label{f9c}
		\end{subfigure}
		\hfill
		\begin{subfigure}[b]{0.48\textwidth}
			\centering
			\includegraphics[width=\textwidth]{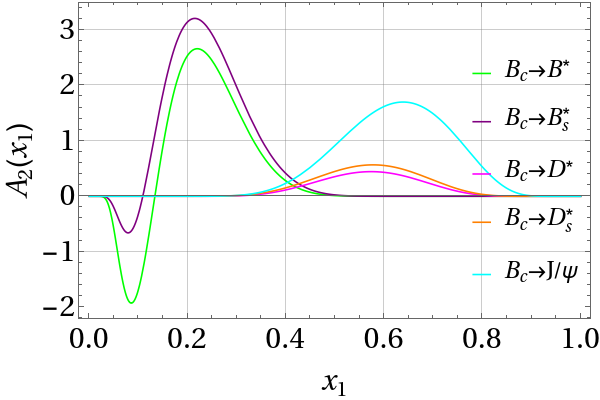}
			\caption{$A_2(x_1)$ versus $x_1$}
			\label{f9d}
		\end{subfigure}
		\caption{Dependence of form factor, $F(x_1)$ on $x_1$ for $B_c\to V$ transition at $\q2 \approx 0~\text{GeV}^2$, in Type-II CLFQM using Eq.~\eqref{e27}.}
		\label{f9}
	\end{figure}
	
	
	\begin{figure}
		\centering
		\begin{subfigure}[b]{0.48\textwidth}
			\centering
			\includegraphics[width=\textwidth]{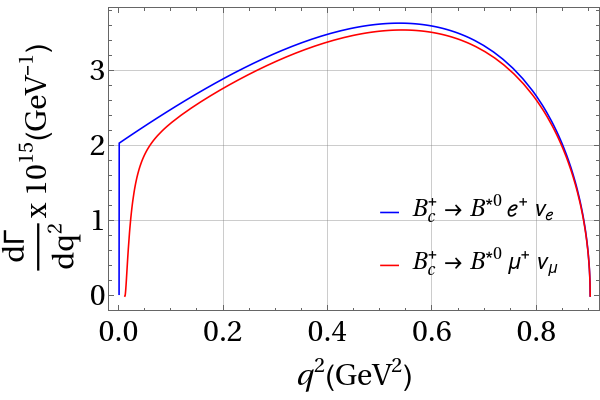}
			\caption{$B_c^+ \to B^* l^+ \nu_{l}$}
			\label{f10a}
		\end{subfigure}
		\hfill
		\begin{subfigure}[b]{0.48\textwidth}
			\centering
			\includegraphics[width=\textwidth]{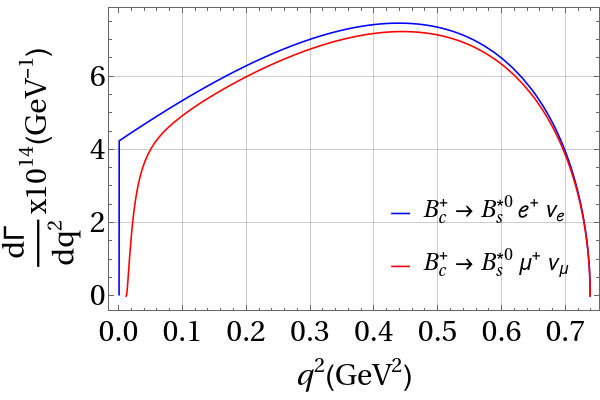}
			\caption{$B_c^+ \to B_s^* l^+ \nu_{l}$}
			\label{f10b}
		\end{subfigure}
		\hfill
		\begin{subfigure}[b]{0.48\textwidth}
			\centering
			\includegraphics[width=\textwidth]{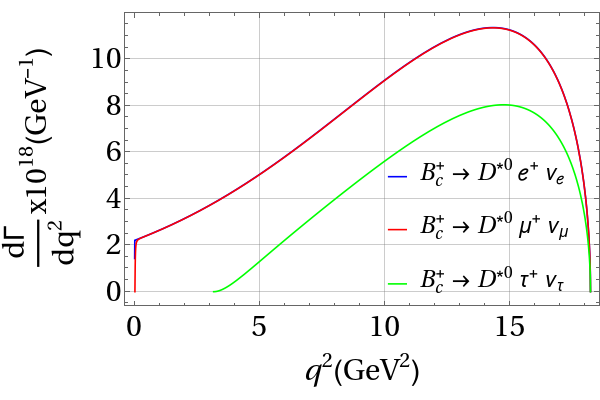}
			\caption{$B_c^+ \to D^* l^+ \nu_{l}$}
			\label{f10c}
		\end{subfigure}
		\hfill
		\begin{subfigure}[b]{0.48\textwidth}
			\centering
			\includegraphics[width=\textwidth]{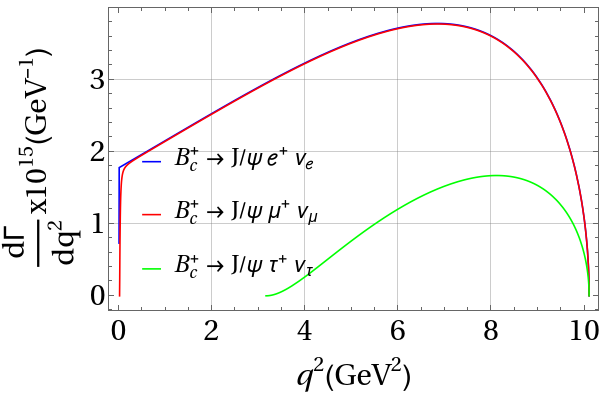}
			\caption{$B_c^+ \to J/\psi l^+ \nu_{l}$}
			\label{f10d}
		\end{subfigure}
		\caption{$\q2$ variation of differential decay rates of $B_c^+ \to V l^+ \nu_{l}$ decays in Type-II CLFQM using Eq.~\eqref{e41}.}
		\label{f10}
	\end{figure}
	
	\begin{figure}
		\centering
		\begin{subfigure}[b]{0.48\textwidth}
			\centering
			\includegraphics[width=\textwidth]{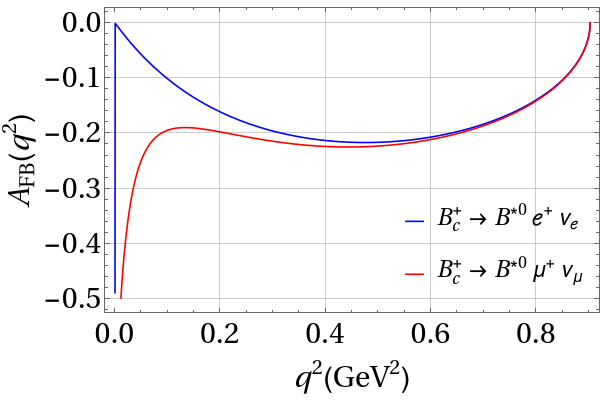}
			\caption{$B_c^+ \to B^* l^+ \nu_{l}$}
			\label{f11a}
		\end{subfigure}
		\hfill
		\begin{subfigure}[b]{0.48\textwidth}
			\centering
			\includegraphics[width=\textwidth]{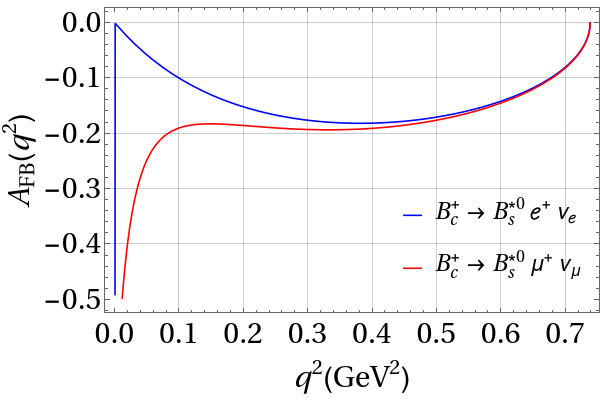}
			\caption{$B_c^+ \to B_s^* l^+ \nu_{l}$}
			\label{f11b}
		\end{subfigure}
		\hfill
		\begin{subfigure}[b]{0.48\textwidth}
			\centering
			\includegraphics[width=\textwidth]{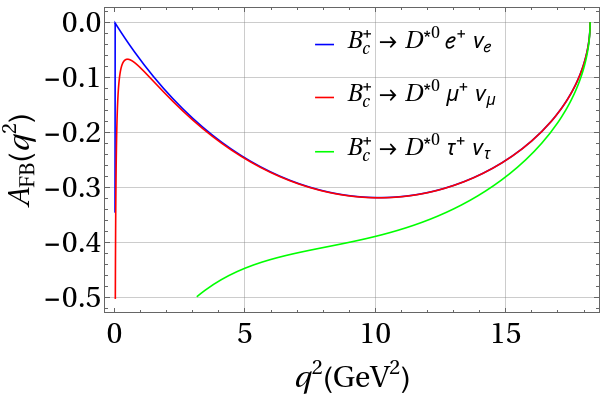}
			\caption{$B_c^+ \to D^* l^+ \nu_{l}$}
			\label{f11c}
		\end{subfigure}
		\hfill
		\begin{subfigure}[b]{0.48\textwidth}
			\centering
			\includegraphics[width=\textwidth]{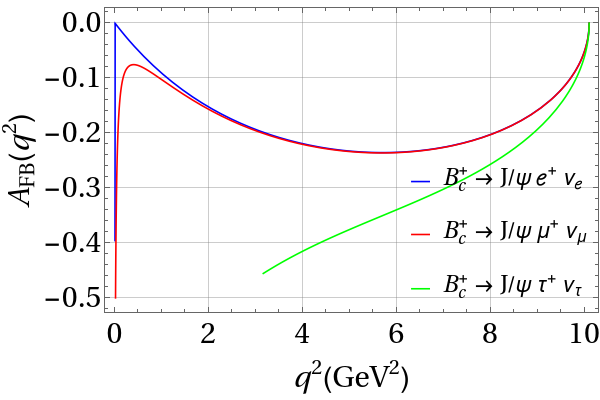}
			\caption{$B_c^+ \to J/\psi l^+ \nu_{l}$}
			\label{f11d}
		\end{subfigure}
		\caption{$\q2$ variation of forward-backward asymmetries of $B_c^+ \to V l^+ \nu_{l}$ decays in Type-II CLFQM using Eq.~\eqref{e50}.}
		\label{f11}
	\end{figure}

\end{document}